\documentstyle[12pt,epsf,epsfig,psfig]{article}
\def\J{$J/\psi$}
\def\j{J/\psi}
\def\X{$\chi_c$}
\def\x{\chi}
\def\P{$\psi'$}
\def\p{\psi'}
\def\U{$\Upsilon$}
\def\u{\Upsilon}
\def\C{c{\bar c}}

\def\b{b{\bar b}}

\def\e{\epsilon}
\def\L{\Lambda_{\rm QCD}}

\def\S{{\sigma}_{\C}}

\def\t{\tau}
\def\L{\Lambda_{\rm QCD}}

\def\be{\begin{equation}}
\def\ee{\end{equation}}

\def\lsim{\raise0.3ex\hbox{$<$\kern-0.75em\raise-1.1ex\hbox{$\sim$}}}
\def\gsim{\raise0.3ex\hbox{$>$\kern-0.75em\raise-1.1ex\hbox{$\sim$}}}

\def\NP{{ Nucl.\ Phys.\ }}
\def\PL{{ Phys.\ Lett.\ }}
\def\PR{{ Phys.\ Rev.\ }}

\def\PRL{{ Phys.\ Rev.\ Lett.\ }}

\def\ZP{{ Z.\ Phys.\ }}
\def\EP{{Eur. Phys. J.}}

\begin{document}

%\noindent 5.\ 7.\ 2000 \hfill 

\thispagestyle{empty}

\centerline{to appear in}

\centerline{\sl Reports on Progress in Physics}

\vskip 2cm

\centerline{\Large \bf COLOUR DECONFINEMENT}

\bigskip

\centerline{\Large \bf IN NUCLEAR COLLISIONS}

\vskip 1.5cm

\centerline{\large \bf Helmut Satz}

\bigskip

\centerline{Fakult\"at f\"ur Physik, Universit\"at Bielefeld}

\par

\centerline{D-33501 Bielefeld, Germany}

\vskip 2.5cm

\centerline{\large \bf Abstract:}

\bigskip

QCD predicts that strongly interacting matter will undergo a
transition from a state of hadronic constituents to a plasma 
of unbound quarks and gluons. We first survey the conceptual 
features of this transition and its description in finite temperature
lattice QCD, before we address its experimental investigation
through high energy nucleus-nucleus collisions. After considering
the conditions achievable in such collisions, we discuss the 
possible probes to check if the produced medium in its early
stages was indeed deconfined. We then elaborate the method 
that has emerged and the results which were obtained using the 
most extensively studied deconfinement probe, the suppression 
of charmonium production. In closing, we discuss possible
supporting information provided through the study of soft
hadronic probes.

\newpage

\thispagestyle{empty}

~~~~

\newpage

\thispagestyle{empty}

~~\vskip1.5cm

\centerline{\large \bf Contents}

\vskip 1.5cm

\leftskip=2.5cm

\noindent 1.\ Introduction

\medskip

\noindent 2.\ States of Matter in QCD

\par

2.1 From Hadronic Matter to Quark-Gluon Plasma

\par

2.2 Critical Behaviour in QCD

\par

2.3 Deconfinement and Chiral Symmetry Restoration

\par

2.4 Deconfinement and Percolation

\medskip

\noindent 3.\ Conditions in Nuclear Collisions

\par

3.1 High Energy Collisions and Nuclear Transparency

\par

3.2 Collision Evolution and Initial Conditions

\par

3.3 The Onset of Deconfinement

\medskip

\noindent 4.\ Probes of Primordial Matter

\par

4.1 Evolution Stages

\par

4.2 Quarkonium Dissociation

\par

4.3 Jet Quenching

\medskip

\noindent 5.\ \J~Suppression in Nuclear Collisions

\par

5.1 The Hadroproduction of Charmonium

\par

5.2 Pre-Resonance Suppression

\par

5.3 Anomalous \J~Suppression

\par

5.4 $P_T$ Dependence

\medskip

\noindent 6.\ Soft Hadronic Probes

\par

6.1 Retrospective Probes

\par

6.2 Hadronisation and Freeze-Out

\par

6.3 Parton Cascade Models

\medskip

\noindent 7.\ Summary

\newpage

\thispagestyle{empty}

~~~~

\newpage

\setcounter{page}{1}

\leftskip=0cm

\vfill \eject

\noindent{\bf 1.\ INTRODUCTION}

\bigskip

Over the past hundred years, our ideas about
the ultimate constituents of matter have undergone a considerable
evolution. Atoms were found to be divisible into electrons and
nuclei. Nuclei in turn consist of nucleons, bound together by strong
short-range forces.  With the advent of the basic theory of strong
interactions, quantum chromodynamics (QCD), has come the conviction that
nucleons - and more generally, all strongly interacting
elementary particles (hadrons) - are bound states of quarks.
Quarks are point-like and confined to ``their'' hadron by a binding
potential $V_0(r)$ which increases linearly with quark separation $r$,
\be
V_0(r)~ \sim~ \sigma r, \label{1.1}
\ee
where the string tension $\sigma$ measures the energy per unit
separation distance. Hence an infinite amount of energy would be needed
to isolate a quark; it cannot exist by itself, and it is therefore not
possible to split an isolated hadron into its quark constituents. The
elementary particles of strong interaction physics remain
elementary in the sense that they are the smallest entities with an
independent existence; however, they have become composite, since they
are bound states of quarks. What effect will this have on the states of
strongly interacting matter?

\par

\begin{figure}[htb]
\centerline{\psfig{file=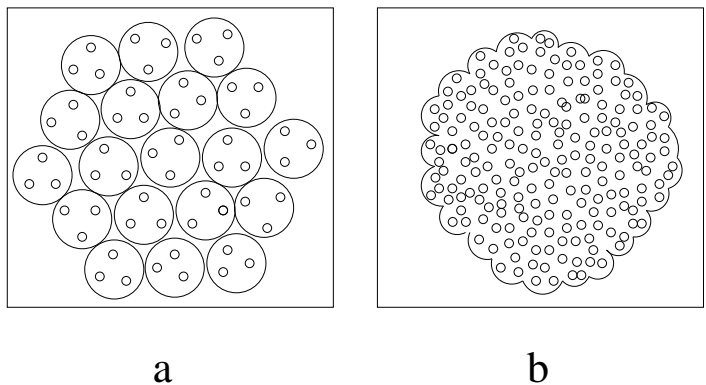,width=5cm,angle= -90}}
\caption{Strongly interacting matter as nuclear matter at a density of
closely packed nucleons (a) and as quark matter at much higher density (b).}
\label{1}
\end{figure}

To get a first idea of what the quark infrastructure of elementary
particles implies for the behaviour of matter at extreme density,
consider a very simple picture. If nucleons, with their intrinsic
spatial extension, were both elementary and incompressible, then a state
of close packing would constitute the high density limit of matter
(Fig.\ \ref{1}a). On the other hand, composite nucleons made up of
point-like quarks will start to overlap with increasing density, until
eventually each quark finds within its immediate vicinity a
considerable number of other quarks (Fig.\ \ref{1}b). It has no way to
identify which of these had been its partners in a specific nucleon at
some previous state of lower density. Beyond a certain point, the
concept of a hadron thus loses its meaning; at extreme density, we are
quite naturally led to a medium whose basic constituents are unbound
quarks.

\par

Ever since the big bang, our universe has been expanding; if we could
reverse time and let the film run backwards, we would find matter of
higher and higher densities. Up to nuclear density, we have
experimental information to guide our understanding of such matter.
What happens beyond this point? This question, the nature of matter
at extreme density, will be the central topic of this report. The
primordial matter in the very early universe provides a particularly
striking instance of such a system, but not the only one, as we shall
see shortly.

\par

On the theoretical side, studies of statistical QCD, in particular
through computer simulations, have confirmed the transition from
hadronic matter to a plasma of unbound quarks and gluons. This
led to the challenge to produce and study the predicted new state of
matter in the laboratory. The only way to achieve this, as far
as we know, is to collide two heavy nuclei and study the resulting
small and short-lived droplets of hot and dense medium. For such studies
it is essential to have viable probes which can determine if the
droplets in the early stages of nuclear collisions indeed consisted of
unbound quarks and gluons: we have to find signatures of colour
deconfinement.

\par

In the next section, we shall survey concepts and theoretical results
concerning the states of matter in strong interaction physics.
Following that, we shall address the thermodynamic conditions which
can be achieved in nuclear collisions. This will set the stage for the
main problem: probing experimentally whether high energy nuclear
collisions have produced deconfined matter and studying the properties
of the new phase. Finally we shall consider the possible support that
soft hadronic observables from later stages can provide to corroborate
the conclusions obtained using early hard probes.

\par

In view of the many-faceted nature of high energy nuclear physics,
it is perhaps appropriate to emphasize here that the aim of this survey
is colour deconfinement and its manifestation in such collisions. Many
other interesting aspects of the field will unfortunately but
unavoidably have to suffer from this restriction.

\bigskip~\medskip

\noindent{\bf 2.\ STATES OF MATTER IN QCD}

\bigskip

\noindent{\bf 2.1 From Hadronic Matter to Quark-Gluon Plasma}

\bigskip

Confinement is a {\sl long-range} feature, which prevents the isolation
of a {\sl single} quark. In the high density situation of Fig.\ \ref{1}a
each quark finds very close to it many others and is thus far from
isolated.
To see how the {\sl short-range} nature of dense matter overcomes
confinement, recall the effect of a dense medium on electric forces. In
a vacuum, two electric charges $e_0$ interact through the Coulomb
potential
\be
V_0(r)~=~ \left({e^2_0\over r}\right), \label{2.1}
\ee
where $r$ again denotes the separation distance. In a dense environment
of many other charges, the potential becomes screened,
\be
V(r)~=~\left({e^2_0\over r}\right)~ \exp(-\mu r); \label{2.2}
\ee
where $r_D=\mu^{-1}$ is the Debye screening radius of the medium; it
decreases as the charge density of the medium increases. Thus the
potential between two test charges a fixed distance apart becomes weaker
with increasing density. This occurs because the other charges in the
medium partially neutralize the test charges and thereby shorten the
range of the interaction. If a bound state, such as a hydrogen atom,
is put into such a medium, the screening radius $r_D$ will for
sufficiently high density become less than the binding radius $r_B$ of
the atom. Once $r_D\ll r_B$, the effective force between proton and
electron has become so short-ranged that the two can no longer bind.
Thus insulating matter, consisting of bound electric charges, will at
sufficiently high density become conducting: it will undergo a phase
transition \cite{Mott}, in which charge screening dissolves the binding
of the constituents, leading to a plasma of unbound charges as a new
state of matter.

\par

\begin{figure}[htb]
\centerline{\psfig{file=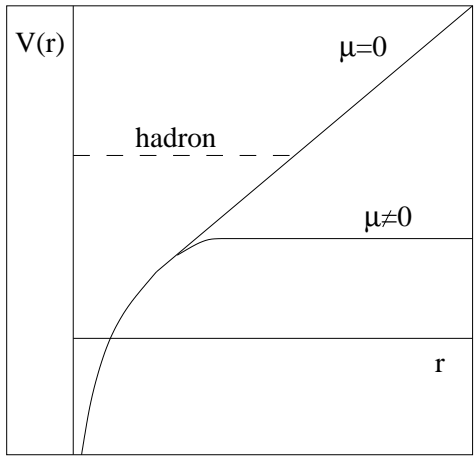,width=6cm,angle= -90}}
\vskip 0.25cm
\caption{Colour screening of the confining potential.}
\label{2_1}
\end{figure}

The interaction of quarks in QCD is based on their intrinsic colour
charge, and in a dense medium this charge can be screened in much the
same way as an electric charge. Hadrons are colour-neutral bound states
of coloured quarks; hence dilute hadronic matter is a colour insulator.
At sufficiently high density, however, we expect colour screening to set
in, so that the potential (\ref{1.1}) becomes \footnote{The functional
form of the screening depends on the form of the unscreened potential
\cite{Dixit}; this leads to the difference between Eqs.\ (\ref{2.2}) and
(\ref{2.3}).}
\be
V(r)~ \simeq~ \sigma r \left[ {1 - \exp(-\mu r) \over \mu r}\right];
\label{2.3}
\ee
as above, the colour screening mass $\mu$ here is also the inverse of
the screening radius for colour charges. The resulting exponential
damping of the binding force will remove all long range effects and in
a sufficiently dense medium `melt' hadrons (see Fig.\ \ref{2_1}) just as
Debye screening dissociated hydrogen atoms. Colour screening will thus
transform a colour insulator into a colour conductor, turning hadronic
matter into a quark plasma \cite{Satz}. The transition from insulator
to conductor by charge screening is a collective effect, so that we
expect a phase transition at the point of plasma formation.

\par

\par

When atomic matter is transformed from an insulator into a conductor,
the effective mass of the conduction electrons is changed. In
insulators, the electrons occur with their physical mass; in the
conducting phase, however, they acquire a different, {\sl effective}
mass, due to the presence of the other conducting electrons,
the periodic field of the charged ions and the lattice vibrations.
All these effects combine to produce a mean background field quite
different from the vacuum. As a result, the insulator-conductor
transition is accompanied by a mass shift for the electrons. Similarly,
the effective quark mass is expected to change between the confined and
the deconfined phase. When confined in hadrons, the basic quarks
`dress' themselves with gluons to acquire an effective constituent
quark mass of about 300 MeV (1/3 of the proton or 1/2 of the
$\rho$-meson mass). On the other hand, the basic bare quarks in the QCD
Lagrangian are almost massless, so that the mass of the constituent
quarks in the confined phase must be generated spontaneously through
the confinement interaction. Hence it is likely that when deconfinement
occurs, this additional mass is `lost' and the quarks revert to their
intrinsic `bare' mass.

\par

A Lagrangian with massless fermions -- the limiting case of the light up
and down quarks in the physical Lagrangian -- possesses chiral symmetry;
this allows a decomposition of the quarks into independent left- and
right-handed massless spin one-half states, which for massive fermions
become mixed. For massless quarks, confinement must thus lead to
spontaneous breaking of chiral symmetry, deconfinement to its
restoration. Hence the mass shift transition in QCD is often
referred to as chiral symmetry restoration. It can, but does not need to
coincide with deconfinement: when the hadrons are dissolved into quark
constituents, the liberated and hence now coloured quarks may still
interact to form coloured bound states. Thus at low temperature and high
density, the quark triplets in nucleons, once deconfined, might choose
to recombine into massive coloured quark pairs (``diquarks")
\cite{diquarks}, similar to Cooper pairs in QED. When the density is
increased further, the diquarks would break up into the massless
basic quarks. This results in a three-phase picture of strongly
interacting matter, with hadronic matter as confined phase,
then deconfinement, followed by a phase consisting of massive coloured
diquark systems, and finally, after chiral symmetry restoration, a
plasma of coloured massless quarks and gluons. Such a three-phase
structure would correspond to the succession of insulator,
superconductor and conductor in atomic matter.

\par

\begin{figure}[htb]
\centerline{\psfig{file=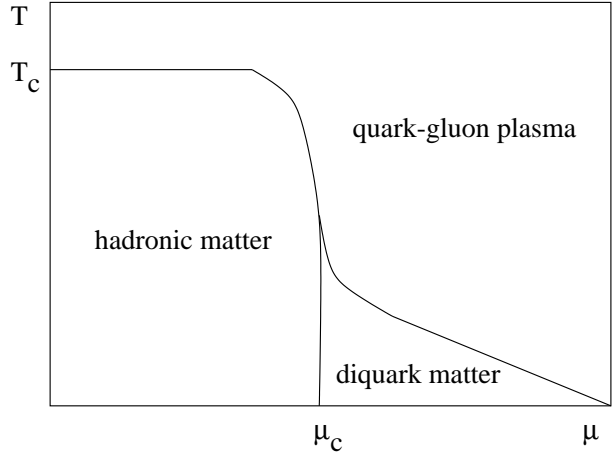,width=6cm,angle= -90}}
\caption{The phase diagram of strongly interacting matter.}
\label{2_2}
\end{figure}

In relativistic thermodynamics, higher densities can be obtained either
by increasing the net baryon number density, or by `heating' the system,
so that collisions between its constituents produce further hadrons.
This leads to the phase diagram shown in Fig.\ \ref{2_2}: for low values
of the temperature $T$ and the baryochemical potential $\mu_B$,
associated
to the net baryon number density, we have confinement and hence
hadronic matter; for high $T$ and/or $\mu_B$, deconfinement sets in and
we get a quark-gluon plasma.  In the compression of cold nuclear matter,
by increasing $\mu_B$ for $T=0$, an intermediate diquark phase could
occur, similar to the superconducting phase in atomic matter. By
increasing $T$ at $\mu_B=0$, we are heating mesonic matter until it
becomes a quark-gluon plasma. Strong interation thermodynamics thus
predicts the existence of new, deconfined states of matter at high
temperatures and densities.

\par

\par

Before turning to the theoretical basis of these conceptual
considerations, we illustrate the transition from hadronic matter to
quark-gluon plasma by a very simple model. For an ideal gas of
massless pions, the pressure as function of the temperature is given by
the Stefan-Boltzmann form
\be
P_{\pi} = 3 {\pi^2 \over 90} T^4 \label{2.3a}
\ee
where the factor 3 accounts for the three charge states of the pion.
The corresponding form for an ideal quark-gluon plasma with two
flavours and three colours is
\be
P_{qg} = \{ 2 \times 8 + {7\over 8}(3 \times 2 \times 2 \times 2) \}
{\pi^2 \over 90} T^4 - B = 37 {\pi^2 \over 90} T^4 - B. \label{2.3b}
\ee
In Eq.\ (\ref{2.3b}), the first temperature term in the curly brackets
accounts for the two spin and eight colour degrees of freedom of the
gluons, the second for the three colour, two flavour, two spin and two
particle-antiparticle degrees of freedom of the quarks, with 7/8 to
obtain the correct statistics. The bag pressure $B$ takes into account
the difference between the physical vacuum and the ground state for
quarks and gluons in a medium.

\par

\begin{figure}[h]
\centerline{\psfig{file=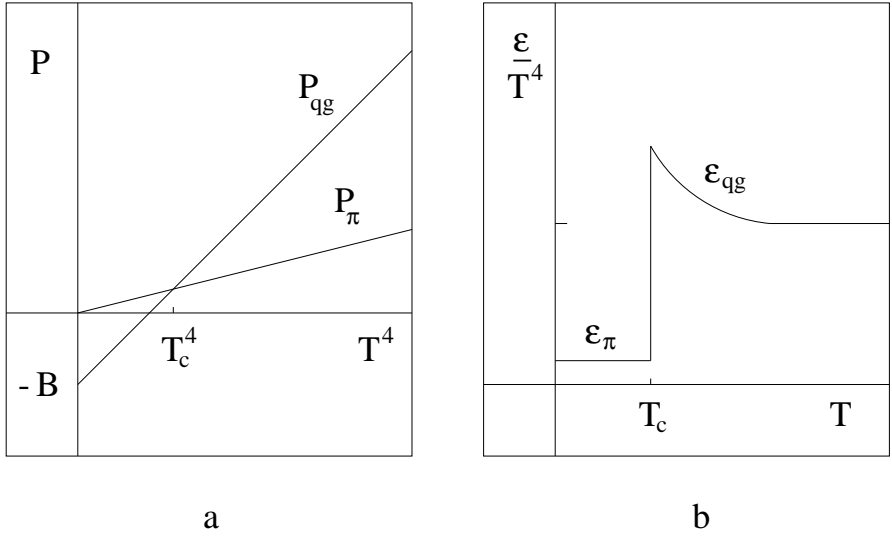,width=7cm,angle= -90}}
\caption{Pressure (a) and energy density (b) in a two-phase ideal gas
model.}
\label{2_3}
\end{figure}

Since in thermodynamics, a system chooses the state of lowest free
energy and hence highest pressure, we compare in Fig.\ \ref{2_3}a the
temperature behaviour of Eq's.\ (\ref{2.3a}) and (\ref{2.3b}). Our
simple model thus leads to a two-phase picture of strongly interacting
matter, with a hadronic phase up to
\be
T_c = \left( {45 \over 17 \pi^2} \right)^{1/4} B^{1/4}
 \simeq 0.72~B^{1/4} \label{2.3e}
\ee
and a quark gluon plasma above this critical temperature. From hadron
spectroscopy, the bag pressure is given by $B^{1/4} \simeq 0.2$ GeV,
so that we obtain
\be
T_c \simeq 145~{\rm MeV} \label{2.3f}
\ee
as the deconfinement temperature. In the next section we shall find
this simple estimate to be remarkably close to the value obtained in
lattice QCD.

%\medskip

\par

The energy densities of the two phases of our model are given by
\be
\e_{\pi} = {\pi^2 \over 10} T^4 \label{2.3g}
\ee
and
\be
\e_{qg} = 37 {\pi^2 \over 30} T^4 + B. \label{2.3h}
\ee
By construction, the transition is first order, and the resulting
temperature dependence is shown in Fig.\ \ref{2_3}b. We note that even
though
both phases consist of massless non-interacting constituents, in the
quark-gluon plasma
\be
(\e - 3P) = 4B \label{2.3i}
\ee
does not vanish, as consequence of the difference between physical
vacuum and in-medium QCD ground state \cite{Asakawa}.

\bigskip

\noindent{\bf 2.2 Critical Behaviour in QCD}

\bigskip

Here we shall derive the conceptual considerations of the last
section from strong interaction thermodynamics, based on QCD as the
input dynamics. QCD is defined by the Lagrangian
\be
{\cal L}~=~-{1\over 4}F^a_{\mu\nu}F^{\mu\nu}_a~-~\sum_f{\bar\psi}^f
_\alpha(i \gamma^{\mu}\partial_{\mu} + m_f
-g \gamma^{\mu}A_{\mu})^{\alpha\beta}\psi^f_\beta
~,\label{2.4}
\ee
with
\be
F^a_{\mu\nu}~=~(\partial_{\mu}A^a_{\nu}-\partial_{\nu}A^a_{\mu}-
gf^a_{bc}A^b_{\mu}A^c_{\nu})~. \label{2.5}
\ee
Here $A^a_{\mu}$ denotes the gluon field of colour $a$ ($a$=1,2,...,8)
and $\psi^f_{\alpha}$ the quark field of colour $\alpha$
($\alpha$=1,2,3) and flavour $f$; the input (`bare') quark masses are
given by $m_f$. With the dynamics thus determined, the corresponding
thermodynamics is obtained from the partition function, which is
most suitably expressed as a functional path integral,
\be
Z(T,V) = \int ~dA~d\psi~d{\bar\psi}~
\exp~\left(-\int_V d^3x \int_0^{1/T} d\tau~
{\cal L}(A,\psi,{\bar\psi})~\right), \label{2.6}
\ee
since this form involves directly the Lagrangian density defining the
theory. The spatial integration in the exponent of Eq.\ (\ref{2.6}) is
performed over the entire spatial volume $V$ of the system; in the
thermodynamic limit it becomes infinite. The time component $x_0$ is
``rotated" to become purely imaginary, $\tau = ix_0$, thus turning the
Minkowski manifold, on which the fields $A$ and $\psi$ are originally
defined, into a Euclidean space. The integration over $\tau$ in Eq.\
(\ref{2.6}) runs over a finite slice whose thickness is determined by
the temperature of the system.

\par

From $Z(T,V)$, all thermodynamical observables can be calculated in
the usual fashion. Thus
\be
\epsilon = (T^2/V)\left({\partial \ln Z \over \partial T}\right)_V
\label{2.7}
\ee
gives the energy density, and
\be
P = T \left({\partial \ln Z\over \partial V}\right)_T
\label{2.8}
\ee
the pressure. For the study of critical behaviour, long range
correlations and multi-particle interactions are of crucial importance;
hence perturbation theory cannot be used. The necessary
non-perturbative regularisation scheme is provided by the lattice
formulation of QCD \cite{Wilson}; it leads to a form which can be
evaluated numerically by computer simulation \cite{Creutz}.
Unfortunately, this method is so far viable only for the case of
vanishing baryon number density, so that results are today available
only for that part of the phase diagram in Fig.\ \ref{2_2}.

\par

Without going into the details of computer simulation of finite
temperature QCD (for a survey, see \cite{Ka/La}), we summarize what has
been obtained so far. The first variable to consider is the
deconfinement measure given by the Polyakov loop \cite{Larry,Kuti}
\be
L(T) \sim \lim_{r \to \infty}~\exp\{-V(r)/T\} \label{2.9}
\ee
where $V(r)$ is the potential between a static quark-antiquark pair
separated by a distance $r$. In the limit of large input quark mass,
$V(\infty)= \infty$ in the confined phase, so that then $L=0$.
Colour screening, on the other hand, makes $V(r)$ finite at large $r$,
so that in the deconfined phase, $L$ does not vanish. It thus becomes
an `order parameter' like the magnetisation in the Ising model;
this is zero at high temperatures, but becomes non-zero below the
Curie point, when the `up-down' $Z_2$ symmetry of the Ising Hamiltonian
is spontaneously broken. In the large quark mass limit, QCD reduces to
pure $SU(3)$ gauge theory, which is invariant under a global $Z_3$
symmetry. The Polyakov loop provides a measure of the state of the
system under this symmetry: it vanishes for $Z_3$ symmetric states and
becomes finite when $Z_3$ is spontaneously broken. Hence the
critical behaviour of the Ising model (and more generally, $Z_N$ spin
systems) as well as that of $SU(N)$ gauge theories is based on the
spontaneous symmetry breaking of a global $Z_N$ symmetry
\cite{Svetitsky}.

\par

For finite quark mass $m_q$, $V(r)$ remains finite for $r \to \infty$,
since the `string' between the two colour charges `breaks' when the
corresponding potential energy becomes equal to the mass $M_h$ of the
lowest hadron; beyond this point, it becomes energetically more
favourable to produce an additional hadron. Hence now $L$ no longer
vanishes in the confined phase, but only becomes exponentially small
there,
\be
L(T) \sim \exp\{-M_h/T\}; \label{2.10}
\ee
here $M_h$ is of the order of the $\rho$-mass, so that $L \sim
10^{-2}$, rather than zero. Deconfinement is thus indeed much like the
insulator-conductor transition, for which the order parameter, the
conductivity $\sigma(T)$, also does not really vanish for $T>0$, but
with $\sigma(T) \sim \exp\{-\Delta E/T\}$ is only exponentially small,
since thermal ionisation (with ionisation energy $\Delta E$) produces
a small number of unbound electrons even in the insulator phase.

\par

\begin{figure}[tbp]
\vspace*{-3cm}
\centerline{\psfig{file=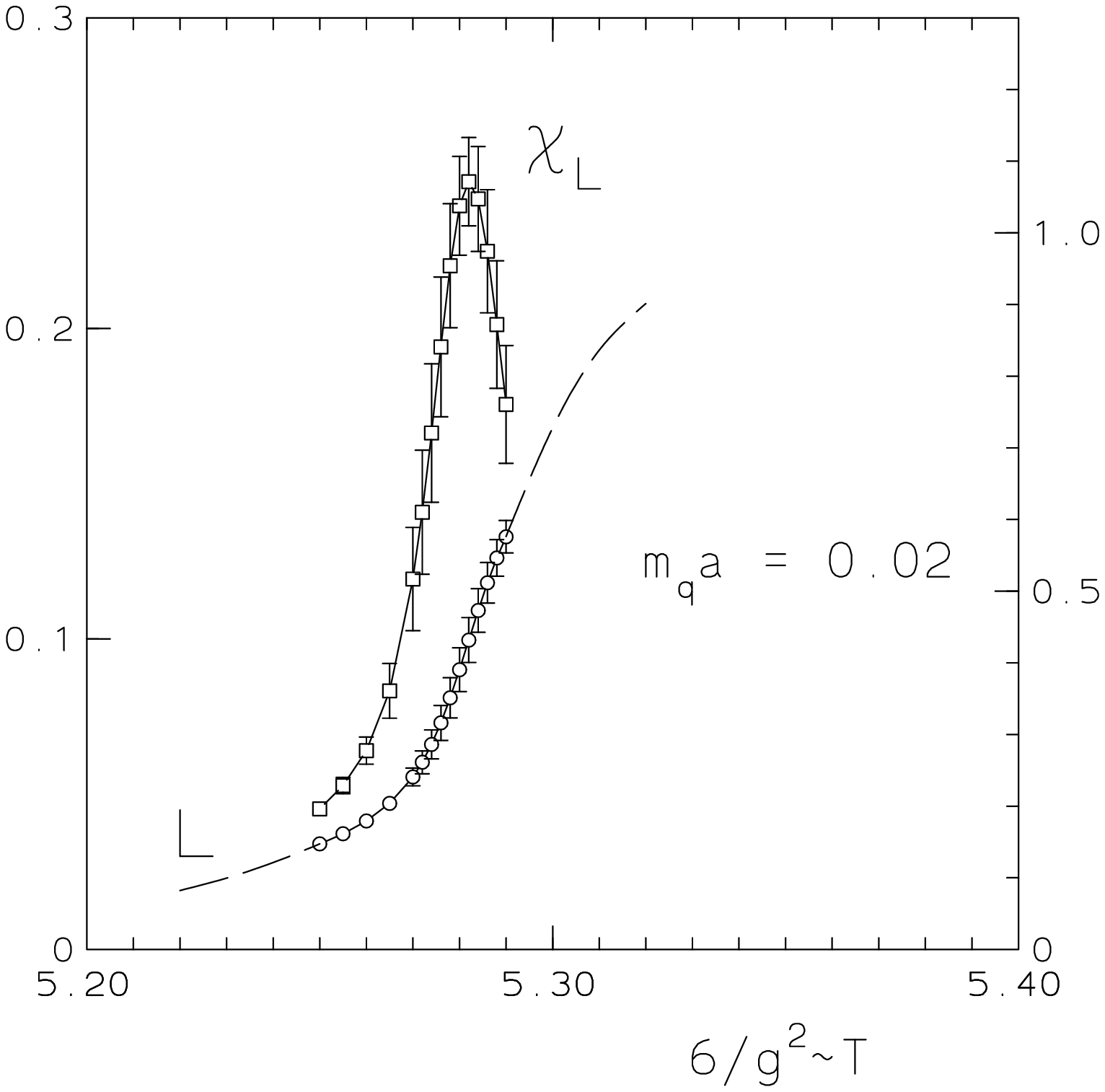,height=13cm,width=10cm}}
\vspace{-2cm}
\caption{The temperature dependence of the Polyakov loop and the associated
susceptibility in two-flavour QCD \cite{K&L}.}
\label{2_4}
%\end{figure}
%\begin{figure}[tbp]
%\vspace{-3cm}
\bigskip
\centerline{\psfig{file=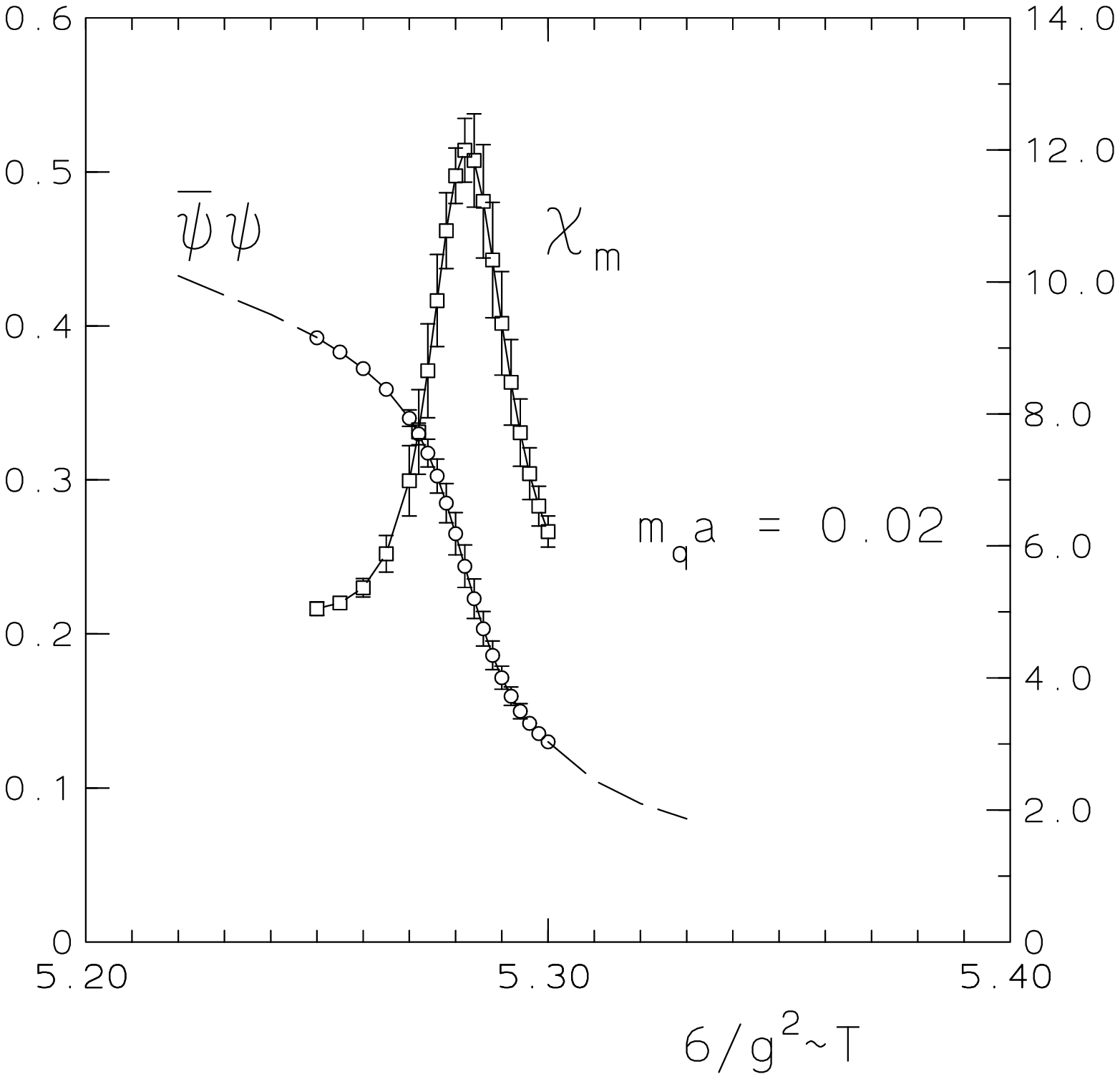,height=13cm,width=10cm}}
\vspace{-2cm}
\caption{The temperature dependence of the chiral condensate and the
associated susceptibility in two-flavour QCD \cite{K&L}.}
\label{2_5}
\end{figure}

Fig.\ \ref{2_4} shows recent lattice results for $L(T)$ and the
corresponding
susceptibility $\x_L(T) \sim \langle L^2 \rangle - \langle L \rangle^2$
\cite{K&L}.
The calculations were performed for the case of two
flavours of light quarks, using a current quark mass about four times
larger than that needed for the physical pion mass \cite{K&L}.
We note that $L(T)$ undergoes the expected sudden increase from a small
confinement to a much larger deconfinement value. The sharp peak of
$\chi_L(T)$ defines quite well the transition temperature $T_c$, which
we shall shortly specify in physical units.

%\par

\begin{figure}[tbp]
\vspace{0.5cm}
\centerline{\psfig{file=Rfig2_6.ps,height=9cm,width=9cm}\hspace{1.5cm}}
\vspace{-0.2cm}
\caption{The temperature dependence of energy density and pressure in
two-flavour QCD \cite{Blum}.}
\label{2_6}
\end{figure}

The next quantity to consider is the effective quark mass; it is
measured by the expectation value of the corresponding term in the
Lagrangian, $\langle {\bar \psi} \psi \rangle(T)$. In the
limit of vanishing current quark mass, the Lagrangian becomes chirally
symmetric and $\langle {\bar \psi} \psi \rangle(T)$ the corresponding
order parameter. In the confined phase, with effective constituent quark
masses $M_q \simeq 0.3$ GeV, this chiral symmetry is
spontaneously broken, while in the deconfined phase, at high enough
temperature, we expect its restoration. In the real world, with finite
pion and hence finite current quark mass, this symmetry is also only
approximate, since $\langle {\bar \psi} \psi \rangle (T)$ now never
vanishes at finite $T$.

\par

The behaviour of $\langle {\bar \psi} \psi \rangle(T)$ and the
corresponding susceptibility $\chi_m \sim \partial \langle {\bar \psi}
\psi \rangle / \partial m_q$ are shown in Fig.\ \ref{2_5} \cite{K&L},
calculated for the same case as above in Fig.\ \ref{2_4}. We note
here the
expected sudden drop of the effective quark mass and the associated
sharp peak in the susceptibility. The temperature at which this occurs
coincides with that obtained through the deconfinement measure. We
therefore conclude that at vanishing baryon number density, quark
deconfinement and the shift from constituent to current quark mass
coincide. In section 2.3, we shall consider why this is so.

\par

We thus obtain for $\mu_B=0$ a rather well defined phase structure,
consisting of a confined phase for $T < T_c$, with $L(T) \simeq 0$ and
$\langle {\bar \psi} \psi \rangle(T) \not= 0$, and a
deconfined phase for $T>T_c$ with $L(T)\not= 0$ and
$\langle {\bar \psi} \psi \rangle(T) \simeq 0$. The
underlying symmetries associated to the critical behaviour at $T=T_c$,
the $Z_3$ symmetry of deconfinement and the chiral symmetry of the quark
mass shift, become exact in the limits $m_q \to \infty$ and $m_q \to
0$, respectively. In the real world, both symmetries are only
approximate; nevertheless, we see from Figs.\ \ref{2_4} and
\ref{2_5} that both
associated measures retain an almost critical behaviour.

\par

Next we turn to the behaviour of energy density $\e$ and pressure $P$ at
deconfinement. In Fig.\ \ref{2_6} it is seen that $\e/T^4$ changes quite
abruptly at the above determined critical temperature $T=T_c$,
increasing from a low hadronic value to nearly that expected for
an ideal gas of quarks and gluons \cite{Blum}. Besides the sudden increase at
deconfinement, there are two further points to note. In the region
$T_c\!<\! T\! <\! 2~T_c$, $\e$ and $3P$ do not yet coincide, as also
found in the simple model of the previous section; much of the
difference is presumably due to the difference between physical vacuum
and QCD in-medium ground state \cite{Asakawa}. Furthermore the
thermodynamic observables do not quite reach their Stefan-Boltzmann
values (marked ``SB" in Fig.\ \ref{2_6}) even at very high temperatures.
These deviations from ideal gas behaviour can be expressed
to a large extent in terms of effective `thermal' masses $m_{\rm th}$
of quarks and gluons, with $m_{\rm th} \sim gT$
\cite{ERSW} - \cite{Patkos}.

\par

Finally we turn to the value of the transition temperature. Since QCD
(in the limit of massless quarks) does not contain any dimensional
parameters, $T_c$ can only be obtained in physical units by expressing
it in terms of some other known observable which can also be calculated
on the lattice, such as the $\rho$-mass, the proton mass, or the string
tension. In the continuum limit, all different ways should lead to the
same result. Within the present accuracy, they define the uncertainty
so far still inherent in the lattice evaluation of QCD. Using the
$\rho$-mass to fix the scale leads to $T_c\simeq 0.15$ GeV, while
the string tension still allows values as large as $T_c \simeq 0.20$
GeV. This means that energy densities of some 1 - 2 GeV/fm$^3$ are
needed in order to produce a medium of deconfined quarks and gluons.

\par

In summary, finite temperature lattice QCD at $\mu_b=0$ shows
\begin{itemize}
%\vspace*{-0.2cm}
\item{that there is a deconfinement transition coincident with an
associated shift in the effective quark mass at $T_c \simeq$ 0.15 -
0.20 GeV;}
%\vspace*{-0.2cm}
\item{that the transition is accompanied by a sudden increase in
the energy density  (``latent heat of deconfinement") from a small
hadronic value to a much larger value near that of an ideal quark-gluon
plasma.}
%\vspace*{-0.2cm}
\end{itemize}

\noindent
From {\sl ab initio} QCD calculations we thus have a relatively good
understanding of the phase structure expected for strongly interacting
matter at vanishing baryon density. Two aspects call for some additional
discussion. What relation between deconfinement and chiral
symmetry restoration makes the two phenomena coincide in finite
temperature QCD? And how does a hadronic medium turn into a
quark-gluon plasma on a microscopic level - do hadrons somehow fuse to
form a deconfined plasma? These questions will be addressed in sections
2.3 and 2.4.

\bigskip

\noindent{\bf 2.3 Deconfinement and Chiral Symmetry Restoration}

\bigskip

The relation between these two transition phenomena has been quite
puzzling all along; why should they (at least for $\mu_B$=0) occur at
the same temperature? We shall here give a somewhat speculative
conceptual answer \cite{Gavai}-\cite{Digal}.

\par

\begin{figure}[tbp]
%\vspace{-1cm}
\centerline{\psfig{file=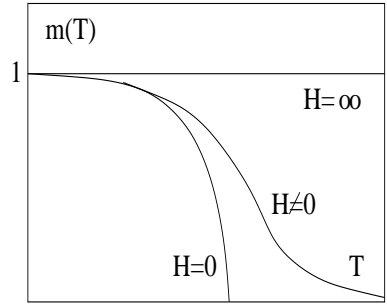,height=5cm,width=4cm,angle= -90}~~}
\vspace{0.2cm}
\caption{Magnetisation in the Ising model with and without
external magnetic field $H$.}
\label{2_7}
\end{figure}

In the presence of light quarks of mass $m_q$, the string between a
static quark-antiquark pair could in principle break when
$V(r)=\sigma r=2m_q$, leading to
\be
L(T) \sim e^{-2m_q/T} \label{2.11}
\ee
for the Polyakov loop. For sufficiently small $m_q$, all
variations of $L(T)$ with $T$ would then be washed out and the
deconfinement transition should disappear \cite{HKS}. Such an effect
would be the counterpart of that caused by an external magnetic field
$H$ in the Ising model. When $H=0$, spontaneous magnetisation sets in
at the Curie point $T_c$; but for $H\! \not=\! 0$, the field always
aligns spins and makes the magnetisation non-zero for all $T$ (Fig.\
\ref{2_7}). We might thus expect that $H \sim 1/m_q$ constitutes an
effective external field in QCD.

\par

\begin{figure}[tbp]
%\vspace{-1cm}
\centerline{\psfig{file=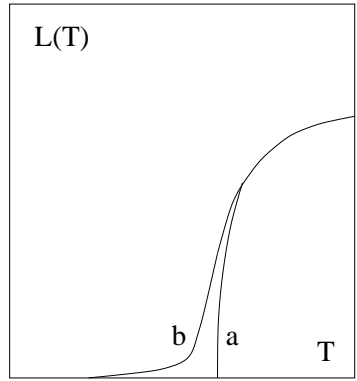,width=5cm,angle= -90}}
%\vspace{1cm}
\caption{The temperature dependence of the Polyakov loop for (a)
pure gauge theory, $m_q=\infty$, and (b) two-flavour QCD with
massless quarks, $m_q=0$.}
\label{2_8}
\end{figure}

However, it turns out that this is not the case; the Polyakov loop shows
strong variations with $T$ even for $m_q\! \to \!0$, as seen in Fig.\
\ref{2_8} \cite{K&L}. Conceptually, there is a good reason for this:
chiral
symmetry breaking in the confined phase keeps the effective constituent
quark mass $M_q$ finite even for vanishing $m_q$, and to break the
string through formation of a normal meson requires $2M_q$, not $2m_q$.
The relevant field should thus be \cite{Gavai}-\cite{Digal}
\be
H \sim 1/M_q, \label{2.12}
\ee
where $M_q$ is the constituent quark mass in the relevant temperature
region. Hence the temperature dependence of $M_q$ plays a crucial role
for the temperature dependence of the Polyakov loop.

\par

\begin{figure}[tbp]
%\vspace{-2cm}
\centerline{\psfig{file=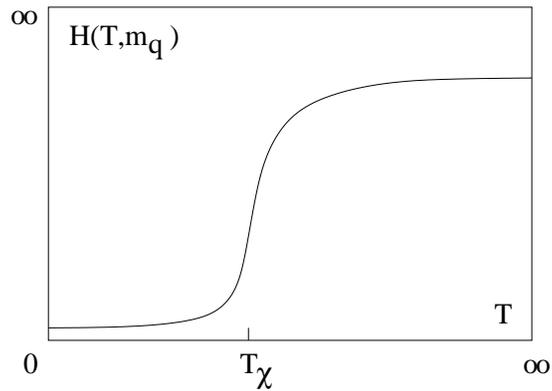,width=5cm,angle= -90}}
\vspace{0.2cm}
\caption{The expected temperature dependence of the effective external
field in full QCD.}
\label{2_9}
\end{figure}

Putting together what we know about the effective quark mass under
different conditions, we expect the external field
\be
H \sim 1/( m_q + c_1 \langle {\bar \psi} \psi \rangle^{1/3} + c_2 g~T),
\label{2.13}
\ee
to be determined by a combination of bare quark mass, chiral
mass and thermal mass (Fig.\ \ref{2_9}).
For $m_q\! \to \! \infty$, we obviously recover the pure gauge case. For
small or vanishing $m_q$, $\langle {\bar \psi} \psi \rangle$ is the
crucial term, since the thermal mass at low $T$ is very small. Below the
chiral symmetry restoration point $T_{\x}$, the large value of $\langle
{\bar\psi}\psi\rangle$ keeps the external field $H$ small, allowing
the Polyakov loop
distribution to remain essentially disordered, so that the system is in
a confined state. At chiral symmetry restoration,
$\langle {\bar \psi} \psi \rangle$ vanishes and hence $H$ suddenly
becomes very large; it aligns the Polyakov loops and thereby explicitly
breaks the $Z_3$ confining symmetry. As a result, $L$ increases sharply
and deconfinement sets in.


We thus have essentially two limiting situations:
\begin{itemize}
\vspace*{-0.2cm}
\item{For large $m_q$, $H$ is always very small, so that deconfinement
is determined by an approximate spontaneous symmetry breaking at
$T=T_c$.}
\vspace*{-0.2cm}
\item{For small $m_q$, $H$ becomes large when $\langle {\bar \psi} \psi
\rangle$ vanishes; $H$ then aligns the Polyakov loops, resulting in
deconfinement at $T=T_{\x}$.}
\vspace*{-0.2cm}
\end{itemize}
In the second case, we have assumed that chiral symmetry restoration
occurs at a temperature $T_{\x}$ less than the deconfinement temperature
in pure gauge theory, as is in fact the case, with $T_c[SU(3)] \simeq
0.26$ GeV and $T_{\x}(N_f=2) \simeq 0.15$ GeV.

\par

Deconfinement is thus
due to {\sl spontaneous} $Z_3$ symmetry breaking only in the large
$m_q$ limit. For small $m_q$, that symmetry is broken {\sl explicitly}
by an effective external field $H$ which becomes large when $\langle
{\bar \psi} \psi \rangle \to 0$. Hence deconfinement and chiral
symmetry restoration coincide.

\bigskip

\noindent{\bf 2.4 Deconfinement and Percolation}

\bigskip

Conceptually, the transition from hadronic matter to quark-gluon plasma
seems quite transparent. As we had seen in the introduction,
deconfinement is expected when the density of quarks and gluons
becomes so high that it no longer makes sense to partition them into
colour-neutral hadrons, since these would strongly overlap. Instead we
have clusters much larger than hadrons, within which colour is not
confined; deconfinement is thus related to cluster formation. This is
the central topic of percolation theory, and hence a connection between
percolation and deconfinement seems very likely
\cite{S-deco,Baym,CKS}. In percolation theory, one considers a
system of extended geometric objects, such as spheres, which can
overlap; percolation is said to occur when, with increasing density, a
connected cluster begins to span the entire system \cite{S&A}. If the
interior of the spheres is coloured, such a cluster signals the onset
of colour conductivity and hence of deconfinement. Moreover, for any
value of the quark mass \cite{S-deco}, the percolation transition is
based on genuine singular behaviour as function of the temperature.

\par

Already the first studies \cite{Baym,CKS} had noted that percolation of
spheres of hadronic size cannot mean deconfinement. The percolation
density $n_p$ for spheres of radius $r$ is given by \cite{Alon}
\be
n_c \simeq {0.34 \over (4\pi/3)~ r^3}. \label{2.14}
\ee
Hence for a nucleon radius of 0.8 fm, the critical value is 0.16
fm$^{-3}$ and thus just standard nuclear density. The percolation of
nucleons therefore corresponds to the formation of nuclear matter;
for deconfinement, the true coloured `interiors' of the hadrons must
overlap. These are more easily defined for mesons.

\par

In QCD, two fundamental triplet colour charges are connected by a
string, with
\be
\sigma \simeq 0.16~ {\rm GeV}^2 = 0.8~{\rm GeV/fm} \label{2.15}
\ee
as string tension. The transverse radius $r_t$ of these strings is
found to be about 0.2 - 0.3 fm in the range $0.5 \leq l \leq 2$ fm
\cite{Alvarez,Schilling}, where $l$ is the length of the string.
A meson is thus like an ellipsoidof major axis $r=l/2$ and minor axis
$r_t$, randomly oriented and oscillating in time about its center of
mass to fill a sphere of radius $r$. The percolation threshold for
ellipsoids of aspect ratio $r/r_t=4$ is found to be \cite{Garboczi}
\be
n_p \simeq {0.21 \over (4 \pi/3)~r~r_t^2}; \label{2.16}
\ee
for $r=1$ fm and $r_t=0.25$ fm, this gives 0.8 fm$^{-3}$ as critical
density. With $m_h = 2r\sigma$ for the `generic' hadron mass in such a
string model, we obtain
\be
\epsilon_p \simeq 1.3~{\rm GeV/fm}^3 \label{2.17}
\ee
as the energy density at the percolation point, in good agreement with
the deconfinement value (1 - 2 GeV/fm$^3$) obtained from lattice QCD
in section 2.2. Thus deconfinement indeed appears to set in at the
percolation point for extended hadronic objects; for mesons, these are
quark-antiquark strings. For nucleons, the quark triplet structure
makes the geometry more complex, but conceptually there should be no
difference.

\par

The connection between percolation and deconfinement is presently also
under investigation on a more formal level, based on the relation
between spin and SU(N) gauge field systems \cite{Svetitsky} and on
the percolation of Polyakov loops \cite{Santo,S-QM}. As mentioned, the
introduction of dynamical quarks in QCD destroys the underlying $Z_N$
symmetry; the expectation value of the Polyakov loop $L(T)$ now never
vanishes, so that $L(T)$ can no longer be taken as an order parameter
to identify the transition. The analogue in the Ising model is, as we saw,
the introduction of an external field, which also breaks the $Z_2$
symmetry and makes the magnetization finite for all temperatures.
Although now no thermal phase transition can take place, the percolation
transition between a region of finite clusters of parallel spins to one
of infinite clusters continues to exist. Its counterpart in SU(N) gauge
theory may therefore constitute a general order parameter for
deconfinement in full QCD, with dynamical quarks \cite{S-deco}.

\bigskip~\medskip

\noindent{\bf 3.\ CONDITIONS IN NUCLEAR COLLISIONS}

\bigskip

\noindent{\bf 3.1 High Energy Collisions and Nuclear Transparency}

\bigskip

Quark matter presumably made up the early universe, and perhaps neutron
stars have quark cores. But the big bang was long ago, and neutron
stars are far away. The rapid growth which the field has experienced in
the past two decades was to a very large extent stimulated by the idea
of using high energy nuclear collisions to produce in the laboratory
droplets of strongly interacting matter large enough and long-lived
enough for a study of statistical QCD.

\par

Normal nuclear matter consists of nucleons of mass 0.94 GeV and has a
density of 0.17 nucleons/fm$^3$; hence its energy density is 0.16
GeV/fm$^3$, i.e., well below the 1 - 2 GeV/fm$^3$ necessary for
deconfinement. We want to collide two heavy nuclei with the aim of
increasing the density of matter as far as possible beyond this value.
To what extent can this be achieved? In a low energy nucleus-nucleus
collision, the two nuclei will remain intact and simply ``bounce off"
each other. With increasing energy, they will penetrate each other more
and more, leading to highly excited nuclear matter, which rapidly
breaks up into nuclear fragments and some additional mesons.

\par

If the collision energy is increased still further, {\sl nuclear
transparency} begins to set in: the two colliding nuclei pass through
each other, leaving behind them a ``vapour trail" of deposited energy,
which eventually decays into hadrons. Different droplets of this trail
move at different speeds, from slow ones in the center of mass to the
fast remnants of target and projectile. We refer to elements in the
trail as `droplets', keeping in mind, however, that there really is a
continuous fluid of such droplets, without any gaps. Since nuclear
transparency clearly limits the local (droplet) energy density available for the medium produced in
the collision, its understanding is necessary in order to estimate the
conditions which can be achieved experimentally. It is a phenomenon
which occurs already in the collision of two {\sl nucleons}, which
at high energy also pass through each other, creating a cascade of
secondary hadrons. Most of these are produced outside the primary
interaction region and long after the primary collision
\cite{Kancheli,Bjorken}.

\par

We therefore first ask why in the more elementary process of a
nucleon-nucleon collision the two collision partners do not stop each
other. Nucleons have a spatial extension of about 1 fm, and
in a nucleon-nucleon collision it therefore takes a certain time $\t_0$
before the entire projectile nucleon has realised that it has hit
something. In the rest-frame of the projectile, a time $\t_0 \sim 1$
fm is required to transmit the information about the collision from one
side of the nucleon to the other. In the rest-frame of the target
nucleon, this time is dilated to
\be
t_0 = \t_0 \left( {P_0\over m} \right),\label{3.1}
\ee
where $P_0= \sqrt{{\bf P}^2 + m^2}$ is the energy of the incident
projectile nucleon, $\bf P$ its lab momentum and $m$ its mass. Thus at
high collision energies, $t_0 \gg 1$ fm; since the target also has a
size of about 1 fm, the projectile has long left the target region
before it fully realises that it was hit. It also takes a time of order
1 fm to form a hadronic secondary of radius $R_h \simeq 1$ fm, like a
pion, and hence only very slow secondaries are produced {\sl inside}
the target region. For faster ones, the formation time is again
dilated, and they are therefore formed {\sl outside} the
target region. The production process in nucleon-nucleon collisions,
beginning about 1 fm after the collision with soft secondaries
produced in the target region and continuing with further production of
faster secondaries later and outside the target, is therefore often
referred to as {\sl inside-outside} cascade.

\par

One thus encounters transparency and a vapour trail of droplets of
deposited energy already in high energy nucleon-nucleon collisions.
Here one moreover observes that the nature of all the droplets of
different rapidities is more or less the same in their respective
rest-frames: they decay into the same types of hadrons, and these have
the same momentum distributions, apart from possible effects due to the
motion of the droplet \cite{Leonidov}. In other
words, the properties of any segment of the vapour trail, measured in
its own rest system, do not depend on its relative velocity with
respect to the others. We will thus assume \cite{Feynman,Bjorken83}
that the vapour trail is invariant under Lorentz transformations along
the collision axis. Evidently there are limits to the validity of such a
picture: it must break down at both ends of the trail, when the
kinematic limit is reached. Moreover, at very high energies,
the overall baryon number is negligible in the main (central) part of
the trail, since the fast target and projectile nucleons emerge
as leading particles at high rapidity. Hence the droplets have a
non-zero baryon number.

\par

Let us make this more precise. A particle of mass $m$, moving with
a momentum $p$ along the beam axis, has a rapidity $y = \sinh^{-1}(p/
m)$; for the moment, we assume vanishing transverse momentum. Since the
transformation from one reference system to another is just a
displacement in rapidity, boost invariance along the beam axis means
that a physical quantity, such as the deposited energy in the restframe
of the droplet, cannot depend on the rapidity $y$. Since the deposited
energy appears later in the form of hadrons, the number of hadronic
secondaries $(dN_h/dy)$ produced in a given rapidity interval $dy$ must
be independent of $y$,
\be
\left({dN_h \over dy}\right) = a(s) ,\label{3.2}
\ee
where $a(s)$ depends on the (squared) collision energy $s$ only.
The central rapidity plateau defined by Eq.\ (\ref{3.2}) is a reasonable
approximation to what one observes in high energy $p-p$ and
$p-{\bar p}$ collisions.

 \par

We now include transverse momenta and note that the transverse momentum
($p_T$) distribution of the produced hadrons must also be
$y$-independent. The full momentum distribution of secondaries of
species $i$ thus becomes (assuming azimuthal symmetry around the
collision axis)
\be
\left( {d^2 N_i \over dy~dp_T^2}\right) = a_i(s) f_i(p_T,s).\label{3.3}
\ee
Here $a_i(s)$ determines the relative abundance of species $i$ at the
given $s$, and $f_i(p_T,s)$ is the corresponding rapidity-independent
transverse momentum distribution, normalized to unity. In principle,
$f_i(p_T,s)$ could depend both on the collision energy
and on the species of hadron involved. One finds, however, that
in very good approximation it is energy-independent, and the
species-dependence enters only through the transverse mass, $m_T(m_i)
\equiv \sqrt{m_i^2 + p_T^2}$,
\be
f_i(p_T, s)= f(p_T, m_i) = \left[{\lambda^2 \over
2(1+m_i\lambda)}\right]
~ \exp\{-\lambda~[m_T(m_i)-m_i]\}, \label{3.4}
\ee
with a universal (i.e., species-independent) constant $\lambda^{-1}
\simeq 0.15$ GeV. Since $[m_T(m_i) - m_i]$ is just the transverse
kinetic energy of species $i$, $f_i(p_T, s) \sim \exp\{-\lambda~E_i\}$
provides a universal cut-off in the transverse kinetic energy for all
species. This phenomenon is generally referred to as $m_T$-scaling.
It is very reminiscent of a Boltzmann distribution and has led to the
formulation of thermodynamic models for such production processes,
with $\lambda^{-1}=T_H$ playing the role of a temperature
\cite{Hagedorn}. A droplet thus freezes out at temperature $T_H$ by
turning into freely flowing (non-interacting) hadrons. From Eq.\
(\ref{3.4}), the average transverse mass of a secondary is
\be
\langle m_T(m_i) \rangle = {2 \over \lambda}\left[ 1 + {(m_i\lambda)^2
\over 2(1+m_i\lambda)}\right]. \label{3.5}
\ee
With $\lambda \simeq 0.15$ GeV, this gives for
pions $\langle m_T(\pi) \rangle \simeq 0.37$ GeV, for nucleons
$\langle m_T(N) \rangle \simeq 1.11$ GeV. The corresponding
average transverse momentum is 0.34 GeV for pions, 0.59 GeV for nucleons.
We thus note that $m_T$ scaling (or thermal production, if we interpret
it in this way) provides higher average transverse momenta for
heavier secondaries.

\par

The possible thermal nature of hadron production in $p-p$ collisions has
recently received further support in analyses of species abundances
\cite{B&H}. Experimentally, the production rates for up to
thirty different species of hadrons and hadron resonances have been
measured, for different collision energies. The simplest thermal
picture has chemical freeze-out occurring when a hadron gas, whose
constituents are all possible hadron and hadron resonance states,
reaches a certain temperature $T_h$ \cite{C&S}. This would mean that the
relative abundances of all species at all (high) collision energies are
determined by just this one parameter. This is in fact not possible for
$p-p$ collisions, because of the so-called strangeness suppression; the
relative abundance of secondaries of non-zero strangeness is
considerably smaller than predicted. A full description does become
possible, however, if one further parameter $\gamma_s$ is introduced to
take this into account \cite{Raf}. It is chosen so that $\gamma_s=1$
implies a fully thermal abundance spectrum, $\gamma_s < 1$ strangeness
suppression. One then finds that for $T_h \simeq 170$ MeV and
$\gamma_s \simeq 0.5$, it is possible to reproduce quite well the
observed secondary hadron abundances in $p-p$ and $p-{\bar p}$
collisions from $\sqrt s=20$ to 900 GeV \cite{B&H} .

\par

\begin{figure}[htb]
%\vspace*{-0.9cm}
%\begin{center}
\mbox{
%\vspace*{1cm}
\hskip1cm\epsfig{file=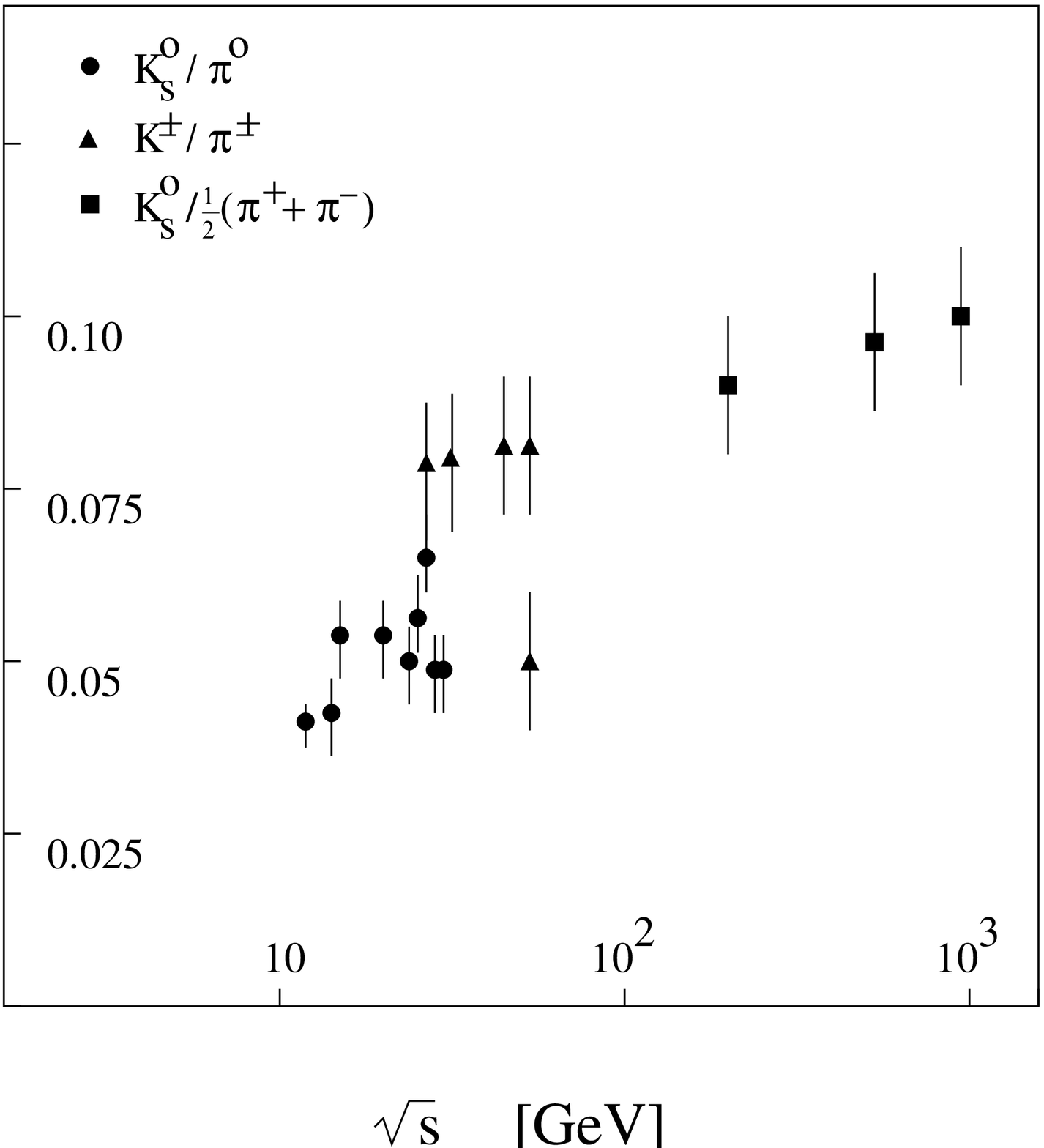,width=6cm}
\hskip1cm
%\vspace*{-1cm}
\epsfig{file=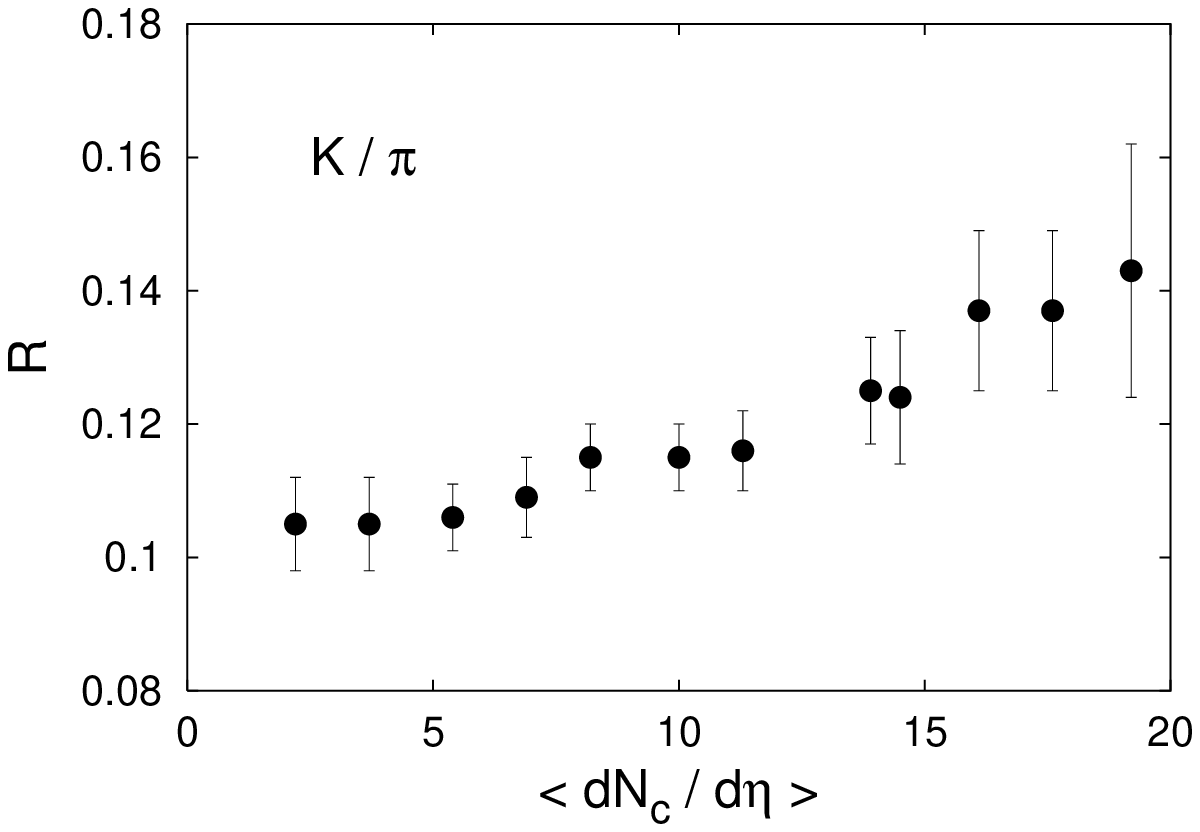,width=7cm}}
\vskip0.5cm
%\end{center}
%\vspace*{-0.8cm}
\caption{The $K/\pi$ ratio (a) in $p-p$ and $p-{\bar p}$  collisions as
function of the cms collision energy \cite{Ansorge} and (b) in
$p-{\bar p}$ collisions at $\sqrt s = 1.8$ TeV as function of the
associated multiplicity \cite{Alexo}.}
\label{3_1}
%\vspace*{-0.8cm}
\end{figure}

The origin of such thermal behaviour remains rather unclear, all the
more so since it holds for hadron production in high energy $e^+e^-$
annihilation as well \cite{Becattini}. It can certainly not arise from
rescattering among the produced secondaries, since neither the life-time
nor the density of the produced medium are high enough for this. It may
be due to a statistical excitation of the vacuum due to the passage of
one or more colour charges, occurring already on the partonic level. The
strangeness suppression factor could then be a consequence of the large
strange quark mass, compared to that of the light $u$ and $d$ quarks.
In a partonic medium of temperature $T > T_h$, the ratio of Boltzmann
weights,
\be
\gamma_s \simeq \exp\{-(m_s - m_u)/T\} \label{3.5a}
\ee
would indeed lead to values around 0.5 for partonic temperatures around
200 MeV. Eq.\ (\ref{3.5a}) implies that with increasing $T$, i.e., with
increasing energy density of the excited vacuum, $\gamma_s$ should
approach unity. One way to check this is to see if the relative
abundance of strange particles increases when one increases either the
collision energy or considers collision fluctuations of higher than
average secondary multiplicity. In both cases, the $K/\pi$ ratio indeed
grows considerably \cite{Ansorge,Alexo}, as seen in Fig.\ \ref{3_1}.

To complete the discussion of nucleon-nucleon collisions, we specify
$a(s)$ in Eq.\ (\ref{3.3}). From $p-p$ and $p-{\bar p}$ experiments over
a large range of collision energies, up to ${\sqrt s} = 1.8$ TeV, one
finds in the central rapidity region \cite{Tevatron} for the sum
over all secondaries (which means predominantly pions),
\be
\left( {dN_h \over dy} \right)_{\! y\simeq 0}
\simeq~ \ln({\sqrt s}/2m), \label{3.6}
\ee
so that $a(s) \simeq~ \ln({\sqrt s}/2m)$; here $m$ denotes the nucleon
mass.

\par

After this brief survey of nucleon-nucleon interactions, we
now return to nucleus-nucleus collisions in more detail. At high
energies, each of the two incoming nuclei (for simplicity, we assume
them to have equal mass numbers $A$) will appear Lorentz-contracted to
a longitudinal thickness of
\be
d_A~=~2R_A\left( {2m\over \sqrt s} \right) \label{3.7}
\ee
in the overall center of mass. Here $s = 2m^2 + 2m\sqrt{{\bf P}^2+m^2}$
denotes the squared nucleon-nucleon collision energy, ${\bf P}$ the
incident lab momentum per nucleon, and $R_A\simeq 1.12~A^{1/3}$ fm
the nuclear radius. For c.m.s. collision energies of 20 GeV or more,
$d_A$ is thus one fermi or less, so that the different nucleon-nucleon
collisions in the course of the nucleus-nucleus collision essentially
occur superimposed in the same space-time region. Consider a central
$A-A$ collision; in the process, each of the $A$ projectile nucleons
will interact one or more times with nucleons from the target nucleus.
When the nuclei have passed through each other, we are thus left with 2
$A$ `wounded' nucleons, each of which eventually emits secondary hadrons
\cite{Bialas}. At present energies it is found that the number of
secondaries is proportional to the number of wounded nucleons; it is not
affected by the number of collisions which the nucleon has experienced
\footnote{Note that this holds for `soft' secondaries; the emission of
`hard' products, such as Drell-Yan dileptons or charmonium states, does
depend on the number of collisions.}.
Since a proton-proton collision results in two wounded nucleons, we
obtain from Eq.\ (\ref{3.6}) that the number of secondary hadrons in a
central high energy $A-A$ collision is given by
\be
\left( {dN_h \over dy} \right)_{\! y\simeq 0}^{AA}
\simeq~ A \ln({\sqrt s}/2m). \label{3.8}
\ee
In effect, this result is modified already in $p\!-A$ collisions, since
the nuclear target stops the projectile nucleon more than a nucleonic
target does. This is taken into account by the form
\be
\left( {dN_h \over dy} \right)_{\! y\simeq 0}^{AA}
\simeq~ A^{\alpha} \ln({\sqrt s}/2m), \label{3.8a}
\ee
with $\alpha \simeq 1.15$ for secondaries close to target rapidity in
$p\!-A$ data \cite{Geist}. The multiplicity predicted by Eq.\
(\ref{3.8a}) agrees quite well with results from $S-S$ to $Pb-Pb$
collisions at the CERN-SPS \cite{WA80,Drapier,NA49}.

\bigskip

\noindent{\bf 3.2 Collision Evolution and Initial Conditions}

\bigskip

Statistical QCD, as described in Section 2, provides predictions for
time-independent systems; we want to apply these to the rapidly evolving
systems produced in nuclear collisions. Hence we have to consider in some
detail the different evolution stages of such collisions. In the period
of nuclear passage, which at high energies lasts a fermi or less, there
is multiple scattering among the quark and gluon constituents of the
colliding nucleons, leading to a rapid generation of entropy and
thereby eventually to thermalisation. The separating nucleons thus
leave behind a more or less equilibrated expanding medium which in its
`hot' early stages we expect to be a quark-gluon plasma. It cools off,
undergoes the confining quark-hadron transition and after that forms
hadronic matter, which finally freezes out into the observed hadronic
secondaries. The aim of this section is to estimate the
energy density of the medium in its early stage.

\par

Thermalisation is presumably attained by multiple scattering of the
incident constituents and the subsequently produced secondaries. At
high energies and early times, these constituents are quarks and
gluons, so that the tool to study the resulting entropy production
must be some form of parton cascade model based on perturbative QCD
\cite{cascade,Wang}. The partons contained in the incident
nucleons collide, produce further partons, and this cascade continues
and eventually leads to a thermal medium. Such partonic cascade models
allow the calculation of entropy production as a function of the local
time after the collision, and the systems produced in high energy
collisions appear to reach thermalisation after about one fermi.
We shall return to parton cascade models in section 6.3.

\par

To estimate the energy density, one has to model the expansion process.
The simplest model \cite{Bjorken83} assumes that the finally observed
hadrons have emerged from an initial interaction volume in free flow,
i.e., without any collective motion. The resulting initial energy
density in an $A-A$ collision is
\be
\e_0 = {p_0 \over \pi R_A^2 \t_0} \left( {dN_h \over dy}
\right)_{y=0}^{AA}, \label{3.9}
\ee
where $p_0$ denotes the average (local) energy of the emitted secondaries
and $(dN_h/dy)$ their multiplicity as given by Eq.\ (\ref{3.8a}). The
initial interaction volume is determined by the transverse size $R_A$ and
the longitudinal extension as obtained from the thermalisation time
$\t_0$, which is generally taken to be
about one fermi. Combining this with Eq.\ (\ref{3.8a}), we get
\be
\e_0 \simeq  {p_0 \over 4} A^{0.48} \ln(\sqrt{s}/2m) ~~[{\rm GeV/fm}^3]
\label{3.10}
\ee
for the initial energy density in a central $A-A$ collision. Since
pions are the dominant secondaries and have an average transverse
momentum of about 0.35 GeV over a wide range of incident energies, we
have $p_0 \simeq 0.5$ GeV. The estimates in Eq's.\ (\ref{3.8a}) and
(\ref{3.10}) give averages over the transverse nuclear area, without
taking into account any details of the nuclear distribution. In the
Glauber formalism, the correct nuclear geometry can be taken fully into
account (see e.g.\ \cite{KLNS}). For central $A-A$ collisions, this
results in little change, but for non-central or $A-B$
collisions with $A \not= B$, the actual geometry becomes important.

Before looking at the attainable energy densities, let us note where
nucleus-nucleus collision experiments reaching the range of interest
for our purposes can actually be performed. The only laboratories
providing sufficiently energetic nuclear beams are the Brookhaven
National
Laboratory (BNL) near New York (USA) and the European Organization for
Nuclear Research (CERN) in Geneva (Switzerland). Both began
experimentation in 1986, using existing accelerators. BNL had the
Alternating Gradient Synchrotron (AGS), designed for 30 GeV/c proton
beams, CERN the Super Proton Synchrotron (SPS) for 450 GeV/c protons.
The injectors available at that time allowed only the acceleration
of nuclei containing equal numbers of protons and neutrons (A=2Z), so
that the beams were restricted to light ions ($A\lsim 40$). Both
laboratories have in the meantime built new injectors, allowing the
acceleration of arbitrarily heavy nuclei in AGS and SPS. Moreover, both
laboratories are presently constructing colliding beam accelerators
which will bring them to very much higher collision energies. At BNL,
the Relativistic Heavy Ion Collider (RHIC) is scheduled to start
operating in 2000; at CERN, the Large Hadron Collider (LHC) is planned
for some five years later.

\par

\begin{figure}[htb]
\centerline{\psfig{file=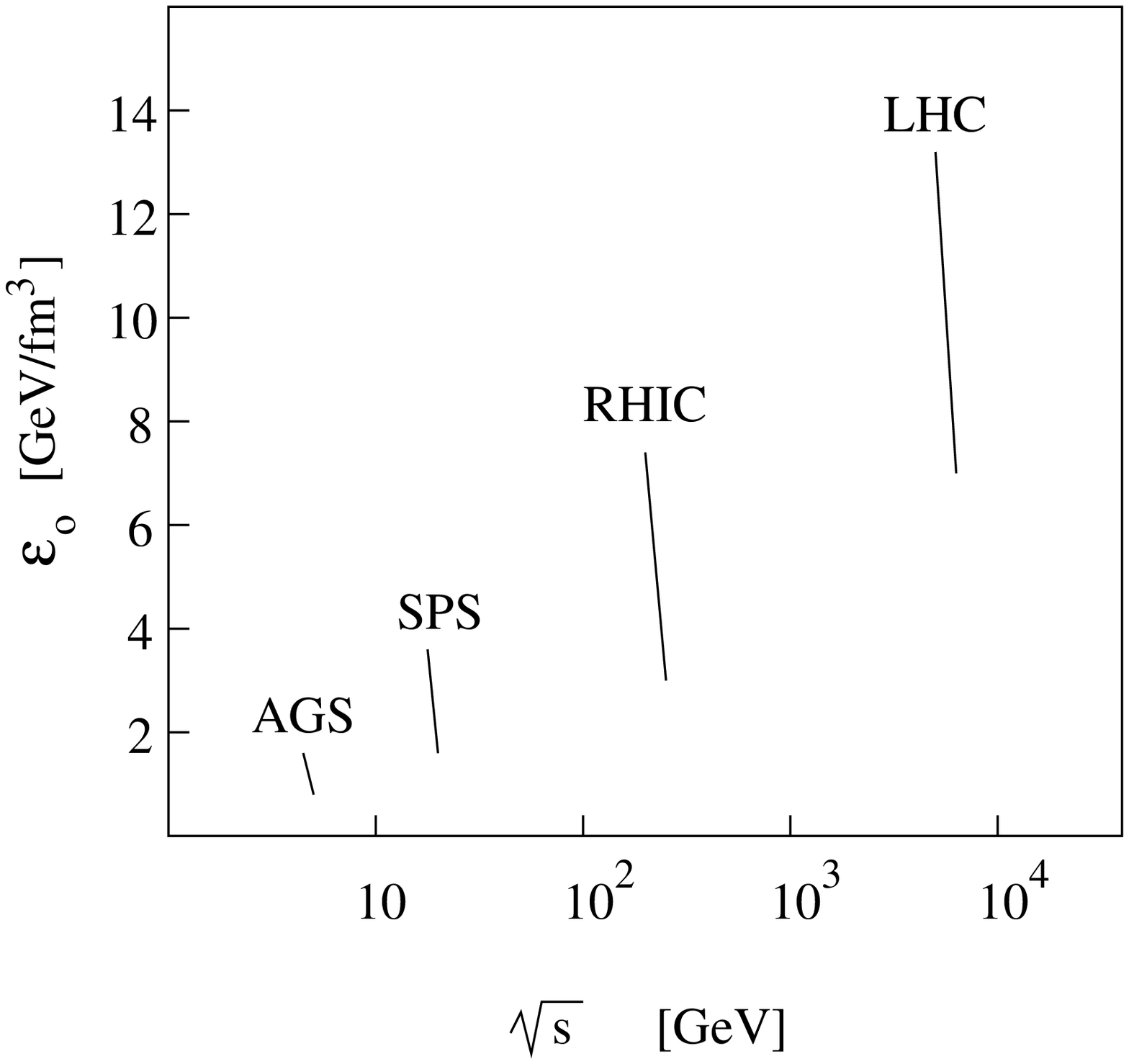,height=9cm,angle= -90}}
\vspace{0.5cm}
\caption{The variation of the initial energy density with incident
collision energy for different experimental facilities.}
\label{3_2}
\end{figure}

The available and planned facilities are summarized in Table 1.
We have there also included the corresponding energy density values;
they were obtained by a Glauber analysis taking into account
the correct nuclear geometry \cite{KLNS}. Note that the listed values
are averages over the nuclear profile; the center of the interaction
region can reach energy densities up to 30 \% higher.
The results using light ion beams are given for heavy targets
($A \simeq 200$), the others for $A-A$ with $A \simeq 200$.
The $\ln \sqrt s$ dependence of the energy density
(Eq.\ \ref{3.10}) makes it rather difficult to achieve significant
changes by varying the beam energy. The use of lighter nuclei is clearly
more efficient. In Fig.\ \ref{3_2}, we show the functional
behaviour of $\e$
for $A=30$ and $A=200$, to give an indication of the ranges achievable
at the different facilities, as well as their relation to the expected
deconfinement threshold.

\begin{table}
$$\vbox{\offinterlineskip
\halign{
\strut\vrule     \hfil $#$ \hfil  &
      \vrule # & \hfil $#$ \hfil  &
      \vrule # & \hfil $#$ \hfil  &
      \vrule # & \hfil $#$ \hfil  &
      \vrule # & \hfil $#$ \hfil  &
      \vrule # & \hfil $#$ \hfil
      \vrule \cr
\noalign{\hrule}
&& && && && &&\cr
~~{\rm Machine}~~&&
~~{\rm Start}~~&&
~~{\rm Type}~~&&
~~~{\rm Beam}~~&&
~~{\sqrt s}~~&&
~~{\e_0^{AB}}~~\cr
&& && && && &&\cr
&& && && &&~~[{\rm GeV/A}]~~&&~~[{\rm GeV/fm}^3]~~\cr
&& && && && &&\cr
\noalign{\hrule}
&& && && && &&\cr
~~{\rm BNL-AGS}~~&&~~1986~~&&~~{\rm Fixed~Target}~~&&
^{28}Si && 5 &&~~0.7 ~~   \cr
&& && && && && \cr
~~{\rm CERN-SPS}~~&&~~1986~~&&~~{\rm Fixed~Target}~~&&
^{16}O,^{32}S && 19 &&~~1.6~~   \cr
&& && && && &&\cr
\noalign{\hrule}
&& && && && &&\cr
~~{\rm BNL-AGS}~~&&~~1992~~&&~~{\rm Fixed~Target}~~&&
^{197}Au && 5 &&~~1.5~~  \cr
&& && && && &&\cr
~~{\rm CERN-SPS}~~&&~~1994~~&&~~{\rm Fixed~Target}~~&&
^{208}Pb && 17 &&~~3.7~~  \cr
&& && && && &&\cr
\noalign{\hrule}
&& && && && && \cr
~~{\rm BNL-RHIC}~~&&~ 2000~~&&~~{\rm Collider}~~&&
^{197}Au && 200 &&~~7.6~~  \cr
&& && && && && \cr
~~{\rm CERN-LHC}~~&&~~\sim 2005~~&&~~{\rm Collider}~~&&
^{208}Pb && 5000 &&~~13~~  \cr
&& && && && &&\cr
\noalign{\hrule}}}$$

\caption{Experimental facilities for high energy
nuclear collisions; the light ion beam results are for heavy ($A=200$)
targets, the others for symmetric ($A-A$) collisions.}
 
\end{table}

\bigskip

\noindent{\bf 3.3 The Onset of Deconfinement}

\bigskip

\begin{figure}[htb]
\centerline{\psfig{file=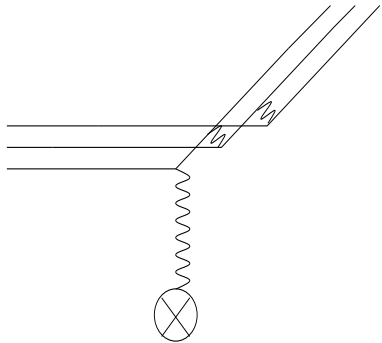,height=5cm,width=4.5cm,angle=-90}}
%\vspace{-0.5cm}
\caption{Scattering of quarks confined to a nucleon.}
\label{3_3}
\end{figure}

At the beginning of a high energy nucleus-nucleus collision, there are
hard interactions between the partonic constituents of the participating
nucleons. These partons are confined: their momentum distribution
is that of partons within a nucleon, as determined e.g.\ in deep
inelastic scattering. In particular, this means that if one quark in the
nucleon scatters, emitting (or absorbing) a gluon, others must follow
the scattered quark to retain colour neutrality (Fig.\ \ref{3_3}).
As a result, high momenta $k$ for
the emitted gluon are suppressed, with a momentum distribution of the
form
\be
xg(x) \sim (1-x)^{1+n}, \label{3.11}
\ee
where $x=k/\sqrt s$ is the fractional momentum of the gluon and $n$
denotes the number of other quarks or antiquarks which must follow the
scattered one.

Deconfinement starts when the given environment leads to partons which
are no longer subject to the confinement constraints just mentioned;
with $n=0$, this results in a hardening of the parton momentum
distribution. The first step towards deconfinement is thus the overlap
of two hard nucleon-nucleon collisions. If there is a secondary
interaction between the two interacting pairs {\sl before} the nucleons
in the primary interactions have had the time to `regroup' into
physical nucleons, a hard gluon exchanged between the two pairs is not
subject to confinement. An `exogamous' system of this kind is therefore
the first step towards a deconfined medium.

\begin{figure}[htb]
\centerline{\psfig{file=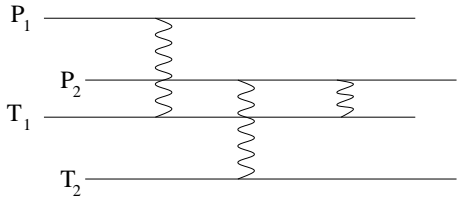,height=6cm,angle=-90}}
%\vspace{-0.5cm}
\caption{`Exogamous' interaction between two colliding nucleon systems
($P_1,T_1$) and ($P_2,T_2$). }
\label{3_4}
\end{figure}

The interaction of two nucleons leads to the emission or absorption of
gluons of transverse momenta $k_T$, with a corresponding transverse area
of radius $r_T \simeq 1/k_T$. In interactions of the type shown in
Fig.\ \ref{3_4}, partons from different collisions overlap in the
transverse plane and can thus no longer be associated to a given
nucleon. To form a connected medium of such deconfined partons, their
density in the transverse plane has to be sufficiently high. This
density increases when the mass number $A$ (more collisions) and/or the
collision energy (more partons) is increased; hence we need high energy
collisions of heavy nuclei for the formation of a quark-gluon plasma.

\bigskip~\medskip

\noindent{\bf 4.\ PROBES OF PRIMORDIAL MATTER}

\bigskip

\noindent{\bf 4.1 Evolution}

\bigskip

We now assume that for sufficiently high energies, the deconfined
partonic system quickly attains equilibrium; this is in accord with
the parton cascade models mentioned above \cite{cascade,Wang}.
Subsequently, it cools off and then hadronizes. For an infinite medium,
the actual transition process depends considerably on the order of the
confinement transition, which so far remains uncertain. The systems
produced in nuclear collisions
are rather small, however, and in addition it is unlikely that any
measurement will fix thermal observables precisely. Hence the effect of
the transition will always appear in somewhat smeared-out form. On the
hadronic side, it is possible for the medium to exist for some time as
a system of interacting hadrons, which eventually freezes out.
However, it is known that an ideal resonance gas leads to critical
behaviour because of the composition law for resonances
\cite{Hagedorn,Fubini}, and that an ideal gas of extended
hadrons does so for geometric reasons \cite{Pomeranchuk}. Hence it is
quite conceivable that the confinement transition leads directly into
an ideal resonance gas.

\par

It is evident that the different evolution stages, from primary
collisions to freeze-out, require very different probes for their
investigation. Consider the two extremes. Pions appear at
the very end, either directly or as decay products of resonances; hence
they can hardly tell us much about the very early phases of the system,
which existed and disappeared long before pions and other large `soft'
hadrons were formed. Hard Drell-Yan dileptons are produced essentially
at the time of the collision and then remain unaffected by the
subsequent history of the system; hence they are not able to provide
information on thermalisation and freeze-out.

\par

The basis for a quark-gluon plasma is the high parton density in the
early stages after the collision; this effectively screens all
confining long range forces, so that only short range interactions
remain. Any deconfinement probe must therefore
\begin{itemize}
\vspace*{-0.2cm}
\item{be hard enough to resolve sub-hadronic scales,}
\vspace*{-0.2cm}
\item{be able to distinguish confinement and deconfinement,}
\vspace*{-0.2cm}
\item{be present in the early stages of the evolution, and}
\vspace*{-0.2cm}
\item{retain the information throughout the subsequent evolution.}
\vspace*{-0.2cm}
\end{itemize}
The last point requires that the probe should not be in thermal
equilibrium with the later evolution stages, since this would lead to a
loss of memory of the previous stages.

\par

So far, two types of probes satisfying these conditions fully or in
part have been considered.
\begin{itemize}
\vspace*{-0.2cm}
\item{{\sl External} probes are produced essentially by primary parton
collisions, before the existence of any medium, and then indicate by
their observed behaviour whether the subsequent medium was deconfined or
not. The production of quarkonium states (\J, \U) provides the best
known example of such a probe. Short distance QCD predicts that the
dissociation of quarkonia is possible only in a deconfined medium,
since it requires harder gluons than those present in hadrons. Another
example, quite likely to gain considerable importance when the high
energy colliders RHIC and LHC come into operation, is the
energy loss or attenuation of hard jets, which is expected to increase
considerably in a deconfined medium.}
\vspace*{-0.2cm}
\item{{\sl Internal} probes are produced by the quark-gluon plasma
itself. Since they must leave the medium without being affected by its
subsequent states, they should undergo only weak or electromagnetic
interactions after their formation; the main candidates here are thermal
dileptons or photons. However, since both can also be produced by a
confined medium, they can serve as thermometer of the medium rather
than as probe for its confinement status. So far, another problem is
their identification: both dileptons and photons are produced
abundantly in hadron decays, and separating out a possible thermal
component appears very difficult.}
\vspace*{-0.2cm}
\end{itemize}
At the present stage of the investigation, quarkonium dissociation and
jet quenching therefore appear the most promising direct signatures for
deconfinement.

\par

Before concentrating on these hard deconfinement probes, we will first
consider electromagnetic probes in somewhat more detail. Real or
virtual photons emitted during the evolution of the collision
subsequently undergo no (strong) interactions with the medium and hence
reflect its state at the time they were produced. On the other hand,
they are emitted during the entire collision evolution, and by different
dynamical mechanisms at different stages:
\begin{itemize}
\vspace*{-0.2cm}
\item Early hard parton interactions produce hard photons and
Drell-Yan dileptons; these provide information about the initial
(primary or pre-equilibrium) stages.
\vspace*{-0.2cm}
\item Thermal photon and dilepton emission by the medium,
through quark or hadron interactions, occur through its entire
evolution, and hence give information about the successive stages,
from quark-gluon plasma to final hadronic freeze-out.
\vspace*{-0.2cm}
\item Hadrons produced at any point of the hadronic stage, from
the quark-hadron transition to freeze-out, can decay and thereby
emit photons or dileptons; depending on the hadron decay time, they
provide information about dense interacting hadronic matter or about
hadrosynthesis at the end of the strong interaction era.
\vspace*{-0.2cm}
\end{itemize}
Each of these formation mechanisms provides information about specific
aspects of the different collision stages.

\par

Drell-Yan dileptons and hard photons are the tools to study the
effective initial state parton distributions; in particular, they will
show any nuclear modifications (shadowing, anti-shadowing, coherence
effects) of these distributions. They also indicate the initial state
energy loss and the initial state $p_T$ broadening suffered by partons
in normal nuclear matter. Since they do not undergo any final state
strong interactions, they moreover provide a reference for
the effect of the produced medium on quarkonium states or jets.

\par

Thermal emission can in principle be used to determine the temperature
at the different evolution stages. The {\it functional form} of thermal
spectra,
\be
dN/dk_{\gamma} \sim e^{-k_{\gamma}/T} \label{4.3}
\ee
for photon momenta, or the corresponding distributions in the
dilepton mass $M_{l^+l^-}$, indicate the temperature $T$ of the medium
at the time the signal was emitted. Since the form (\ref{4.3}) for
thermal production is the same for a hadronic medium and for a
quark-gluon plasma, it cannot specify the nature of the emitting medium,
only its temperature; this feature finds support in various cascade
model calculations \cite{cascade,Wang}. The crucial thermometric problem
is to find a ``thermal window", since the measured spectra are dominated
at high photon momenta or dilepton masses by hard primary reactions and
at low momenta or masses by hadron decay products. As mentioned,
thermal photon or dilepton emission has so far not been identified,
perhaps because of the dominance of the hadronic stage at present
energies.

\par

The dileptons produced by hadron decays can probe hadronic in-medium
modifications, provided the hadrons actually decay within the
me\-dium. The $\rho$, with a half-life of about a fermi, appears to be
the best candidate for such studies. Chiral symmetry restoration is
expected to change the properties of hadrons as the temperature of the
medium approaches the restoration point \cite{chiral}; hence such
in-medium changes are of particular interest, since they might be the
only experimental tool to address the chiral aspects of deconfinement.
If at the onset of chiral symmetry restoration, the mass of the $\rho$
decreases sufficiently much
\cite{Brown},
\be
{m_{\rho}(T)\over m_{\rho}(T=0)}~ \to~ 0~~{\rm as} ~~T~\to~T_c,
\label{4.4}
\ee
then this should lead to an observable enhancement of the low mass
dilepton spectrum. There are first indications for such an enhancement
\cite{Helios} - \cite{NA38dilep}; however, so far alternative
explanations based on in-medium hadron interactions, leading to a
change of resonance widths, can also account for the effect
\cite{Wambach}. Moreover, a temperature dependence of the form
(\ref{4.4}) has so far not been corroborated by lattice studies of
hadron masses \cite{Boyd}.

\par

In the following two subsections, we shall discuss how the behaviour of
quarkonium states and of hard jets can provide information about the
confinement status of a given medium. In order to apply the results as
probes in actual nuclear collision experiments, the production of the
probes in the collision process must be understood as well, which
clearly complicates matters. In section 5, we shall carry out a full
analysis of this type for charmonium production and suppression in
nuclear collisions; for this, there now exists a wealth of beautiful
data \cite{Kluberg}.

\bigskip

\noindent{\bf 4.2 Quarkonium Dissociation}

\bigskip

We begin our study of deconfinement probes by restating the basic
question: Given a box of unidentified matter, determine if the
quarks and gluons which make up this matter are in a confined or
deconfined state. It is thus not evidence for quarks and gluons that we
look for, but rather the confinement/deconfinement status of these
elementary building blocks of all forms of matter.

\par

As quarkonium prototype, consider the \J; it is the $1S$ bound state
of a charm quark ($m_c \simeq 1.4$ GeV) and its antiquark, with $M_{\j}
\simeq 3.1$ GeV. Its usefulness as deconfinement probe is easily seen.
If a \J~is placed into a hot medium of deconfined quarks and gluons,
colour screening will dissolve the binding, so that the $c$ and the
$\bar c$ separate. When the medium cools down to the confinement
transition point, they will therefore in general be too far apart to
see each other. Since thermal production of further $\C$ pairs is
negligibly small because of the high charm quark mass, the $c$ must
combine with a light antiquark to form a $D$, and the $\bar c$ with a
light quark for a $\bar D$. The presence of a quark-gluon plasma will
thus lead to a suppression of \J~production \cite{Matsui}.

\par

We shall now first consider this \J~suppression in terms of colour
screening, i.e., as consequence of global features of the medium, and
then turn to a microscopic approach, in which the bound state is
assumed to be broken up by collisions with constituents of the medium.

\medskip

\noindent{\bf 4.2a.\ \J~Suppression by Colour Screening}

\medskip

Because of the large charm quark mass, the charmonium spectrum can be
calculated with good precision by means of the Schr\"odinger equation
\cite{Schroedinger}
\be
[2m_c + {1 \over m_c}\nabla^2 + V(r)] \Psi_{n,l} = M_{n,l} \Psi_{n,l},
\label{3}
\ee
where the potential $V(r)=\sigma r - \alpha/r$ contains a confining
long-distance part $\sigma r$ and a Coulomb-like short-distance term
$\alpha/r$. For different values of the principal quantum number $n$
and the orbital quantum number $l$, the masses $M_{n,l}$ and the wave
functions $\Psi_{n,l}(r)$ of different charmonium states
$\j,~\chi, ~\psi',~...$ in vacuum are given in terms of the
constants $m_c,~\sigma$ and $\alpha$.

\par

In a medium, the potential becomes screened, as we had seen above in
Eq.\ \ref{2.3}; we thus obtain
\be
V(r,\mu) = {\sigma \over \mu}[1 - e^{-\mu r}]  - {\alpha \over r}
e^{-\mu r}, \label{4}
\ee
where $\mu$ is the screening mass and $r_D = \mu^{-1}$ the `Debye'
colour screening radius. Screening is a global feature of the medium,
shortening the range of the binding potential. Once $\mu$ becomes
sufficiently large,
the bound states begin to disappear, starting with the most weakly
bound; hence for $\mu \geq \mu_d^i$, the bound state $i$ is no longer
possible \cite{KMS}.

\medskip

\begin{center}
\begin{tabular}{|c||c|c|c||c|c|c|c|c|}
\hline
&  &  & & & & & & \\
 state & $\psi$(1S) & \P(2S) & $\chi_c$(1P) & \U(1S) & \U$^\prime$(2S) 
& \U$^{\prime\prime}$(3S)  & $\chi_b$(1P)  & $\chi_b'$ (2P) \\
&  &  & & & & & & \\
\hline
\hline
&  &  & & & & & & \\
$\mu_d$~[GeV] & 0.68 & 0.35  & 0.35  & 1.52 & 0.64 & 0.43 & 0.59 & 0.42 \\
&  &  & & & & & & \\
\hline
&  &  & & & & & & \\
$T_d/T_c$ & 1.2  & 1 & 1 & 2.7 & 1.1 & 1 & 1.1 & 1 \\
&  &  & & & & & & \\
\hline
\hline
&  &  & & & & & & \\
$\Delta E$~[GeV] & 0.64 & 0.06 & 0.24 & 1.10 & 0.54 & 0.20 & 0.67 &
0.31 \\
&  &  & & & & & & \\
\hline
\end{tabular}\end{center}

\centerline{Table 2: Quarkonium Dissociation by Colour Screening}

\bigskip

With the help of finite temperature lattice QCD, we can now determine
the screening mass $\mu(T)$ as function of the temperature $T$. With the
approximate relation \cite{HKR} $\mu \simeq 3.3~T$ and $T_c \simeq 0.17$
GeV, one obtains the melting pattern for the most important $\C$ and
$\b$ states summarized in Table 2; we see that while both \P~and
\X~melt at the critical deconfinement point, the \J, being smaller and
more tightly bound, survives to about 2$T_c$ and hence about twice the
critical energy density. A similar pattern holds for the bottonium
states; the dissociation values of the screening masses for $\chi_b'$
and \U''~have not yet been determined in potential theory. With
increasing temperature, a hot medium will thus lead to successive
quarkonium melting, so that the suppression or survival of specific
quarkonium states serves as a thermometer for the medium, in much the
same way as the relative intensity of spectral lines in stellar
interiors measure the temperature of stellar matter \cite{Kajantie}.
Note, however, that other possible sources of quarkonium dissociation
have to be considered before the method becomes unambiguous. This
problem will be addressed in detail in section 5.

\medskip

\noindent
{\bf 4.2b.\ \J~Suppression by Gluon Dissociation}

\medskip

The binding energy of the \J, i.e., the energy difference between the
\J~mass and the open charm threshold, 
$\Delta E_{\j}=2M_D - M_{\j} \simeq 0.64~{\rm GeV}$ is
considerably larger than the typical non-perturbative hadronic scale
$\L \simeq 0.2$ GeV. As a consequence, the size of the \J~is much
smaller than that of a typical hadron, $r_{\j} \simeq 0.2~{\rm fm} \ll
\L^{-1} = 1~{\rm fm}$. Hence the \J~is a hadron with characteristic
short-distance features; in particular, rather hard gluons are
necessary to resolve or dissociate it, making such a dissociation
accessible to perturbative calculations. \J~collisions with ordinary
hadrons made up of the usual $u,d$ and $s$ quarks thus probe the local
partonic structure of these `light' hadrons, not their global hadronic
aspects, such as mass or size. It is for this reason that \J's can be
used as a confinement/deconfinement probe.

\par

This can be illustrated by a simple example. Consider an ideal pion gas
as a confined medium. The momentum spectrum of pions has the Boltzmann
form $f(p) \sim \exp-(|p|/T)$, giving the pions an average momentum
$\langle |p| \rangle = 3~T$. With the pionic gluon distribution function
$xg(x) \sim (1-x)^3$, where $x=k/p$ denotes the fraction of the pion
momentum carried by a gluon, the average momenta of gluons confined to
pions becomes
\be
\langle |k| \rangle_{\rm conf}  \simeq 0.6~T. \label{5}
\ee
On the other hand, an ideal QGP as prototype of a deconfined medium
gives the gluons themselves the Boltzmann distribution $f(k) \sim
\exp-(|k|/T)$ and hence average momenta
\be
\langle |k| \rangle_{\rm deconf} = 3~T. \label{6}
\ee

\begin{figure}[tbp]
\vspace*{-0mm}
\centerline{\psfig{file=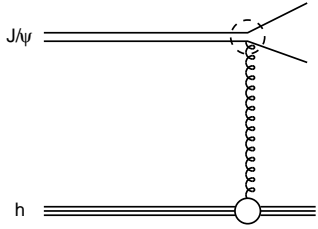,height=30mm}}
%\vspace*{-20mm}
\caption{\J~dissociation by hadron interaction.}
\label{4_1}
\end{figure}
\begin{figure}[tbp]
\vspace*{-0mm}
\centerline{\psfig{file=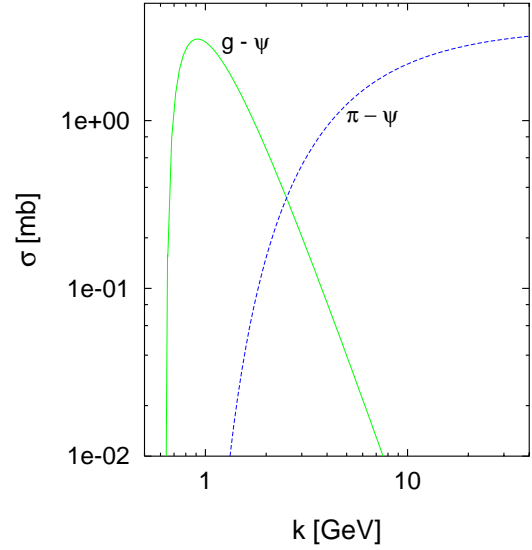,height= 75mm,angle= -90}}
%\vspace*{-20mm}
\caption{\J~dissociation by gluons and by pions; $k$ denotes the
momentum of the projectile incident on a stationary \J.}
\label{4_2}
\end{figure}
\begin{figure}[tbp]
%\vspace*{-0mm}
\centerline{\psfig{file=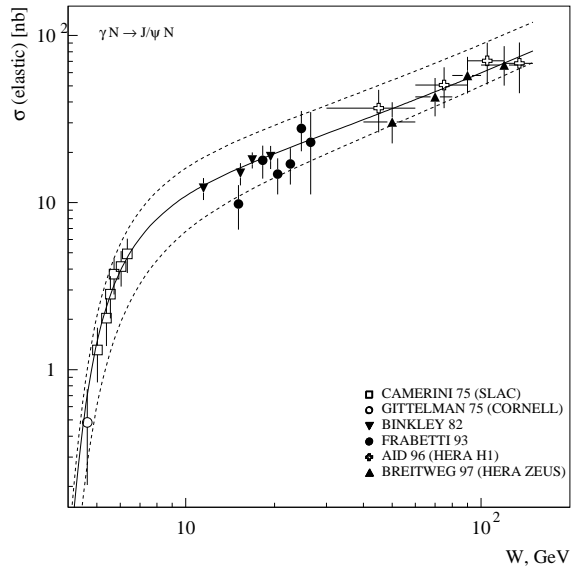,height=75mm}}
%\vspace*{-20mm}
\caption{Elastic \J~photoproduction data, compared to the prediction
obtained from short-distance QCD and VDM \cite{Arkady}.}
\label{4_2a}
\end{figure}

Deconfinement thus results in a hardening of the gluon momentum
distribution. More generally speaking, the onset of deconfinement will
lead to parton distribution functions which are different from those
in vacuum, as determined by deep inelastic scattering experiments.
Since hard gluons are needed to resolve and dissociate \J's, one can use
\J's to probe the in-medium gluon hardness and hence the confinement
status of the medium.

\par

These qualitative considerations can be put on a solid theoretical basis
provided by short-distance QCD \cite{Peskin} -- \cite{KS3}. In
Fig.\ \ref{4_1} we show the relevant diagram for the calculation
of the inelastic
\J-hadron cross section, as obtained in the operator product expansion
(essentially a multipole expansion for the charmonium quark-antiquark
system). The upper part of the figure shows \J~dissociation by gluon
interaction; the cross section for this process,
\be
\sigma_{g-\j} \sim (k-\Delta E_{\psi})^{3/2}  k^{-5}, \label{7}
\ee
constitutes the QCD analogue of the photo-effect. Convoluting the \J~
gluon-dissociation with the gluon distribution in the incident hadron,
$xg(x) \simeq 0.5(1-x)^{1+n}$, we obtain
\be
\sigma_{h-\j} \simeq \sigma_{\rm geom} (1 - \lambda_0/\lambda)^{n+3.5}
\label{8}
\ee
for the inelastic \J-hadron cross section, with $\lambda \simeq
(s-M_{\psi}^2)/M_{\psi}$ and $\lambda_0 \simeq (M_h + \Delta E_{\psi}$);
$s$ denotes the squared \J-hadron collision energy. In Eq.\ (\ref{8}),
$\sigma_{\rm geom} \simeq {\rm const}.\ r_{\psi}^2 \simeq 2 - 3$ mb is
the geometric cross section attained at high collision energies with the
mentioned gluon distribution. In the threshold region and for
relatively low collision energies, $\sigma_{h-\j}$ is very strongly
damped because of the suppression $(1-x)^{1+n}$ of hard gluons in
hadrons, which leads to the factor $(1 - \lambda_0/\lambda)^{n+3.5}$ in
Eq.\ (\ref{8}). In Fig.\ \ref{4_2}, we compare the cross sections
for \J~dissociation by gluons (``gluo-effect") and by pions ($n=2$), as
given by Eq's (\ref{7}) and (\ref{8}).
Gluon dissociation shows the typical photo-effect form, vanishing until
the gluon momentum $k$ passes the binding energy $\Delta E_{\psi}$;
it peaks just a little later and then vanishes again when
sufficiently hard gluons just pass through the much larger charmonium
bound states. In contrast, the \J-hadron cross section remains
negligibly small until rather high hadron momenta (3 - 4 GeV). In a
thermal medium, such momenta correspond to temperatures of more than one
GeV. Hence confined media in the temperature range of a few hundred MeV
are essentially transparent to \J's, while deconfined media of the
same temperatures very effectively dissociate them and thus are
\J-opaque.

\par

The formalism leading to Eq.\ (\ref{8}) is also applicable to
\J~photoproduction, where it can be compared directly to data \cite{KS3}.
An alternative relation between \J~photoproduction and
\J~break-up by hadrons is obtained through a combination of the
vector meson dominance model and the short-distance QCD approach
\cite{Arkady}. In both cases one finds clear evidence for the strong
threshold suppression due to the form of the gluon distribution
function; an illustration is given in Fig.\ \ref{4_2a}.
In contrast, several phenomenological models based on effective hadronic
Lagrangians \cite{Blaschke} - \cite {Ko}
suggest a $\sigma_{g-\j}$ which is very much larger than Eq.\ (\ref{8})
in the threshold region. It seems unlikely that such models can survive
a comparison to \J~photoproduction.

\par

The situation for \X's is quite similar, except for an earlier gluon
dissociation threshold due to the lower \X~binding energy of about 250
MeV. Corresponding differences occur for the various bottonium states.
The difference in binding energies thus provides the micro\-scopic basis
for the successive melting of different quarkonium states noted above.
The binding energies for the quarkonium states considered above are
included in Table 2. We note in particular that for
the \P,  $\Delta E$ is almost negligible (about 40 MeV),
so that there cannot really be any difference in its
dissociation by confined or deconfined media. In other words, any
form of strongly interacting matter is expected to be \P-opaque.

\par

We can thus define a schematic test to determine the confinement status
of a box of unidentified matter. We first shine a \P~beam at it: if it
is transparent to this, the box does not contain strongly interacting
matter; otherwise it does. In that case we repeat the procedure with a
\J~beam: if this is unaffected, the medium in the box is confined; if
the beam is attenuated, it is deconfined. The problem of an unambiguous
deconfinement test is thus in principle solved: there is \J~dissociation
if and only if the medium is deconfined. However, the test pre-supposes
the existence of a prepared strongly interacting medium and the
availability of \J~and \P~beams as probes, and in nuclear interactions,
both have to be prepared by the collision process. The actual
application of such a test is therefore more complex; it will be taken
up in detail in section 5.

%\bigskip
\newpage

\noindent{\bf 4.3 Jet Quenching}

\bigskip

One important aspect in the use of quarkonium states as deconfinement
test is that in a collision with usual `big light' hadrons, the small
quarkonia probe only the local partonic structure of their collision
partner, not global features such as mass or size. This aspect is also
the basis for using hard jets to study the confinement status of a given
medium.

\par

\begin{figure}[tbp]
\vspace*{-0mm}
\centerline{\psfig{file=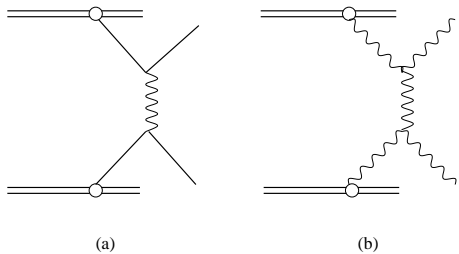,height=60mm,angle=-90}}
%\vspace*{-20mm}
\caption{Some lowest order diagrams for jet production in hadron collisions.}
\label{4_3}
\end{figure}

Hard jets are produced through hard partonic interactions in the very
early stages of nucleon-nucleon collisions; some typical processes are
illustrated in Fig.\ \ref{4_3}. In nuclear collisions, the produced
gluons
or quarks leave their formation point with a very high momentum transverse
to the collision axis, and in doing so, pass through the primary
nuclear matter and through whatever secondary medium is produced by the
collision. The effect of such a passage through a quark-gluon plasma has
in recent years been studied in considerable detail, leading to rather
well-defined predictions \cite{G-W} - \cite{Zak}.

\par

An electric charge, passing through matter containing other bound or
unbound charges, loses energy by scattering. For charges of low incident
energy $E$, the energy loss is largely due to ionization of the target
matter. For sufficiently high energies, the incident charge scatters
directly on the charges in matter and as a result radiates photons of
average energy $\omega \sim E$. Per unit length of matter, the
`radiative' energy loss due to successive scatterings,
\be
-{dE \over dz} \sim E \label{9}
\ee
is thus proportional to the incident energy.

\par

This probabilistic picture of independent successive scatterings breaks
down at very high incident energies \cite{LPM}. The squared amplitude
for $n$ scatterings now no longer factorizes into $n$ interactions;
instead, there is destructive interference, which for a regular medium
(crystal) leads to a complete cancellation of all photon emission except
for the first and last of the $n$ photons. This
Landau-Pomeranchuk-Migdal (LPM) effect greatly reduces the radiative
energy loss.

\par

\begin{figure}[tbp]
\vspace*{-0mm}
\centerline{\psfig{file=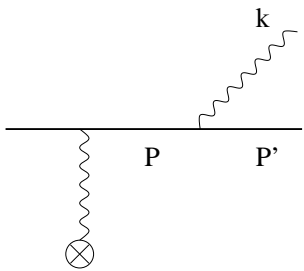,height= 40mm,angle=-90}}
%\vspace*{-20mm}
\caption{Gluon emission after scattering.}
\label{4_4}
\end{figure}

The physics of the LPM effect is clearly relevant in calculating the
energy loss for fast colour charges in QCD media. These media are not
regular crystals, so that the cancellation becomes only partial. Let us
consider the effect here in a heuristic fashion; for details of the
actual calculations, see \cite{Baier,Zak}. The time $t_c$ needed for the
emission of a gluon after the scattering of a quark (see Fig.
\ref{4_4}) is given by
\begin{figure}[tbp]
%\vspace*{-0mm}
\centerline{\psfig{file=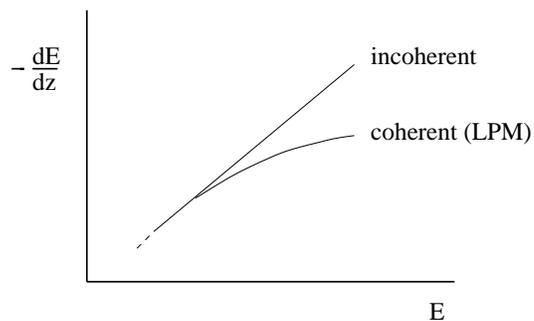,height=70mm,angle=-90}\hspace*{3mm}}
%\vspace*{-20mm}
\caption{Energy loss in incoherent and coherent interactions.}
\label{4_5}
\end{figure}
\begin{figure}[tbp]
%\vspace*{-0mm}
\centerline{\psfig{file=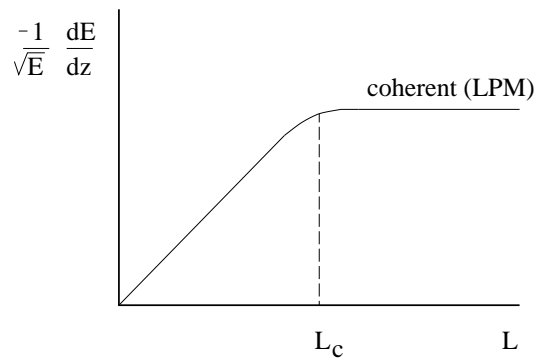,height=70mm,angle=-90}}
%\vspace*{-20mm}
\caption{Energy loss in coherent interactions
as function of the thickness $L$ of the medium}
\label{4_6}
\end{figure}

\noindent
\be
t_c = {1 \over \sqrt{P^2}}{E \over \sqrt{P^2}} =
{E\over 2P'k}, \label{9a}
\ee
in the rest system of the scattering center, where $P^2$ measures 
how far the intermediate quark state
is off-shell; on-shell quarks and gluons are assumed to be massless, and
$E/\sqrt{P^2}$ is the $\gamma$-factor between the lab frame and the
proper frame of the intermediate quark. For gluons with $k_L >> k_T$,
we thus get
\be
t_c \simeq {\omega \over k_T^2}. \label{9b}
\ee
If the passing colour charge can interact with several scattering
centers during the formation time of a gluon, the
corresponding amplitudes interfere destructively, so
that in effect after the passage of $n$ centers over the
coherence length $z_c$, only one gluon is emitted, in contrast to
the emission of $n$ gluons in the incoherent regime. Nevertheless,
in both cases each scattering leads to a $k_T$-kick of the charge,
so that after a random walk past $n$ centers, $k_T^2 \sim n$. Hence
\be
k_T^2 \simeq \mu^2 {z_c\over \lambda}, \label{9c}
\ee
where $\lambda$ is the mean free path of the charge in the medium, so
that $z_c/ \lambda >1$ counts the number of scatterings. At
each scattering, the transverse kick received is measured by the mass
of the gluon exchanged between the charge and the scattering center, i.e., by
the screening mass $\mu$ of the medium. From Eq.\ \ref{9b} we have
\be
z_c \simeq {\omega \over k_T^2}, \label{9d}
\ee
so that the formation length in a medium characterized by $\mu$ and
$\lambda$ becomes
\be
z_c \simeq \sqrt{{\lambda \over \mu^2} \omega}. \label{9e}
\ee
For the validity of Eq.\ (\ref{9e}), the mean free path has to be larger
than the interaction range of the centers, i.e., $\lambda > \mu^{-1}$.

\par

The energy loss of the passing colour charge is now determined by
the relative scales of the process. If $\lambda >z_c$, we have
incoherence, while for $\lambda < z_c$ there is coherent scattering with
destructive interference. In both cases, we have assumed that the
thickness $L$ of the medium is larger than all other scales. When
the coherence length reaches the size of the system, $z_c = L$,
effectively only one gluon can be emitted. This defines a critical
thickness $L_c(E)=(E \lambda / \mu^2)^{1/2}$ at fixed incident energy
$E$, or equivalently a critical $E_c=\mu^2 L^2 /\lambda$ for fixed
thickness $L$; for $L > L_c$, there is bulk LPM-behaviour, below $L_c$
there are finite-size corrections.

\par

We are thus left with three regimes for radiative energy loss. In case
of incoherence, $z_c < \mu^{-1}$, there is the classical radiative loss
\be
-{dE \over dz} \simeq {3 \alpha_s \over \pi} {E \over \lambda},
\label{9f}
\ee
where $\alpha_s$ is the strong coupling.
In the coherent region, $\lambda >z_c$, the energy loss is given by the
LPM bulk expression when $L > L_c$ \cite{Baier},
\be
-{dE \over dz} \simeq {3 \alpha_s \over \pi} \sqrt{{\mu^2 E \over \lambda}}.
\label{10}
\ee
The resulting reduction in the radiative energy loss $dE/dz$ is
illustrated in Fig.\ \ref{4_5}. Note that in earlier estimates
the energy loss due to interactions of the gluon cloud accompanying the
passing colour charge had been neglected \cite{G-W}; this led to a
considerably smaller energy loss, proportional to $\ln E$ instead of
$\sqrt E$. Finally, in a medium of thickness $L < L_c$, there is
less scattering and hence still less energy loss. Eq.\ (\ref{10}) can be
rewritten as
\be
-{dE \over dz} \simeq {3 \alpha_s  \over \pi}{\mu^2 \over \lambda} L_c(E),
\label{11}
\ee
and for $L<L_c$, this leads to
\be
-{dE \over dz} \simeq {3 \alpha_s \over \pi} {\mu^2 \over \lambda} L
\label{12}
\ee
as the energy loss in finite size media with $L \leq L_c$.
The resulting variation of the radiative energy loss with the
thickness of the medium is shown in Fig.\ \ref{4_6}, with
saturated (i.e., bulk) LPM behaviour setting in for $L\geq L_c$.

\par

Eq.\ (\ref{12}) has been used to compare the energy loss in a deconfined
medium of temperature $T =0.25$ GeV to that in cold nuclear matter of
standard density \cite{Schiff}. For the traversal of a medium of 10 fm
thickness, estimates give for the total energy loss
\be
\Delta E = \int_{0~\rm fm}^{10~\rm fm} dz {dE \over dz} \label{12a}
\ee
in a quark-gluon plasma
\be
-\Delta E_{qgp}  \simeq 30~~{\rm GeV},
\label{13}
\ee
corresponding to an average loss of 3 GeV/fm. In contrast, cold
nuclear matter leads to
\be
-\Delta E_{cnm} \simeq 2~~{\rm GeV}
\label{14}
\ee
and hence an average loss of 0.2 GeV/fm. A deconfined medium thus leads
to a very much higher rate of jet quenching than confined hadronic
matter, as had in fact been suggested quite some time ago \cite{Bj-jet}.

\par

To convert this into an operational signature for deconfinement,
considerably more work is needed \cite{Dok,cone}.
Some further basic ingredients needed
for this effort will become clear in the next section, where the
corresponding task is carried out for \J~suppression.

\bigskip~\medskip

\noindent
{\bf 5.\ \J~SUPPRESSION IN NUCLEAR COLLISIONS}

\bigskip

In this section, we shall present a realistic deconfinement study, based
on the analysis of \J~production in nuclear collisions. On one hand,
this provides a nice application of the general ideas presented above;
on the other hand, it illustrates the steps necessary in any
other, future deconfinement study. First, the production of the
probe must be understood in hadronic collisions, both theoretically and
experimentally. We therefore begin with the hadroproduction of
charmonium. Next, one has to understand the modifications arising
when the production occurs in a confined medium, for which $p\!-\!A$
collisions provide the experimental reference. With the probe thus
(a) prepared and (b) gauged to confined matter, it can be applied in an
environment in which there might be deconfinement.

\bigskip

\noindent{\bf 5.1\ The Hadroproduction of Charmonium}

\bigskip

\begin{figure}[tbp]
\centerline{\psfig{file=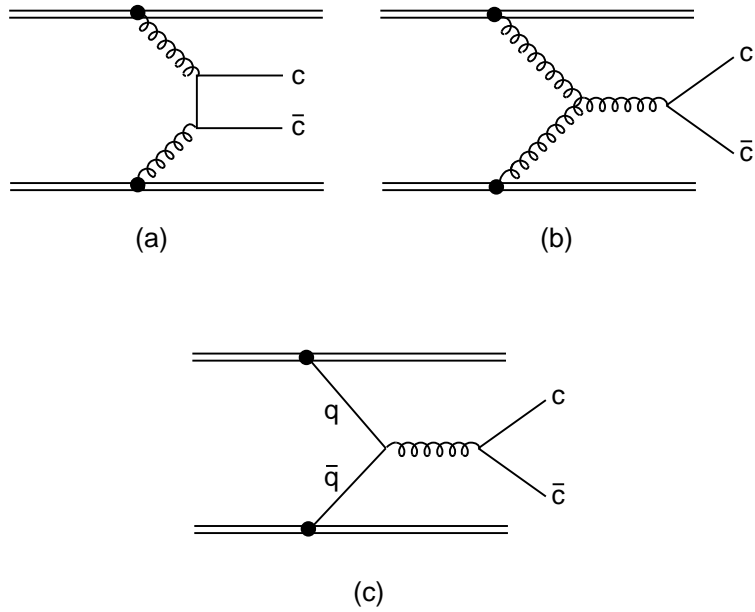,height=80mm}}
\caption{Lowest order diagrams for $\C$ production in hadronic
collisions, through (a,b) gluon fusion and (c) quark-antiquark
annihilation.}
\label{5_1}
\end{figure}

The first stage of charmonium formation in hadronic collisions is the
production of a $\C$ pair; because of the large quark mass, this
process can be described by perturbative QCD (Fig.\ \ref{5_1}).
A parton from the projectile interacts with one from the target; the
(non-perturbative) parton distributions within the hadrons are
determined empirically in other reactions, e.g.\ by deep inelastic
lepton-hadron scattering. Initially, the $\C$ pair will in general be in
a colour octet state. It subsequently neutralises its colour and binds
to a physical resonance, such as \J, $\chi_c$ or \P. Colour
neutralisation occurs by interaction with the surrounding colour field;
this and the subsequent resonance binding are presumably of
non-perturbative nature. The colour evaporation model \cite{CE}
provides the simplest and most general phenomenological approach to
colour neutralisation. In the evaporation process, the $\C$ pair can
either combine with light quarks to form open charm mesons ($D$ and
$\bar D$) or bind with each other to form a charmonium state.

\begin{figure}[tbp]
%\vspace{-1cm}
\centerline{\psfig{file=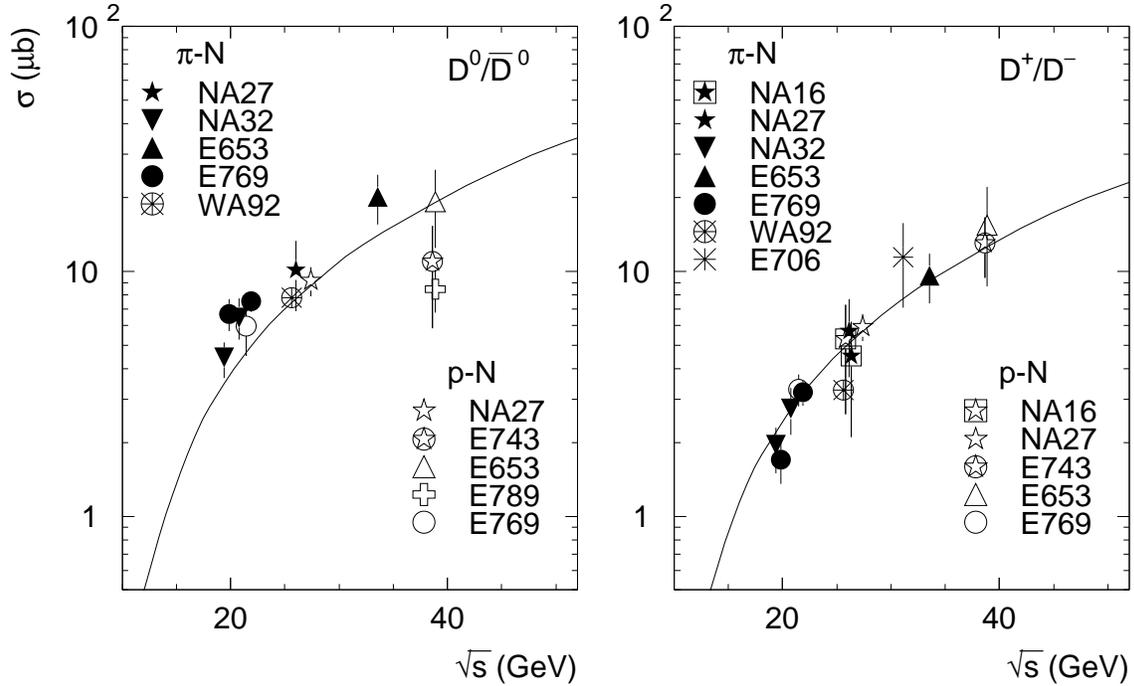,height=90mm}}
%\vspace{-1.2cm}
\caption{Open charm hadroproduction at different cms energies, compared
to PYTHIA calculations \cite{PBM-charm}.}
\label{5_2}
\end{figure}

The basic quantity in this description is the total sub-threshold
charm cross section, obtained by integrating the perturbative $\C$
production cross section
over the mass interval from $2m_c$ to $2m_D$. At high
energy, the dominant part of $\S$ comes from gluon fusion (Fig.\
\ref{5_1}a), so that we have
\be
\S(s) = \int_{2m_c}^{2m_D} d\hat s \int dx_1 dx_2~g_p(x_1)~g_t(x_2)~
\sigma(\hat s)~\delta(\hat s-x_1x_2s), \label{5.1}
\ee
with $g_p(x)$ and $g_t(x)$ denoting the gluon densities, $x_p$ and $x_t$
the fractional momenta of gluons from projectile and target,
respectively; $\sigma$ is the $gg \to \C$ cross section. In
pion-nucleon collisions, there are also significant quark-antiquark
contributions (Fig.\ \ref{5_1}c), which become dominant at low
energies. The essential claim of the colour evaporation model is
that the production cross section of any charmonium state $i$
is given by
\be
\sigma_i(s)~=~f_i~\S(s), \label{5.2}
\ee
where $f_i$ is an energy-independent constant to be determined
empirically. It follows that the energy dependence of any charmonium
production cross section is predicted to be that of the perturbatively
calculated sub-threshold charm cross section. As a further consequence,
the production ratios of different charmonium states
\be
{\sigma_i(s)\over \sigma_j(s)} = {f_i\over f_j} = {\rm const.}
\label{5.3}
\ee
must be energy-independent. We note that only a small part of the total
subthreshold cross section $\S$ goes into charmonium formation. In
accord with perturbative open charm calculations, the remainder (more
than 90 \%) leads to $D{\bar D}$ production; the missing energy
needed to reach the $D{\bar D}$ threshold is obtained by interaction
with the colour field. The cross section for open charm production is
thus directly predicted by the form of Eq.\ (\ref{5.1}), with
the integration essentially running to $\sqrt s$, instead of ending at
$2m_D$. In Fig.\ \ref{5_2} we see that the result agrees well
with experiment
\cite{PBM-charm}.


\begin{figure}[tbp]
\centerline{\psfig{file=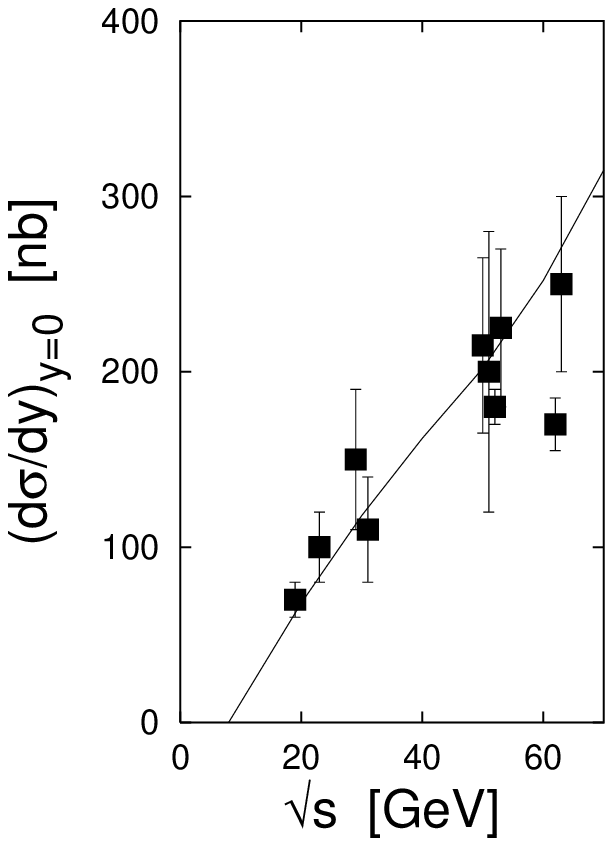,height=80mm}}
\vspace{-0.5cm}
\caption{The energy dependence of \J~hadroproduction based on MRS
D-' parton distributions, compared to data \cite{Quarko}.}
\label{5_3}
\end{figure}
\begin{figure}[tbp]
\centerline{\psfig{file=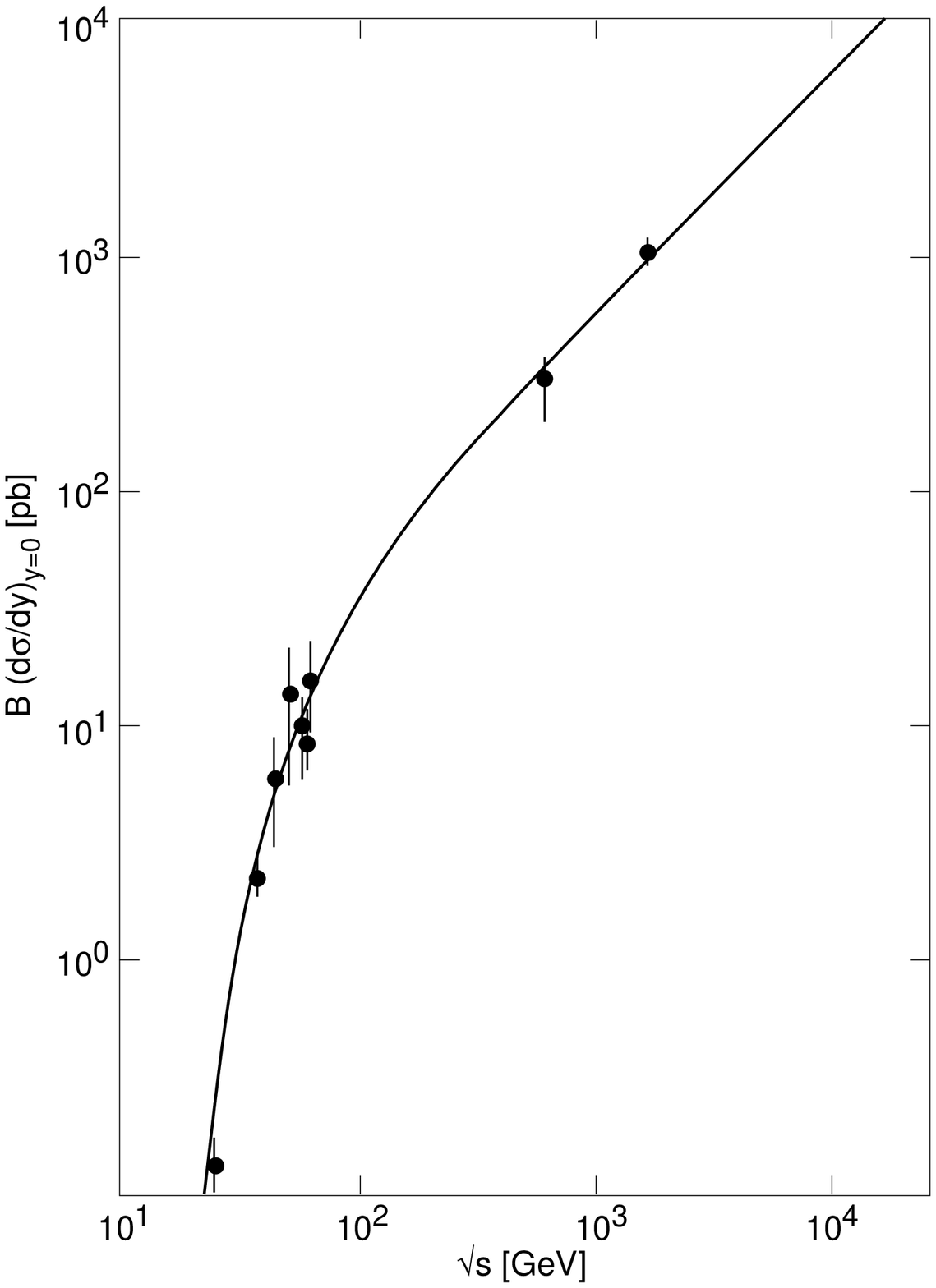,height=100mm}}
%\vspace{-0.5cm}
\caption{The energy dependence of bottonium hadroproduction based on MRS
D-' parton distributions, compared to data \cite{Quarko}.}
\label{5_4}
\end{figure}

%\medskip

The predictions of the colour evaporation model have been compared in a
comprehensive survey \cite{Quarko} to the available data, using parton
distribution functions which include the new HERA results
\cite{MRS,GRV}. In Figs.\ \ref{5_3} and \ref{5_4}, we
see that the energy-dependence is well described for both \J~and
\U~production; for \J~production, the normalisation coeffficient is
$f_{\j}$=0.025. The \U~results are obtained for the sum of \U, \U' and
\U'' decaying into dimuons, with $Bf_{\u}=1.6\times 10^{-3}$ for the
normalisation; here the branching ratios cannot be directly removed. In
the fixed-target/ISR energy range, the results from the two different
sets of parton distributions coincide; for the \J~at LHC energies,
there is some spread due to scale uncertainties in the parton
distributions, which hopefully can be removed by more precise DIS data.
For the \U~production, we have already now data up to 1.8 TeV, and in
Fig.\ \ref{5_4} they are seen to agree very well with the
prediction obtained using the ``low energy" value $Bf_{\u}=1.6\times
10^{-3}$.

\par

\begin{figure}[tbp]
\mbox{
\psfig{file=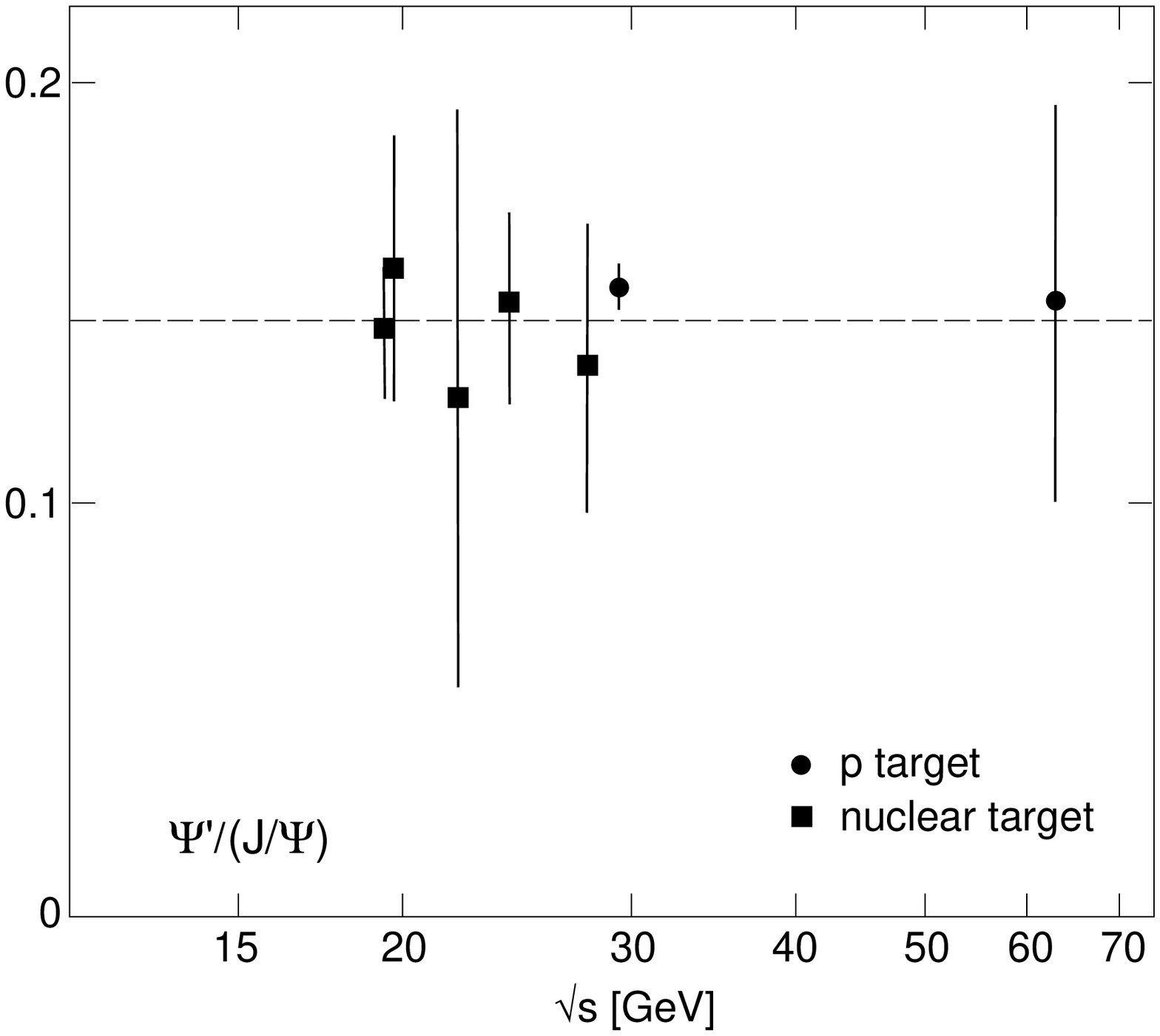,height=70mm,width=7.2cm}
\hskip1cm
\psfig{file=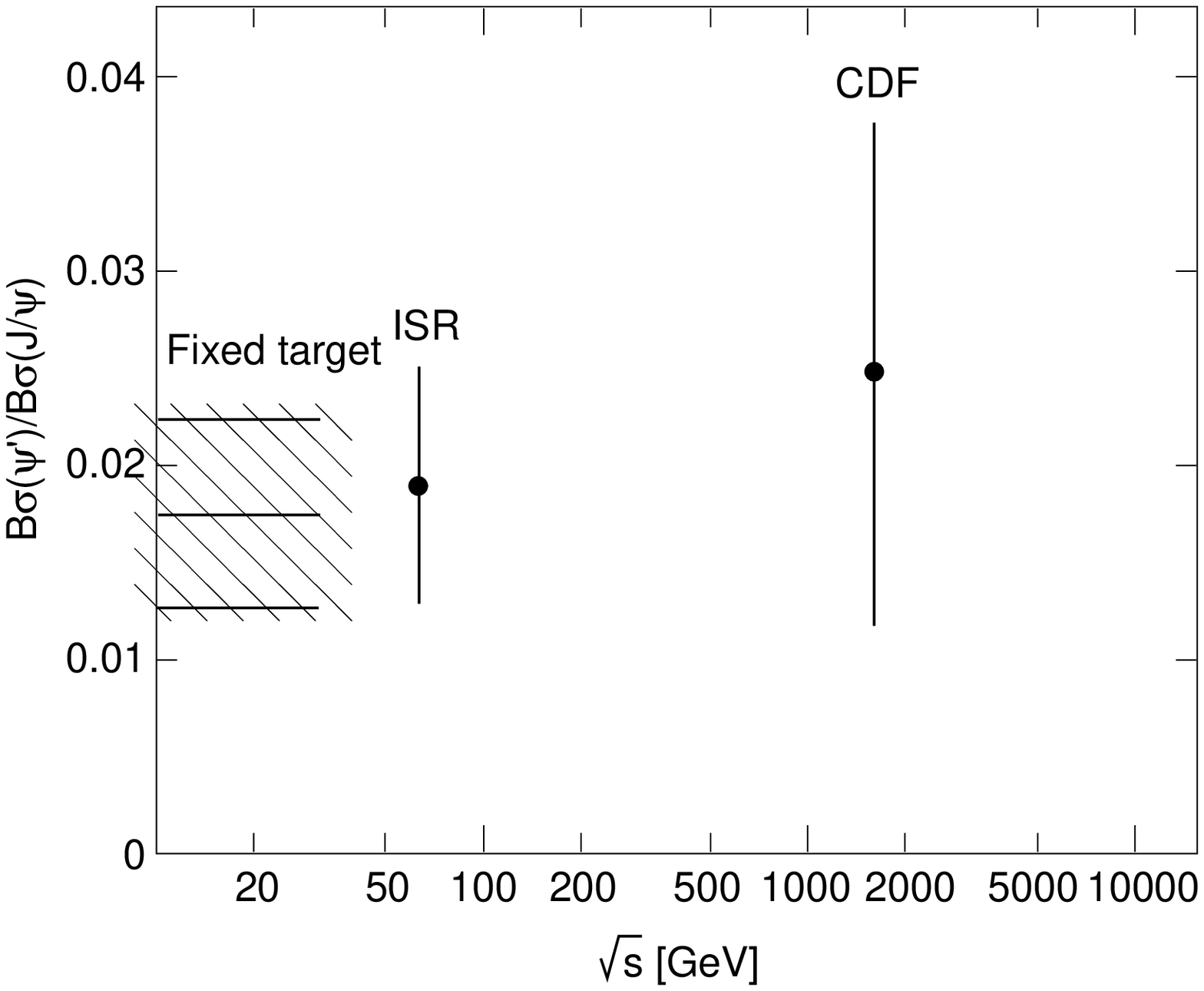,height=70mm,width=7.2cm}}

\caption{The ratio of \J~to \P~hadroproduction as function of the cms
energy \cite{Quarko}.}
\label{5_5}
\end{figure}

\begin{figure}[tbp]
\centerline{\psfig{file=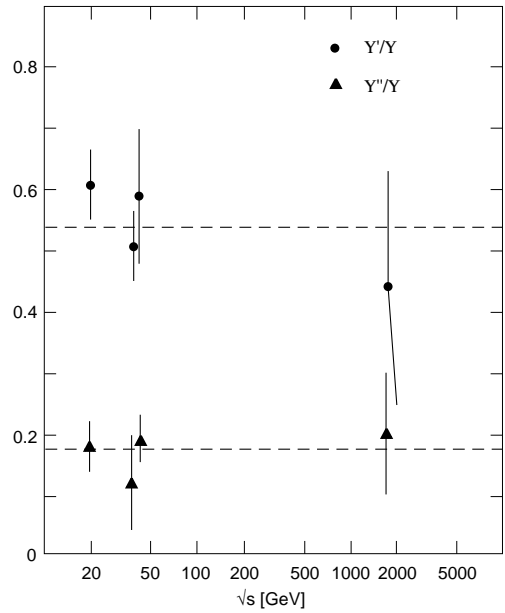,height=80mm}}

\caption{The ratio of \U' to \U~and \U'' to \U~hadroproduction as
function of the cms energy \cite{Quarko}.}
\label{5_5c}
\end{figure}

In Figs.\ \ref{5_5} and \ref{5_5c}, the predicted
energy-independence of production ratios is found to hold as well,
again up to Tevatron energy. Here it should be noted that
the CDF data for the ratio \P/(\J) are taken at large transverse momenta
($5\leq p_T\leq 15$ GeV), while the lower energy data are integrated
over $p_T$, with the low $p_T$ region dominant. Hence colour evaporation
appears to proceed in the same way at both small and large $p_T$.

\par

With the relative fractions $f_i$ for the different quarkonium states
determined, colour evaporation also predicts the longitudinal and
transverse momentum distributions for the hadroproduction of these
states. In both cases, the agreement is quite good \cite{Quarko,Ramona}.

\par

The colour evaporation model thus provides a viable phenomenological
description of the hadroproduction of quarkonia, leading to quantitative
predictions up to the highest energies under consideration. However, it
does not give a theoretical basis for the fractions $f_i$ of the hidden
charm cross section, and hence one cannot use it to determine the
space-time evolution of quarkonium production. For the production in
nuclear collisions, this is crucial, as we shall see shortly, and so we
have to consider more detailed dynamical mechanisms for hadroproduction.
The first such attempt, the colour singlet model, assumed that the
colour neutralisation of the $\C$ is perturbatively calculable
\cite{B-R}. The resulting well-defined predictions were found to
disagree by orders of magnitude with high energy data from FNAL
\cite{CDF}, so that non-perturbative features seem to be essential for
quarkonium production \cite{Bodwin,Braaten}. This has led to extensive
studies of possible mechanisms; see \cite{Mangano} for recent reviews.

\par

We shall here concentrate on one resulting proposal, the colour octet
model \cite{Bodwin,Braaten}, keeping in mind, however, that only
rather general features of such a model will enter the subsequent
applications to nuclear collisions. The basis of the colour octet model
is that any quark-antiquark bound state is in reality a superposition
of the basic $q\bar q$ state and fluctuations involving in adddition one
or more gluons (``higher Fock space components").
We can thus decompose the \J~state $|\psi \rangle$
\be
|\psi\rangle = a_0 |(\C)_1\rangle + a_1 |(\C)_8 g\rangle
+ a_2 |(\C)_1 gg \rangle + a'_2|(\C)_8 gg \rangle + ... \label{5.4}
\ee

\noindent
into a pure $\C$ colour singlet component ($^3S_1$), into a component
consisting of $\C$ colour octet ($^1S_0$ or $^3P_J$) plus a gluon,
and so on. The higher Fock space coefficients correspond to an
expansion in the relative velocity $v$ of the charm quarks. As we shall
see below, this corresponds in coordinate space to an expansion in
terms of components of decreasing spatial extension. For the wave
function of the \J, the higher components correspond to (generally
small) relativistic corrections. In high energy \J~production, however,
the higher Fock space components of charmonium states play an important
role. The short time available before confinement constraints appear,
favours colour neutralisation of the $(\C)_8$ by gluons already
present. Since $(\C)$ production at high energies occurs at small $x$,
the density of such comoving gluons is in fact high, and so the
higher Fock space components become dominant.

\par

Analogous Fock space decompositions hold for the other charmonium states
\cite{Bodwin,Braaten}. In all cases, the first higher state consists of
a colour octet $\C$ plus a gluon. For the \P, the next-to-leading terms
again contain a colour octet ($^3P_J$ or $^1S_0$ $\C$) plus a gluon;
for the $\chi$'s, a $^3S_1$ colour octet $\C$ is combined with a gluon.

\par

\begin{figure}[htb]
\vspace*{-0mm}
\epsfysize=6cm
\centerline{\epsffile{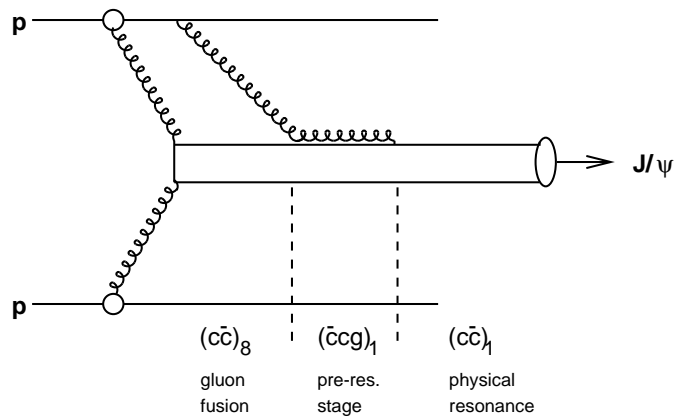}}
\caption{\J~production in $p-p$ collisions.}
\label{5_6}
\end{figure}

This sheds some light onto the unspecified colour evaporation process.
When the colour octet $\C$ leaves the field of the nucleon in which
it was produced, it will in general neutralise its colour by combining
non-perturbatively with an additional collinear gluon, thus producing
the $(\C)_8 g$ component of the \J~or the other charmonium states
(Fig.\ \ref{5_6}). After a ``relaxation time" $\t_8$, the $(\C)_8
g$ will absorb the accompanying gluon to revert to the dominant $(\C)_1$
charmonium mode. Note that we are here considering those
$\C$ pairs which will later on form charmonia. The $(\C)_8$ could as
well neutralise its colour by combining with a light quark-antiquark
pair, but this would result predominantly in open charm production.

\par

\par

Charmonium production in hadronic collisions thus inherently involves
different space-time scales \cite{Bodwin}. The first step is the
creation of a heavy $\C$ pair, which takes place in a very short
time, $\t_{\rm pert} \simeq 1/(2m_c)$. The colour octet $\C$ state
combines with an additional gluon to form a pre-resonance $(\C)_8 g$
to neutralise its colour and yield a resonance state of the correct
quantum numbers. The lifetime $\t_8$ of this state is determined by
the virtuality of the intermediate $\C$ state. In the rest frame of
the $\C$, this is approximately \cite{KS1}
\be
\t_8 \simeq  {1\over \sqrt \Delta}, \label{5.5}
\ee
where $\Delta \equiv [(p+k)^2-m_c^2]= 2pk$, with $p$ denoting the four
momentum of a quark in the final \J, $k$ that of the accompanying gluon.
Because of the large charm quark mass $m_c$, the time uncertainty
associated with gluon absorption becomes
\be
\t_8 \simeq {1 \over \sqrt{2m_c k_0}} \label{5.6}
\ee
where $k_0$ is the gluon energy. In a confining medium, $k\geq \L$,
with $\L \simeq 0.20 - 0.25$ GeV. We thus encounter a new scale
$(2m_c\L)^{1/2}$, with $\L \ll (2m_c\L)^{1/2} \ll m_c$; it makes
$\t_8 \gg \t_{\rm pert}$. For \J~production at mid-rapidity of a
nucleon-nucleon collision, the colour neutralisation time becomes
\be
t_8 \simeq \t_8 [1+(P_A/2m_c)^2]^{1/2} \label{5.7}
\ee
in the rest frame of target or projectile nucleon, with $P_A$
denoting the momentum of the $\C$ pair in this frame. As seen from the
nucleon, colour neutralisation of fast $\C$ pairs will thus take a
long time. Equivalently, the $(\C)_8 g$ travels in the time $t_8$ a long
distance,
\be
d_8 \simeq \t_8 (P_A/2m_c), \label{5.8}
\ee
in the rest frame of the nucleon.

\par

Moreover, the production of high energy charmonia implies the production
of composite and hence extensive states $(\C g)_i\equiv |(\C)_8g
\rangle_i$; here the subscript $i$ again refers to the specific
charmonium state in question (\J, $\chi_c$, \P). Although the intrinsic
transverse size of the $(\C g)_i$ depends in principle on the quantum
numbers of the state $i$, confinement constraints on the gluon lead to a
universal $(\C g)$ size $r_8$. Let us consider this in more detail.

\par

The formation process of the $\C$ pair inside the $(\C g)$ state (see
Fig.\ \ref{5_6}) makes this pair very compact, with a spatial
extension of
about $1/2m_c$. The size of the $(\C g)$ is thus essentially determined
by the softness of the gluon, which is restricted by $k \geq \L$. The
effect of this can be estimated by noting that in the time $\t_8$
obtained above, a gluon can propagate over a distance $r_8 \simeq (2m_c
\L)^{-1/2}$. Equivalently, we can consider energy conservation in the
passage from $(\C g)$ to the leading Fock space component $(\C)_1$;
the non-relativistic form for heavy quarks leads to
\be
{p^2\over 2m_c} \simeq  k_0, \label{5.9}
\ee
where $p$ is the quark three-momentum in the charmonium cms.
>From the lowest allowed gluon energy $k_0=\L$ we thus obtain the
intrinsic size
\be
r_8 \simeq {1\over p} \simeq
{1\over \sqrt{2m_c\L}} \simeq 0.20 - 0.25~{\rm fm}.
\label{5.10}
\ee
Since this size is determined only by the $(\C)_8 g$ composition of the
next-to-leading Fock space state, the compactness of the produced
$(\C)_8$ in this state and the gluon momentum cut-off in confined
systems, it is essentially the same for the different charmonium states.
The Fock space state $(\C g)$ thus constitutes something like a
gluonic hard core present in all charmonium states. Its size is seen to
be approximately that of the ground state charmonium \J, while the
higher excited charmonium states are larger. The hard gluon core in
these just corresponds to the fact that in extended bound states the
emission (or absorption) of gluons with momenta bigger than $\L$ can
occur only in a sufficiently small inner region. In other words: while
the basic $(\C)_1$ state gives for different quantum numbers quite
different spatial distributions, the much more localised higher Fock
space states are of universal size $(2m_c\L)^{-1/2}$. In the large
quark mass limit, however, the bound state radii become smaller than
$(2m_Q\L)^{-1/2}$, so that the universality would then be lost.

\par

In summary: charmonium hadroproduction is based on a hard perturbative
process, such as gluon fusion into a $\C$ pair. The subsequent colour
neutralisation brings in non-perturbative features, leading to the
formation of a pre-resonance state, such as the first higher Fock state
component $(\C)_8 g$. Confinement constraints result in a universal
pre-resonance charmonium radius of some 0.25 fm and an associated life
time of the same size. In this evolution era of charmonium production,
the different resonant states (\J, $\chi_c$, \P) thus appear essentially
indistinguishable. This has striking consequences for charmonium
production on nuclear targets.

\bigskip

\noindent{\bf 5.2 Pre-Resonance Suppression}

\bigskip

\begin{figure}[htb]
\vspace*{-0mm}
\centerline{\psfig{file=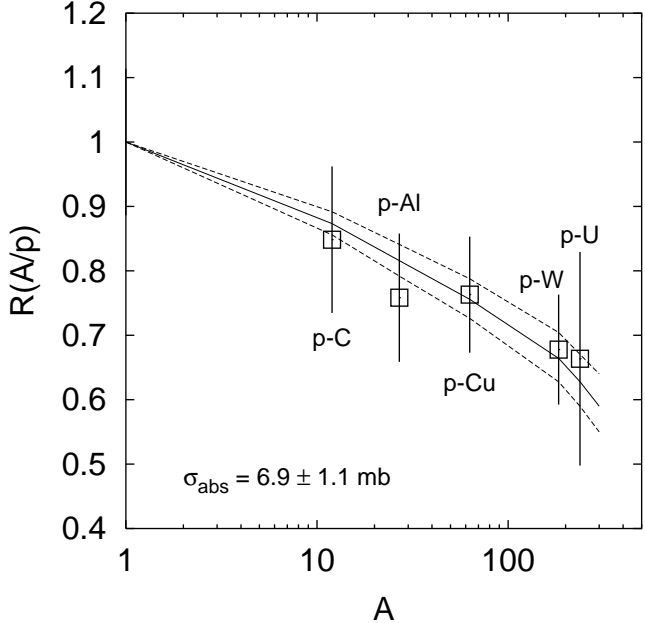,height= 90mm}}
%\vspace*{-20mm}
\caption{\J~production in $p-A$ vs. $p-p$ collisions \cite{NA38-f}.}
\label{5_7}
\end{figure}

The colour evaporation model alone, without further dynamical input,
predicts that the charmonium production rate in high energy $p-A$
collisions is $A$ times that in $p-p$ collisions at the same energy,
just as it is for open charm production. We assume here that the hard
process is dominated by gluon fusion; hard processes involving quarks
would lead to a small modification due to isopin effects.
Experimentally, however, the presence of the nuclear target medium is
known to reduce \J~production rates in $p-A$ collisions significantly,
up to about 40\% relative to those in $p-p$ interactions
\cite{NA3}-\cite{NA38pA}; in Fig.\ \ref{5_7} we show the most
recent
data \cite{NA38-f}. The suppression in $p-A$ collisions was the first
indication
that the presence of a strongly interacting medium influences the rate
of \J~production. The nuclear medium introduces space-time scales, and
hence we have to consider the space-time structure involved in the
production process.

\par

\par

\begin{figure}[htb]
\vspace*{-0mm}
\centerline{\psfig{file=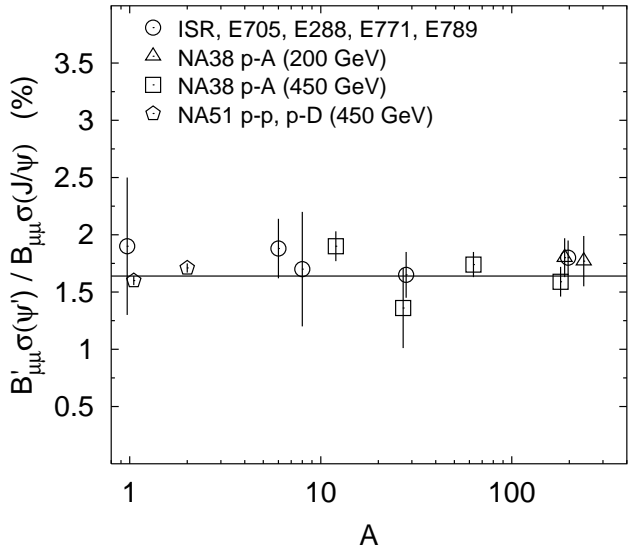,height= 80mm}}
%\vspace*{-20mm}
\caption{The relative $A$-dependence of \J~and \P~production
\cite{NA38-psi'}.}
\label{5_8}
\end{figure}

The mentioned experiments were carried out with incident protons with
momenta between 200 and 800 GeV/c. This gives the nascent \J~momenta of
30 GeV/c or more in the target rest frame. As a result, the transition
$\C g \to \j,~\x_c$ or \P~occurs outside the target nucleus; the nuclear
medium of the target sees only the passage of the corresponding
pre-resonance states. Since these have essentially the same size and
life-time for all charmonium states, the observed attenuation of the
production rates should be the same for \J~as for \P. This prediction
is indeed found to be quite well satisfied \cite{NA38-psi'} (see
Fig.\ \ref{5_8}). Earlier attempts to explain charmonium suppression in
$p-A$ interactions in terms of the absorption of physical \J~states
\cite{Huefner} had encountered difficulties precisely because of this
feature. The equal attenuation of \J~and \P~is a natural consequence
of pre-resonance absorption; it can never be obtained for the
physical \J~and \P~states with their very different geometric sizes.

\par

\par

The cross section for the dissociation of the $\C g$ state through
collisions in nuclear matter can be estimated theoretically \cite{KS6};
it can also be determined directly from $p-A$ data \cite{KLNS}.
Since the components of the $\C g$ are colour octets, the gluon
interaction with this state will be 9/4 times stronger than the gluon
interaction with the fundamental $\C$, whose components are
colour triplets. Hence we expect the $\C g$ dissociation cross section
to be about 9/4 times the high energy \J~dissociation cross section,
i.e., around 6 mb.

\par

For an experimental determination, we note that a $\C g$ pair formed at
point $z_0$ in the target nucleus has a survival probability
\be
S^A_{\C g} = \exp \left\{ -\int_{z_0}^{\infty} dz~ \rho_A(z)~
\sigma_{\C g-N} \right\},
\label{5.11}
\ee
where the integration covers the path remaining from $z_0$ out of the
nucleus. The traversed medium of nucleus $A$ is parametrized through a
Woods-Saxon density distribution $\rho_A(z)$, and by comparing $S^A_{\C
g}$ with data for different targets $A$, the dissociation cross
section for $\C g-N$ interactions is found to be \cite{Marzia}
\be
\sigma_{\C g-N} = 6.9 \pm 1.1~{\rm mb}. \label{5.12}
\ee
In Fig.\ \ref{5_7} it is seen that pre-resonance absorption with
this
cross section agrees well with all presently available $p-A$ data on
\J~production.

%\par
\begin{figure}[tbp]
%\vspace{-20mm}
\centerline{\psfig{file=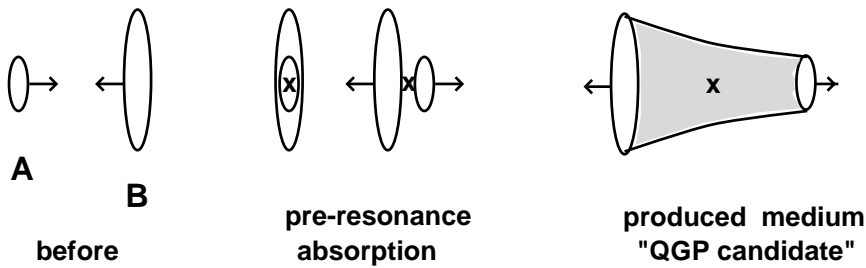,width=40mm,angle=90}\hspace{12cm}}
\vspace*{5mm}
\caption{Schematic view of \J~production in $A-B$ collisions.}
\label{5_9}
\end{figure}
\begin{figure}[tbp]
\vspace*{-0mm}
\centerline{\psfig{file=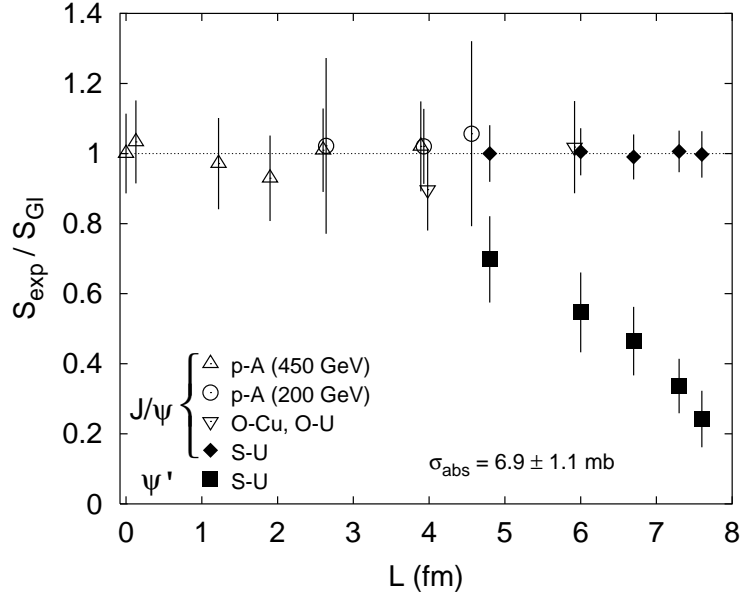,height= 80mm}}
%\vspace*{-20mm}
\caption{\J~and \P~production in nuclear collisions, compared
to pre-resonance absorption in nuclear matter \cite{NA38-f,NA38}.}
\label{5_10}
\end{figure}

We now turn to nucleus-nucleus collisions; here there will certainly
also be pre-resonance absorption in nuclear matter. However, in addition
to the target and projectile nuclei, there could now be a substantial
amount of produced `secondary' medium (Fig.\ \ref{5_9}). Our
ultimate
aim is to look for such a medium and test its confinement status.

\par

The survival probability of a pre-resonance charmonium state in an $A-B$
collision at impact parameter $b$ is given by
\be
S^{AB}_{\C g}(b) = \exp \left\{ -\int_{z_0^A}^{\infty} dz~ \rho_A(z)~
\sigma_{\C g-N} ~\int_{z_0^B}^{\infty} dz~ \rho_B(z)~ \sigma_{\C g-N}
\right\},
\label{5.13}
\ee
in extension of Eq.\ (\ref{5.11}). Here $z_0^A$ specifies the formation
point
of the $\C g$ within nucleus $A$, $z_0^B$ its position in $B$. Since
experiments cannot directly measure the impact parameter $b$, we have
to specify how Eq.\ (\ref{5.13}) can be applied to data.

\par

The Glauber formalism allows us to calculate the number
$N^{AB}_w(b)$ of participant (`wounded') nucleons for a given collision.
The number of secondary hadrons produced in association with the
observed \J~is at present energies (up to SPS energy)
found to be proportional to $N^{AB}_w$. The transverse energy $E_T$
carried by the secondaries is measured experimentally, together with the
\J's. We thus have
\be
E_T(b) = q~N_w^{AB}(b); \label{5.14}
\ee
the proportionality constant $q$ has to be determined on the basis of
the given experimental acceptance. Once it is fixed, we have to check
that the collision geometry (the measured relation between $E_T$ and the
number of spectator nucleons, the measured $E_T$-distribution) is
correctly reproduced \cite{KLNS}. When this is assured, we can check if
the \J~production in $O-Cu,~O-U$ and $S-U$, as measured by the NA38
experiment at CERN over the past ten years \cite{NA38-f,NA38}, shows
anything
beyond the expected pre-resonance nuclear absorption with the cross
section $\sigma_{\C g-N} = 6.9 \pm 1.1$ mb determined from $p-A$
interactions.

\par

The answer is clearly negative, as seen in Fig.\ \ref{5_10} for
the integrated cross sections as well as for the centrality dependence
of $S-U$ collisions. We consider here the \J~survival probability, i.e.,
the measured production rate $S_{\rm exp}$, normalised to the
predicted rate $S_{\rm Gl}$, including pre-resonance suppression with
the given absorption cross section. To compare all data in one figure,
we have parametrized the different collision configurations in terms of
the path length $L$ of the charmonium state in the nuclear environment.
As a rough estimate,
$L\simeq (3/4)R_A$ in $p-A$ collisions, with $R_A=1.13~A^{1/3}$ for the
nuclear radius; the factor (3/4) takes into account the average over
centrality. Similar estimates can be obtained for $A-B$ collisions;
however, the Glauber formalism allows a direct determination of this
path length at each impact parameter, and we shall use the values of $L$
thus determined and then related to $E_T$ through Eq.\ (\ref{5.14}).

\begin{figure}[tbp]
\vspace*{-0mm}
\centerline{\psfig{file=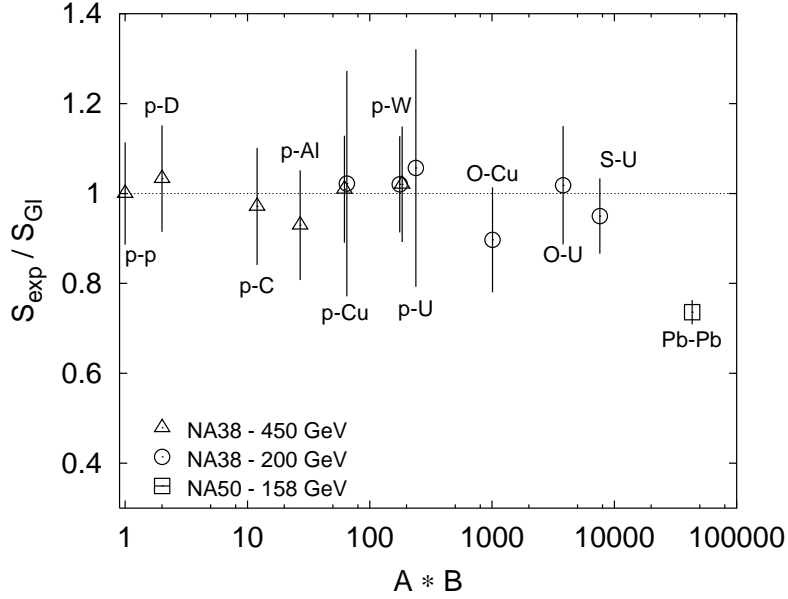,height= 80mm}}
%\hspace{2.5cm}}
%\vspace*{-20mm}
\caption{Centrality integrated rates for \J~production in $A-B$
and $Pb-Pb$ collisions,
normalized to pre-resonance absorption in nuclear matter
\cite{NA38-f,NA50}.}
\label{5_11}
\end{figure}
\begin{figure}[tbp]
\vspace*{-0mm}
\centerline{\psfig{file=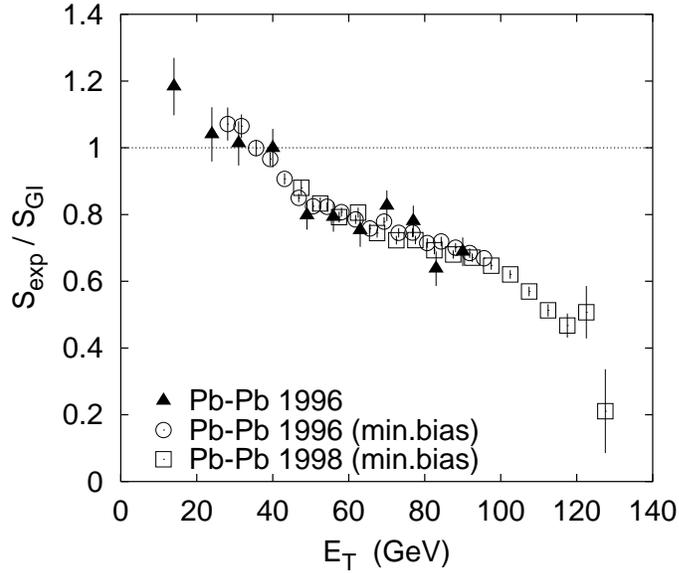,height= 80mm}}
%\hspace{2.5cm}}
%\vspace*{-20mm}
\caption{The $E_T$-dependence of \J~production in $Pb-Pb$ collisions,
normalized to pre-resonance absorption in nuclear matter
\cite{NA50last}.}
\label{5_12}
\end{figure}

All nuclear collisions measured by NA38 thus show
only what is now called `normal' \J~suppression, i.e., the pre-resonance
suppression already observed in $p-A$ interactions
\cite{KLNS,Huefner,KS6}. We thus have to ask if the $A-B$ collisions
studied by NA38 lead to any produced secondary medium at all. The
behaviour of \P~production (included in Fig.\ \ref{5_10}) shows
that this is indeed the case \cite{NA38-f,NA38-psi'}: in $S-U$
collisions,
there is \P~suppression beyond the expected pre-resonance nuclear
absorption. The collisions thus do form a secondary medium, and this
can distinguish a \P~(which is suppressed) from a \J~(which remains
unaffected). Since the former is suppressed, the latter unaffected, the
medium does not signal colour deconfinement.

\par

The observed \P~suppression should thus be accountable in terms of a
medium consisting of hadrons, and such a `hadronic comover' description
\cite{B-M} is indeed found to work \cite{KLNS}. It involves as
basic quantities the \P-hadron dissociation cross section and the
transverse density of hadronic comovers. Since the \P~is very loosely
bound (it lies only 60 MeV below the open charm threshold, see Table
2), it seems reasonable to assume that the cross section attains its
geometric value of about 10 mb soon after the threshold. The density 
of wounded nucleons
is given as function of collision centrality by the mentioned Glauber
formalism, and the observed rate of about 1.5 secondary hadrons per
wounded nucleon determines the comover density. The resulting form for
\P~suppression is found to agree quite well with the existing data
\cite{KLNS}.

\bigskip

\noindent{\bf 5.3 Anomalous \J~Suppression}

\bigskip

The suppression of \J~production in nuclear collisions, from $p-A$ to
central $S-U$ collisions, is thus understood in terms of `normal'
absorption in nuclear matter: the pre-resonance charmonium state is
dissociated by collisions with the primary target or projectile
nucleons. The \J~as probe of the produced environment indicates ``no
deconfinement".

\par

\begin{figure}[htb]
\vspace*{-0mm}
\centerline{\psfig{file=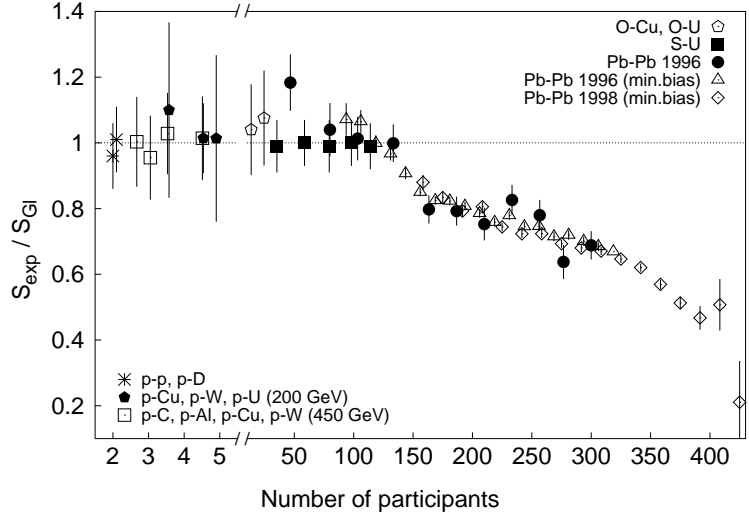,height= 70mm}}
%\hspace{2.5cm}}
%\vspace*{-20mm}cp
\caption{\J~production in $A-B$ and $Pb-Pb$ collisions,
normalized to pre-resonance absorption in nuclear matter, as function
of the number of participating nucleons \cite{NA50last,Nardi}.}
\label{5_13}
\end{figure}

This situation came to an end with the advent of truly {\sl heavy} ion
experiments. In $Pb\!-\!Pb$ collisions at the CERN-SPS, the NA50
Collaboration \cite{NA50,NA50last}
observed a further `anomalous' suppression of about 30\%
for the overall \J~production rate (Fig.\ \ref{5_11}). Looking in more
detail, it was observed that peripheral $Pb\!-\!Pb$ collisions show
only normal suppression (Fig.\ \ref{5_12}). Then, at a transverse
hadronic energy $E_T \simeq 40$ GeV (corresponding to an impact
parameter $b$ of about 8 fm), the anomalous \J~suppression sets in quite
suddenly, with a drop of 20\% for an $E_T$ increase of less than 20
GeV, or a decrease in $b$ of about 1 fm. For very central collisions,
$E_T \geq 100$ GeV or $b \leq 2$ fm, something like a second drop
appeared. The most central collisions thus show a suppression by more
than a factor two beyond the normal pre-resonance absorption. The
general suppression pattern, combining all data from $p\!-\!p$ to
central $Pb\!-\!Pb$ collisions, is illustrated in Fig.\
\ref{5_13},
where the \J~survival probability is shown as function of the number of
participant nucleons. What conclusions can be drawn from this behaviour?

\bigskip

\noindent{\bf 5.3a Deconfinement vs.\ Hadronic Comovers}

\bigskip

Over the past decade, quarkonium suppression mechanisms have become
classified into two categories. The suppression could be due to colour
deconfinement, as originally predicted \cite{Matsui}. To verify this,
one has to exclude what are generally called
`conventional' mechanisms, covering all possible ways to suppress
\J~production without invoking deconfinement. There are a considerable
variety of possible conventional suppression schemes: absorption in
nuclear matter \cite{Capella}, absorption by produced secondary hadrons
(i.e., by the `hadronic comovers' mentioned above \cite{B-M,comovers}),
nuclear shadowing of parton distribution functions \cite{Gupta1},
medium-induced energy loss of the incident partons leading to $\C$
production \cite{Hwa} or more general in-medium modifications of the
$\C$ production process \cite{Qiu}; for an extensive review,
see \cite{Ramona-PR}. Since all models involve several
more or less adjustable parameters, the task of confirming one and
ruling out all others seems quite difficult indeed. It is greatly
simplified, however, by two basic {\sl qualitative} features of
deconfinement, which distinguish it from all conventional approaches.

\par

Deconfinement is a critical phenomenon, with a well-defined onset
for colour conductivity specified by theory. There is a thermodynamic
region for confinement, a critical point, and then a region of
deconfinement. All other approaches are always present in smoothly
varying degrees. To illustrate the point, we consider two simple
models, one for hadronic comover absorption, as representative and most
intensively studied conventional model \cite{Gavin}, the other for
colour deconfinement \cite{Gupta2}.

\par

In the comover picture, the \J~is
dissociated with a cross section $\sigma_{\rm com}$ in
a medium of produced hadronic secondaries; the density $n(\e)$ of this
medium is determined by the initial energy density $\e$ of the
produced environment. The medium subsequently cools off and stops
affecting the \J's when its density drops to a freeze-out value $n_f$.
The resulting \J~survival probability becomes \cite{comover}
\be
S_{\j}^{\rm com}(\e) = \exp\{-\sigma_{\rm com} n(\e) \tau_0
\ln[n(\e)/n_f]\},
\label{5.15}
\ee
where $\tau_0\simeq 1$ fm denotes the formation time of the medium.
According to Eq.\ (\ref{8}), the cross section for \J~dissociation in
the comover medium is expected to be very small, ruling out significant
contributions from comover absorption. An essential ingredient in all
comover models is thus the assumption that the short distance QCD
arguments leading to Eq.\ ({8}) become applicable only for much heavier
quarks and cannot be used for charmonium states.\footnote{The behaviour
of $\sigma^{\rm in}(\j-h)$ for low collision energies can be measured
directly either by shooting a nuclear beam at a hydrogen target
(`inverse kinematics' experiment \cite{KS5}) or by annihilating
antiprotons in a nuclear target \cite{Dima}. Hence this question can be
settled empirically.} The cross section $\sigma_{\rm com}$ is then
treated as adjustable parameter, as is the
freeze-out density $n_f$. Typical fit values \cite{Gavin} lead to
$\sigma_{\rm com} \simeq$ 4 - 5 mb. We shall return shortly to the
latest and most detailed comover fits.

\par

In contrast to comover absorption, deconfinement sets in at a critical
point $\e_c$; a simple model \cite{Gupta2} leads to
\be
S_{\j}^{\rm dec}(\e) = \Theta(\e_c - \e) + \Theta(\e - \e_c)
\left({\e \over \e_c}\right)^{9/4}.
\label{5.16}
\ee
The power of the variable $\e/\e_c$ is determined by the energy density
profile in the collision, since the hot interior melts \J's, while they
survive in the cool rim. The specific form (\ref{5.16}) arises for
nuclei of constant nuclear density \cite{Karsch-S}. More recent
calculations of this type \cite{B&O} have led to a very similar
suppression pattern, while percolation arguments, as will be seen
below, lead to a still more abrupt onset of deconfinement suppression.
The abrupt onset of \J~suppression in Eq.\ (\ref{5.16}) shows a
fundamental qualitative difference to the smooth monotonic decrease of
the comover suppression pattern Eq.\ (\ref{5.15}); an explicit
illustration will be given shortly. It should also be noted that
hadronic comover absorption would have to be present already in $S-U$
collisions, thus ruling out pure pre-resonance absorption here. The
behaviour seen in Fig.\ \ref{5_10} seems difficult to reconcile
with this.

So far, we have addressed the fate of a well-defined \J~in
a strongly interacting environment. However, sufficiently energetic
nuclear collisions produce the entire spectrum of charmonium states;
below the open charm threshold, they produce
$\eta_c(1S)$, $\j(1S)$, $\chi_{c0}(1P)$, $\chi_{c1}(1P)$,
$\chi_{c2}(1P)$ and $\p(2S)$.
All higher excited states have significant decay rates into
the \J, and so the observed \J's can be either directly produced, or
they can come from the decay of higher excitations. Since the decay
times of these excitations are very long (more than 1000 fm for the
$\chi_c$), the medium affects the intermediate higher excitations, not
the final decay product \J.

\par

\begin{figure}[p]
\setlength{\unitlength}{1mm}
%\vspace*{-20mm}
\centerline{\epsfig{file=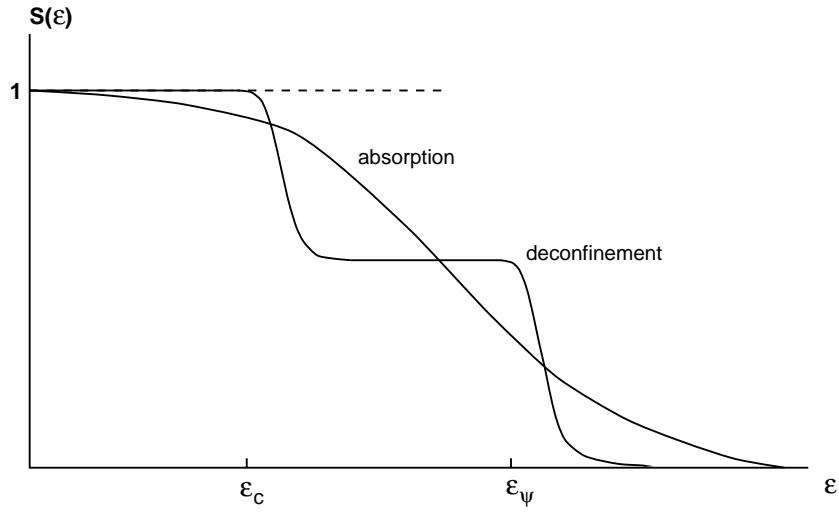,height= 110mm,angle=-90}}
%\hspace{2.5cm}}
%\vspace*{-20mm}cp
\caption{\J~suppression by deconfinement compared to that by
hadronic comover absorption \cite{Gupta2}.}
\label{5_15}
\end{figure}

In nucleon-nucleon collisions, about 60\% of the observed \J's are
produced directly, about 32\% come from $\chi_c$ decays, the remaining
8\% from \P~decays. To simplify matters, we neglect the \P~component for
the moment and assume 2/3 direct production and 1/3 $\chi_c$ decay. Both
colour screening and gluon dissociation indicate that the $\chi_c$ will
be destroyed at a lower energy density than the direct \J's.
As second basic qualitative feature, colour screening thus
predicts sequential \J~suppression, with a multi-step survival
probability \cite{KMS,Karsch-S}. For \J~suppression, the simple
analytic form (\ref{5.16}) now becomes
\be
S_{\j}^{dec}(\e) = {1 \over 3} \left\{ \Theta(\e_{\chi} - \e) +
\Theta(\e - \e_{\chi}) \left({\e \over \e_{\chi}}\right)^{9/4}
\right\}
 + {2 \over 3} \left\{ \Theta(\e_{\psi} - \e) + \Theta(\e -
\e_{\psi}) \left({\e \over \e_{\psi}} \right)^{9/4} \right\},
\label{5.17}
\ee
where $\e_{\chi}\simeq \e_c$ and $\e_{\psi} > \e_c$ denote the
thresholds for $\chi$ and direct $\psi$ dissociation, respectively.
This form of the resulting functional behaviour of deconfinement
suppression (Fig.\ \ref{5_15}) was given eight years ago
\cite{Gupta2}; it clearly illustrates the quantitative difference to
suppression by comover absorption. Today we find it to be remarkably
similar to the suppression pattern now observed in $Pb-Pb$ interactions
(Figs.\ \ref{5_12} and \ref{5_13}).

\begin{figure}[p]
%\vspace*{-0mm}
\centerline{\psfig{file=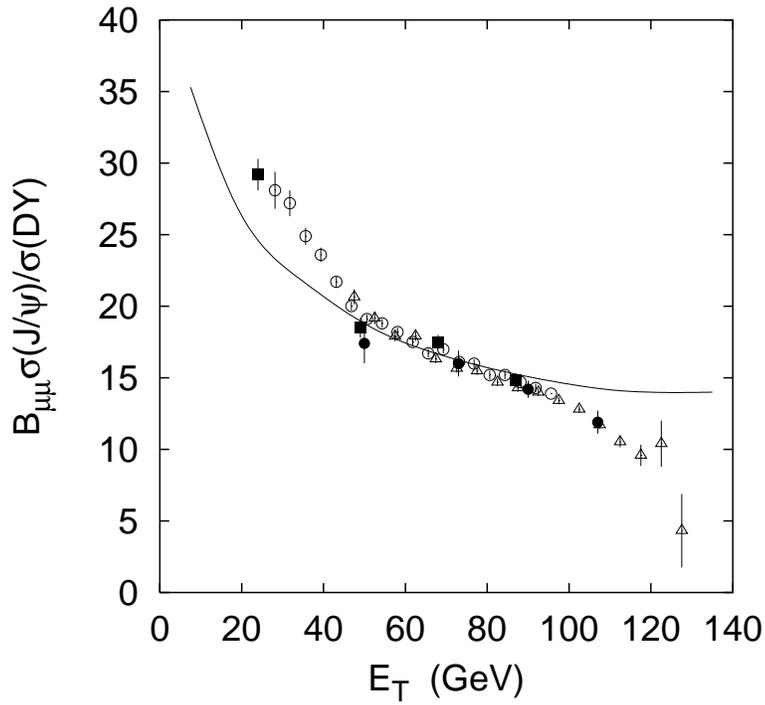,height= 100mm}}
%\hspace{2.5cm}}
%\vspace*{-20mm}cp
\caption{\J~production in $Pb-Pb$ collisions
\protect\cite{Kluberg,Cicalo},
compared to the dual parton model comover approach \protect\cite{DPM}.}
\label{5_16}
\end{figure}

The first $Pb-Pb$ results provided by NA50 \cite{NA50-95} were of much
lower precision than those now available. This stimulated various
comover approaches to attempt a unified description of $p-A$, $S-U$ and
$Pb-Pb$ data. The monotonic onset of any comover suppression requires
the presence of some anomalous suppression also for $S-U$, and it
excludes any thresholds. Hence even the first, relatively pronounced
onset of suppression in the $Pb-Pb$ data posed problems. Nevertheless,
qualitatively the comover pattern also leads to a decrease with
increasing energy density, and so more precise data were needed for any
reliable conclusion. Sucessive NA50 runs have provided this
\cite{NA50,NA50last}, and the given experimental situation today, with a clear
two-step suppression pattern, excludes all conventional scenarios. In
Fig.\ \ref{5_16} we show as illustration the results of one of
the most general comover approaches, based on the dual parton model
\cite{DPM}; it includes comover production dependent on the
number of collisions as well as on the number of participant nucleons,
and it also takes into account \J-\P/\P-\J~transitions. In spite of the
resulting number of adjustable parameters, it is clearly incompatible
with the latest data. All other conventional models share this fate:
none can account for all the available data, including $S - U$ and
$Pb - Pb$ collisions; in particular, none can produce a mechanism
leading to any kind of non-monotonic step structure. This provides the
basis for the conclusion reached by the NA50 collaboration
\cite{NA50last}: {\sl The onset of the anomalous \J~suppression is the
first clear observation of a threshold effect in heavy ion collisions
and can be considered as a strong indication of the production of a
deconfined quark-gluon plasma phase in central Pb-Pb collisions.}

\bigskip

\noindent{\bf 5.3b Towards an Understanding of Deconfinement}

\bigskip

The observed \J~suppression thus rules out conventional explanations;
moreover, it is qualitatively in accord with the behaviour expected
from colour deconfinement. Nevertheless, the details of the onset of
deconfinement are not yet clear, and in particular the crucial
collision variable triggering this onset is not yet unambiguously
identified. We present in this section one possible approach, based on
parton percolation \cite{S-QM,Nardi&S}.

\par

In a high energy nuclear collision viewed in the overall center of mass,
the two Lorentz-contracted nuclei quickly pass through each other. After
about 1 fm at the SPS and only about 0.1 fm at RHIC, they have separated
and left behind a partonic medium as a potential QGP candidate. Inside
this medium, primary nucleon-nucleon collisions have left charmonium
states as probes, whose fate can provide information about the nature of
the partonic medium.

\par

The average number of produced partons will depend on the number of
nucleon-nucleon collisions and/or the number of participating
nucleons. Up to SPS energies, the number of wounded nucleons
appears to be the main determining factor. This is understandable in
terms of the dilated and hence large soft parton formation time as seen
in the rest frame of either nucleus. Interference and cancellation
effects of the Landau-Pomeranchuk type thus prevent soft parton
emission at each nucleon-nucleon collision. At higher energies, large
additional contributions can come from hard partons (minijets or jets)
with short formation time, and these will be proportional to the number
of collisions. Even at SPS energy, nucleon stopping and secondary
multiplicities at mid-rapidity seem to increase somewhat with
increasing mass number $A$, and this could be the onset of collision
dependent effects. The $s$-dependence of the mid-rapidity multiplicity
is presumably a combination of the $s$-dependence of the parton
distribution function and the onset of significant hard hadron
production. We shall try to include these dependences here in rather
phenomenological terms, without specifying their origin.

\par

The number of gluons of transverse momentum $k_T$ emitted by a wounded
nucleon can be obtained from the nucleonic gluon distribution function
$g(x)$ from deep inelastic scattering studies,
\be
\left( {dN_g\over dy dk_T^2} \right)_{y=0} \simeq xg(x) f(k_T^2)
\label{5.18}
\ee
where $x\simeq k_T/\sqrt s$. Here $\sqrt s$ is the incident collision
energy, and $f(k_T^2)$ is normalized to unity. Using the MRS-H form of
the gluon distribution function \cite{MRS-H} and integrating the
resulting Eq.\ (\ref{5.18}) over $k_T$ then gives us $(dN/dy)_{y=0}
\simeq 2$ for SPS and 4 for RHIC energy.

\par

Each gluon of transverse momentum $k_T$ has an effective transverse
size $r \simeq k_T^{-1}$, and nucleus-nucleus collisions provide many
such gluons overlapping in the transverse plane. We want to study their
clustering behaviour. To simplify matters, we shall consider `average'
gluons (rather than averaging results with $f(k_T^2)$). From
the transverse momentum dependence of Drell-Yan dilepton production
through quark-antiquark annihilation, or from that of \J~production
through gluon fusion, one finds that the effective intrinsic transverse
momentum of gluons is $\langle k_T \rangle \simeq 0.75$ GeV/c, leading
to an average transverse parton radius $r \simeq 0.27$ fm. We now want
to consider the deconfinement pattern for such systems.

\par

\begin{figure}[htb]
\vspace*{-0.9cm}
%\begin{center}
\mbox{
\psfig{file=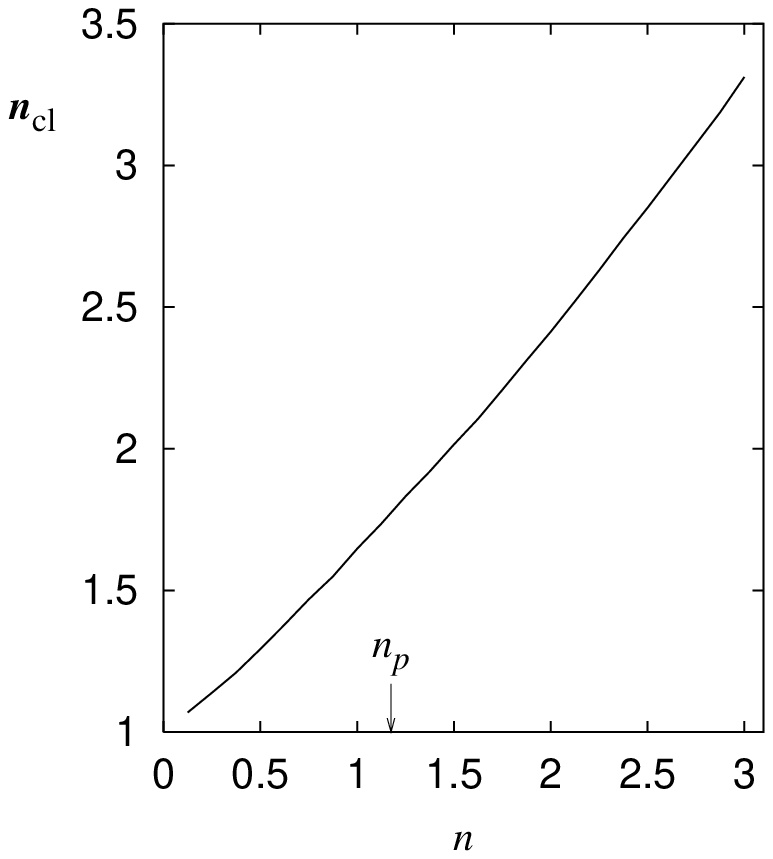,width=8cm}\hskip-.6cm
\psfig{file=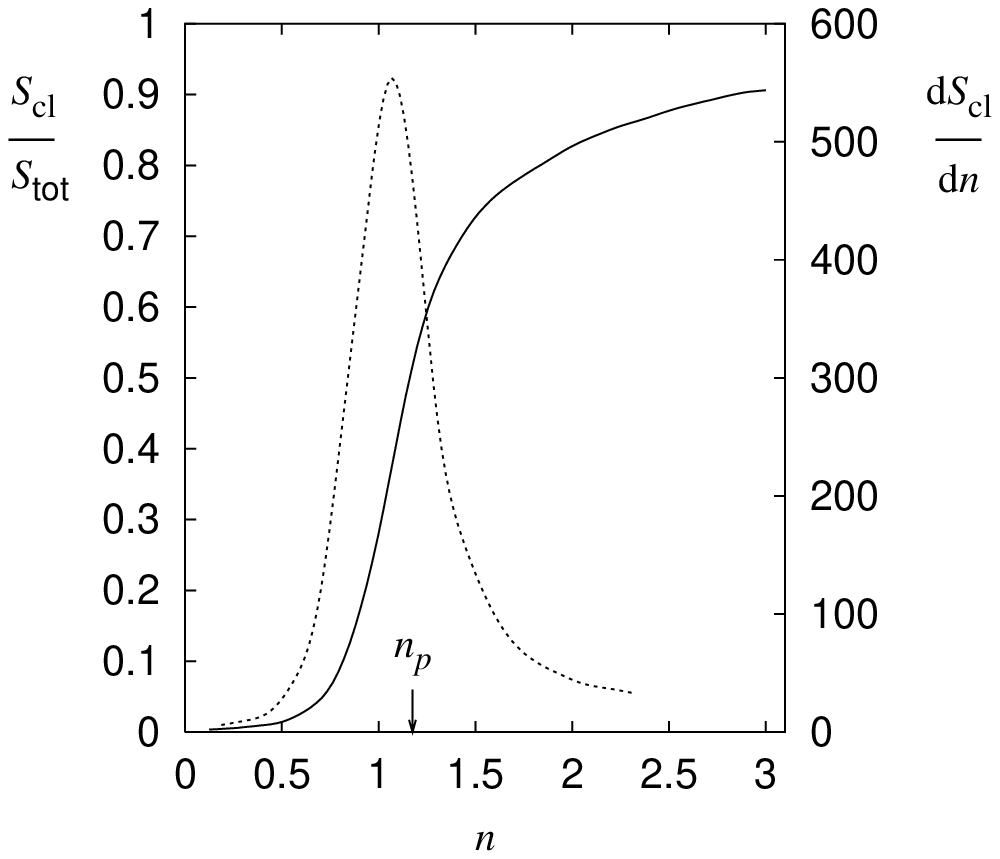,width=8cm}}
%\end{center}
%\vspace*{-0.8cm}
\caption{Average cluster density $n_{\rm cl}(n)$ (left) and average
fractional cluster size $S_{\rm cl}(n)/S_{\rm total}$ (right) as function of
the overall density $n$ of discs, for $r/R=1/20$; in (b), the derivative
of $S_{\rm cl}(n)/S_{\rm total} $ with respect to $n$ is also shown
(dotted line). The percolation point in the limit $r/R \to 0$ is
indicated by $n_p$.}
\label{T2_1}
%\vspace*{-0.8cm}
\end{figure}

The collisions distribute the gluons over the
transverse area according to the nucleon distribution within the nuclei.
We thus have to distribute discs of radius $r \simeq 0.27$ fm in the
transverse plane and determine the overall nucleon density for
which colour conductivity sets in, i.e., for which there appear clusters
which reach the dimension of the entire system. Hence we again encounter
percolation as crucial feature for deconfinement; here it occurs as a
two-dimensional phenomenon, since dynamics aligns the gluon motion
along the collision axis. If we denote with $n_c$ the critical gluon
density for the production of a deconfined medium, then as in section
2.4, it appears natural to identify the onset of quark-gluon plasma
formation with the percolation point, which is \cite{Alon}
\be
n_c = {1.175 \over \pi r^2}, \label{5.19}
\ee
where $\pi r^2$ is the transverse gluon area. If we consider
gluons of transverse momentum 
$\langle k_T \rangle = 0.75$ GeV/c as characteristic for
nucleon-nucleon collisions at present CERN-SPS energies, then the
threshold obtained from Eq.\ (\ref{5.19}) is $n_c\simeq 5$ fm$^{-2}$.
The deconfined medium thus obtained consists of gluons freed of their
confinement constraints, but it does not yet need to be in thermal
equilibrium.

\par

As a prelude, we assume $N$ discs of radius $r$ to be randomly
distributed on a flat surface of radius $R>>r$ and study the average
cluster density $n_{\rm cl}(n)$ and the average cluster size $S_{\rm
cl}(n)$ as a function of the overall density $n=N/\pi R^2$. The result
is shown in Fig.\ \ref{T2_1}, where we have normalized all
quantities to the inverse disc area $1/\pi r^2$ in order to make them
dimensionless.
We note that the derivative of the cluster density peaks sharply at a
certain density; in the `thermodynamic' limit $R \to \infty$, this
percolation point is known to be $n_p=1.175$ .
We see in particular that if we want to reach a certain cluster density
$n_{\rm cl}$, attained at a certain overall $n$, then at this point the
average cluster $S_{\rm cl}(n)$ also has a certain finite size. In other
words, requiring a specific density for the onset of a new phase
implies that this onset occurs for a specific, finite size of the
system. The percolation point, for example, is reached when about 50\%
of the surface is covered by discs.

\par

Turning now to nuclear collisions, we have to distribute the partonic
discs not on a flat surface, but instead in the transverse plane
according to the nucleon distribution determined by the profiles of the
colliding nuclei \cite{W-S}. Moreover, we have to allow collisions at
different impact parameters, and we want to consider different $A\!-\!B$
collisions.

\par

\begin{figure}[tbp]
  \begin{center}
    \mbox{
      \begin{minipage}{7.5cm}
        \psfig{file=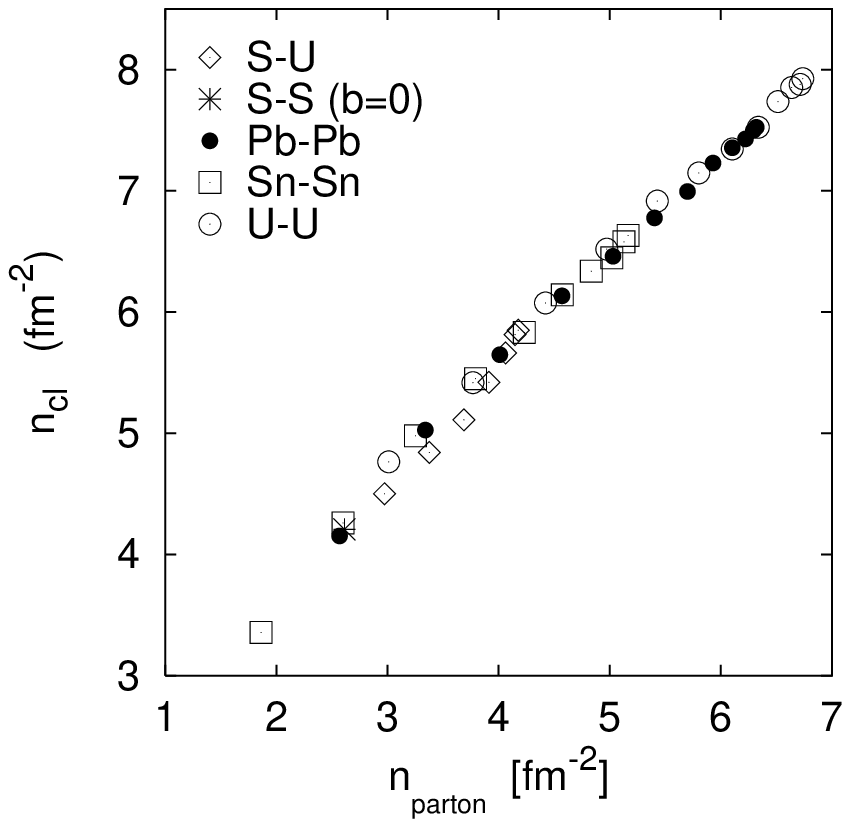,width=7.5cm}
%\vspace*{-1.5cm}
        \caption{Cluster density vs.\ parton density for different
          centralities and different $A\!-\!B$ configurations at SPS energy.}
       \label{T2_2}
     \end{minipage}
%      \hfill
      \hskip.5cm
      \begin{minipage}{7.5cm}
        \psfig{file=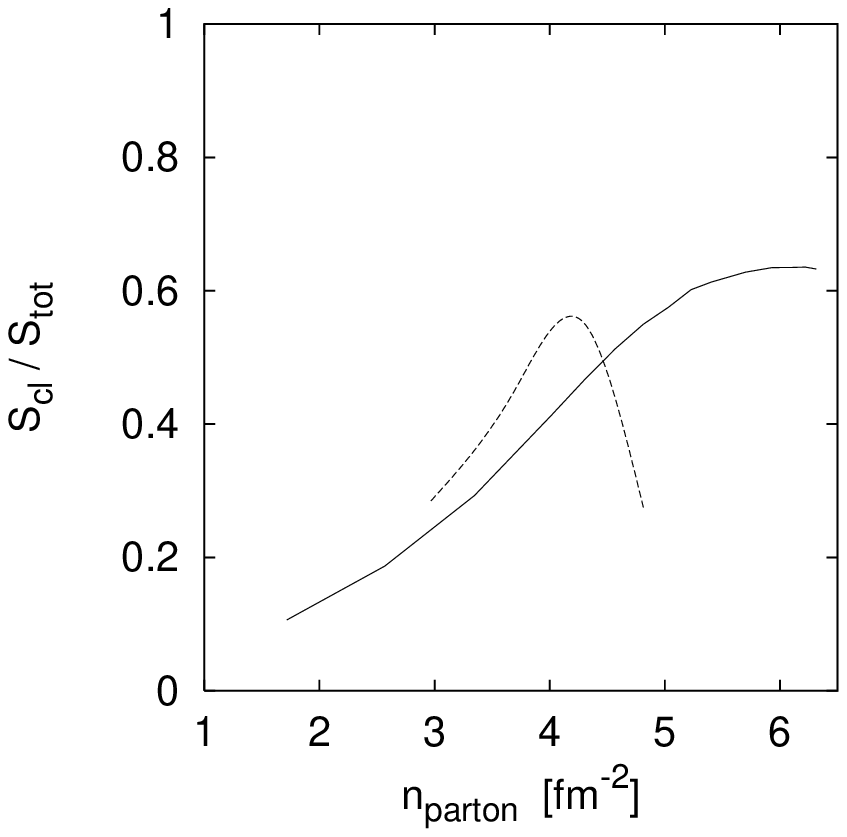,width=7.5cm}
%\vspace*{-1.5cm}
        \caption{Fractional cluster size vs.\ parton density, together
          with its derivative, for $Pb\!-\!Pb$ collisions at SPS energy.}
        \label{T2_3}
     \end{minipage}
      }
  \end{center}
\end{figure}

In Fig.\ \ref{T2_2}, the cluster density $n_{\rm cl}$ is shown as
function of the overall density $n_{\rm parton}$ of `average' partons
of radius $r=0.27$ fm, for different centralities of various $A\!-\!B$
combinations at SPS energy, with 2.1 partons per wounded
nucleon. For each $A\!-\!B$, the point at highest $n_{\rm parton}$
corresponds to central (impact parameter $b=0$) collisions, the one at
lowest $n_{\rm parton}$ to the most peripheral collisions. We
note that by varying $A\!-\!B$ and centrality we produce an essentially
universal cluster density $n_{\rm cl}(n_{\rm parton})$. Now we assume
that deconfinement occurs at the percolation point, which can be
determined by studying the average fractional cluster size $S_{\rm
cl}/S_{\rm total}$. In Fig.\ \ref{T2_3}, we see the result for
$Pb\!-\!Pb$ collisions. At the density for which the derivative peaks,
$n_{\rm parton} \simeq 4.2$ fm$^{-2}$, the cluster density reaches its
critical value $n_{\rm cl} \simeq 6$ fm$^{-2}$. The same critical value
is found for $Sn\!-\!Sn$ and $U\!-\!U$ collisions at SPS energy, even
though this value is attained at quite different centralities in the
different configurations ($b \simeq 4.5$ fm for $Sn$, 8.5 fm for $Pb$
and 11.5 fm for $U$ collisions).

\par

\begin{figure}[tbp]
  \begin{center}
    \mbox{
      \begin{minipage}{7.5cm}
        \psfig{file=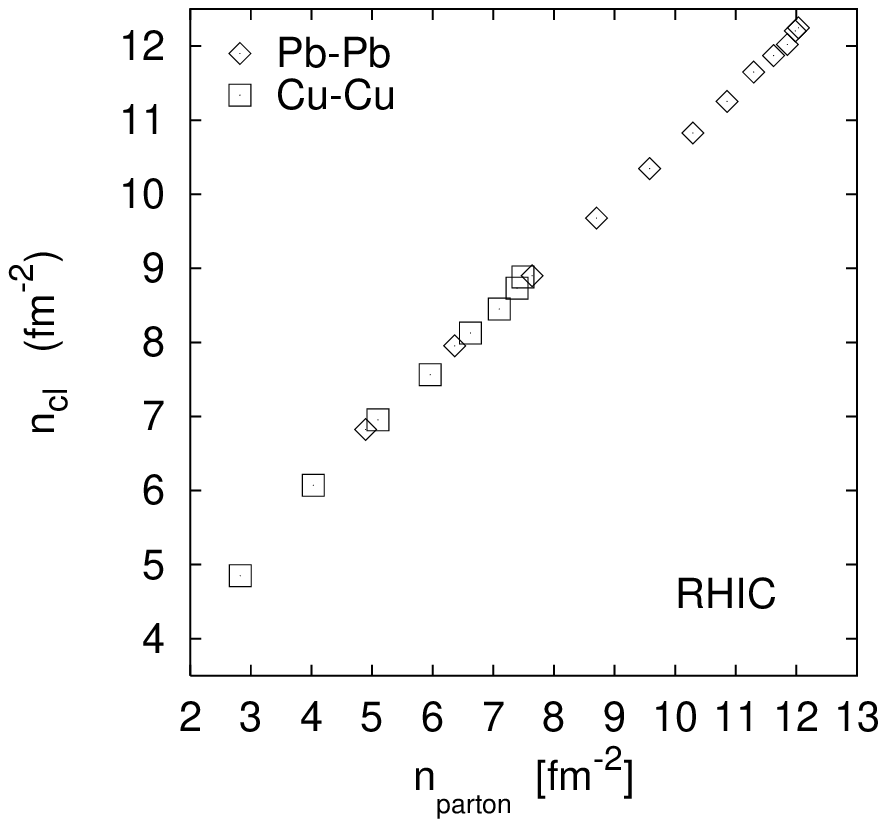,width=7.5cm}
%\vspace*{-1.5cm}
         \caption{Cluster density vs.\ parton density for different
          centralities and different $A\!-\!B$ configurations at RHIC energy.}
   \label{T2_4}
    \end{minipage}
      \hskip.5cm
      \begin{minipage}{7.5cm}
        \psfig{file=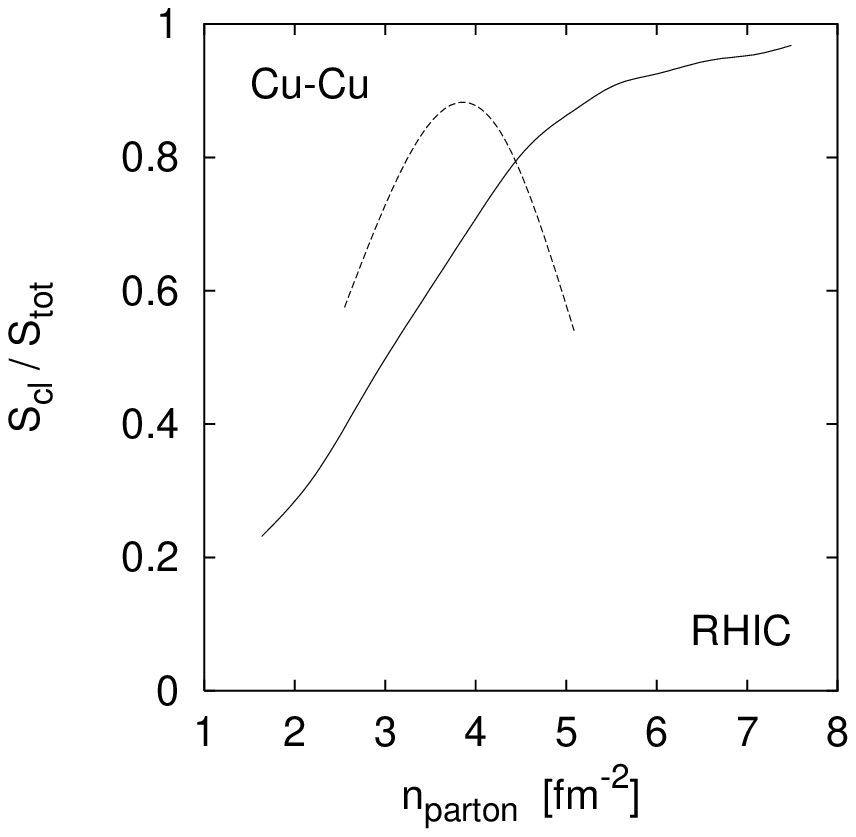,width=7.5cm}
%\vspace*{-1.5cm}
        \caption{Fractional cluster size vs.\ parton density, together
          with its derivative, for $Cu\!-\!Cu$ collisions at RHIC energy.}
      \label{T2_5}
 \end{minipage}
      }
%\vspace*{-0.5cm}
  \end{center}
\end{figure}

In Fig's.\ \ref{T2_4} and \ref{T2_5}, we show the result of this
procedure for RHIC energy, where we have 4.0 gluons per wounded
nucleon. Again the critical cluster density at deconfinement is found
to be $n_{\rm cl} \simeq 6$ fm$^{-2}$. Here we have to consider
$Cu\!-\!Cu$ collisions in order to study the onset of deconfinement,
since for $Pb-Pb$ all possible centralities are above the threshold.

\par

We thus conclude that the critical cluster density is, as expected, a
universal quantity, independent of the choice of $A\!-\!B$, of
centrality, and of incident energy. The fractional size of the
deconfined cluster at threshold does, however, depend on the collision
configuration: for $Pb\!-\!Pb$ at SPS, we have $S_{\rm cl}/S_{\rm
total} \simeq 0.45$, while $Cu\!-\!Cu$ at RHIC gives 0.70. In other
words, at the higher RHIC energy, deconfinement sets in for a larger
bubble than at the lower SPS energy.

\par

We further note that at SPS energy, $S\!-\!S$ and $S\!-\!U$ are below
the deconfinement threshold even for the most central collisions, while
at RHIC even the most peripheral $Pb\!-\!Pb$ collisions are above the
threshold. To study the threshold, we thus require heavy nuclei
(such as $Pb$) at the SPS, lighter nuclei (such as $Cu$) or lower
incident energy at RHIC.


\begin{figure}[tbp]
\vspace*{-0mm}
\centerline{\psfig{file=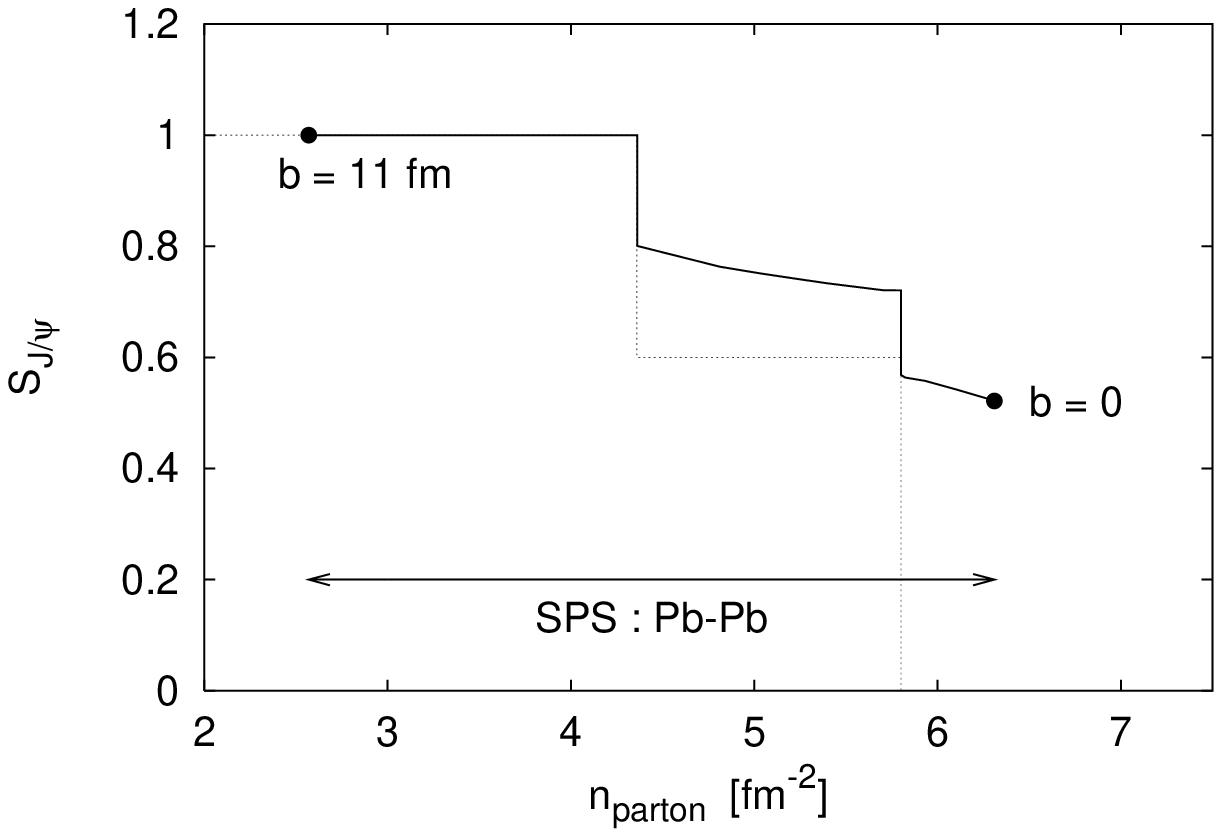,height= 80mm}}
%\vspace*{-1.1cm}
\caption{The \J~survival probability as function of the parton density for
$Pb\!-\!Pb$ collisions at SPS energy; the dotted line corresponds to a
medium of uniform parton density, the solid line to collisions with
parton densities determined by the profiles of the colliding nuclei.}
\label{T2_6}
%\vspace*{-0.5cm}
\end{figure}
\begin{figure}[tbp]
  \begin{center}
    \mbox{
        \psfig{file=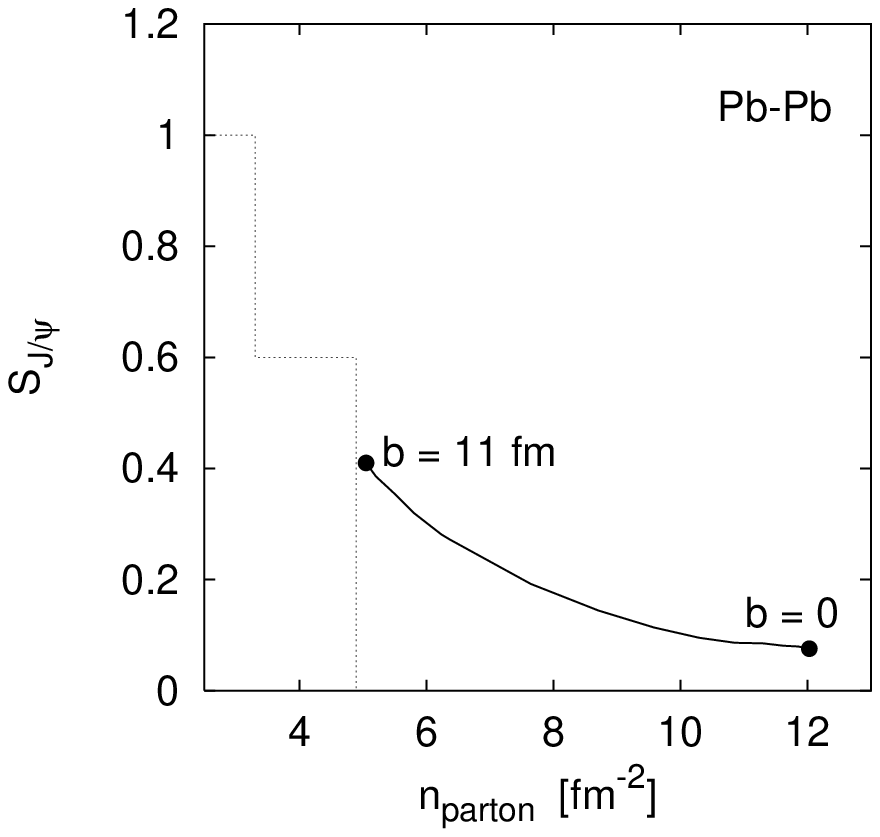,width=7.5cm}
      \hskip.5cm
        \psfig{file=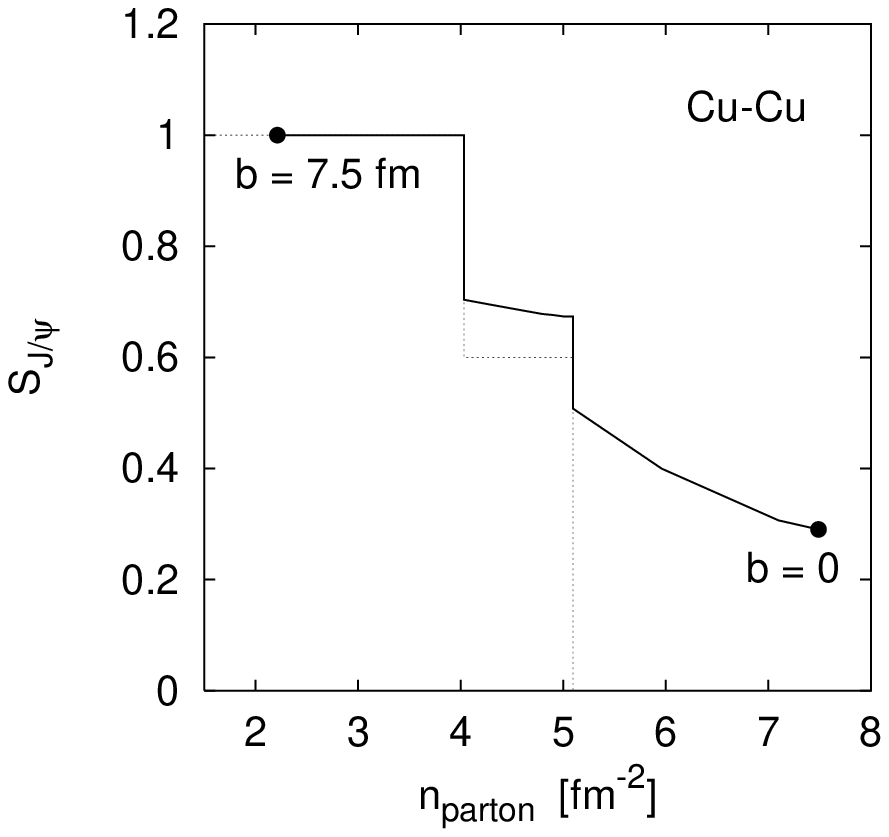,width=7.5cm}
}
%\vspace*{-1.2cm}
        \caption{The \J~survival probability as function of the parton
density for
$Pb\!-\!Pb$ (left) and for $Cu\!-\!Cu$ (right) collisions at RHIC
energy; the dotted line
corresponds to a medium of uniform parton density, the solid line to
collisions with parton densities determined by the profiles of the
colliding nuclei.}
      \label{T2_7}
  \end{center}
%vspace*{-0.8cm}
\end{figure}

Having specified deconfinement through percolation, we now turn to
\J~suppression as a probe for its onset. As noted above, the \X~is
dissociated essentially at deconfinement, so when the cluster density
reaches $n_{\rm cl}^{\rm crit}\simeq 6$ fm$^{-2}$, all \X's which are
inside the percolating cluster disappear. The amount of \X~suppression
thus depends on the fractional cluster size at deconfinement, which is
about 45 \% for $Pb-Pb$ at the SPS, about 70 \% for $Cu-Cu$ at RHIC and
effectively 100 \% for $Pb-Pb$ at RHIC. At deconfinement, the
corresponding fractions of \J~production through intermediate \X~states
should thus be suppressed.

As already noted, the dissociation of directly produced \J's requires a
a larger screening mass and thus higher density than available at the
point of deconfinement. On a microscopic level, \J~break-up becomes
possible only for harder than average gluons at deconfinement. In
principle, the relevant quantities can be determined by lattice QCD
studies; however, this requires computer capabilities which are being
reached only now. For the moment, we therefore choose $r^{\j}=0.22$ fm
and $n_{\rm cl}^{\j}=7.8$ fm$^{-2}$; this corresponds to a ratio of
transverse dissociation energy densities $\e_T(\j)/\e_T(\x_c) \simeq
1.6$, in accord in studies combining potential theory and lattice
results \cite{KMS,Karsch-S}. With these parameters, one obtains 20\%
direct \J~suppression in $Pb\!-\!Pb$ collisions at SPS energy, once the
required cluster density is reached. The corresponding fractions for
RHIC are 30\% for $Cu\!-\!Cu$ and 35\% for $Pb\!-\!Pb$ collisions.

\par

Schematically, one thus obtains a suppression pattern as illustrated in
Fig.\ \ref{T2_6}, where it is assumed that 40\% of the observed
\J's come from \X~decay and 60\% are produced directly. For clarity
purposes, we ignore here the small fraction (about 8\%) coming from
\P~decay. The dotted curve corresponds to the suppression which would
take place in a uniform medium of precisely specified parton density:
at the deconfinement point, all \X' are dissociated, so that the
corresponding fraction of decay \J's is gone; when the density for
direct \J~melting is reached, these disappear as well, leading to
complete \J~suppression. In actual nuclear collisions, the medium is
not uniform, with denser `inner' and less dense `outer' regions in the
transverse plane. Suppression now occurs only in the fraction
$S_{\x_c}/S_{\rm total}$ for the \X~part and $S_{\j}/S_{\rm total}$ for
the direct \J~part, where $S_{\x_c}$ and $S_{\j}$ denote the clusters in
which the respective dissociation density is reached. Since these
fractions depend on energy and nuclear geometry, the suppression curves
are different for different experimental configurations. In Fig.\
\ref{T2_6}, the result is shown for $Pb\!-\!Pb$ collisions at the SPS,
from impact parameter $b=11$ fm to $b=0$ fm. The difference between
this curve and the one for a uniform medium thus reflects the surviving
\J's produced in the less dense outer regions.

\par

In Fig.\ \ref{T2_7}, similar calculations are shown for $Cu\!-Cu\!$
and $Pb\!-\!Pb$ at RHIC energy. Here we note in particular that
$Pb\!-\!Pb$ collisions are for all meaningful centralities ($b \leq 11$
fm) above both the \X~and the direct \J~threshold, so that we get a
smooth anomalous suppression increasing from about 60\% at $b=11$ fm
to about 90\% at $b=0$. Combining this suppression with the `normal'
pre-resonance absorption in nuclear matter, we thus predict for central
$Pb\!-\!Pb$ collisions at RHIC a \J~production rate of less than 5\% the
corresponding unsuppressed rate (excluding possible $B$ decay
contributions). It should also be noted that this result is based on
twice the number of gluons per wounded nucleon at RHIC, compared to the
SPS value. A larger increase, based on a possible larger
hadron multiplicity at RHIC, would lead to more \J~suppression.
Similarly, we assume an average number $(dN_g/dy)_{y=0}$ of gluons
per wounded nucleon. It is conceivable that very central collisions
reach into the tail part of the multiplicity distribution, with a larger
number of hadrons and hence also gluons. This would lead to a larger
suppression for very central collisions. At the SPS, the NA50
collaboration can check if the basis for this exists. By combining
measurements of hadron multiplicity, transverse energy $E_T$ and forward
energy $E_{\rm ZDC}$, it is possible to study the number of hadrons per
wounded nucleon as function of $E_T$ and check if there is an increase
at highest $E_T$ values.

\par

\begin{figure}[htb]
\vspace*{-0mm}
\centerline{\psfig{file=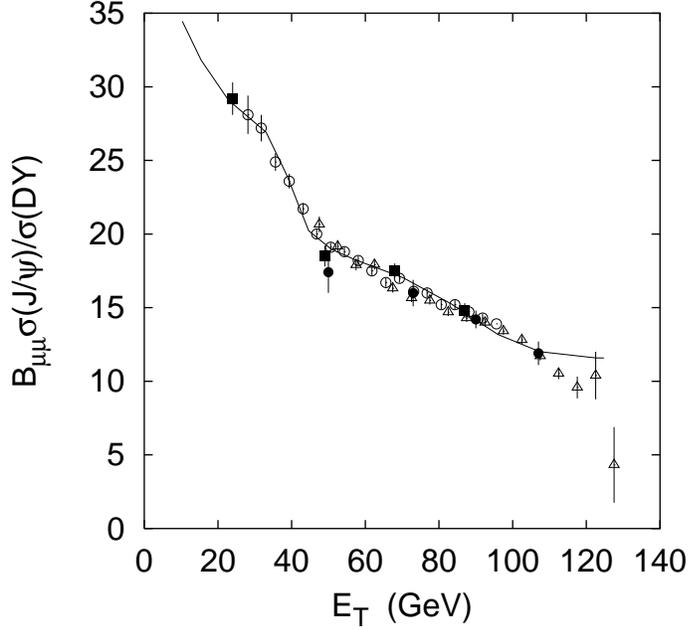,height= 90mm}}
%\hspace{2.5cm}}
%\vspace*{-20mm}cp
\caption{\J~production in $Pb-Pb$ collisions
\protect\cite{Kluberg,Cicalo},
compared to colour deconfinement in the parton percolation model
\protect\cite{Nardi&S}.}
\label{T2_9}
\end{figure}

Finally we can then turn to a comparison of our results on parton
percolation to the $E_T$-dependence of the actual data, which contain
an additional smearing due to the fact that a given $E_T$ bin
corresponds to a range of impact parameters and hence parton densities.
Including this effect in the standard way \cite{KLNS}, we obtain the
result shown in Fig.\ \ref{T2_9}. Included in this figure are the 1997
and 1998 data \cite{Cicalo,Kluberg} with minimum bias
determined Drell-Yan reference. While reproducing the overall
trend, our results clearly show deviations from the data in detail. As
mentioned, the levelling-off of our curve at high $E_T$ would have to be
modified if an increase in the multiplicity per wounded nucleon
at high $E_T$ should be observed. Another possible modification could
enter through a density-dependent charmonium dissociation. We have here
assumed that all \X's melt once the critical density is reached.
Allowing a partial survival chance, which decreases with increasing
density, would lead to a steeper drop of the suppression with $n_{\rm
parton}$ as well as with $E_T$. One possible source for such an effect
could be the finite life-time of the deconfining medium in actual
nuclear collisions.

\par

\par

Summarizing this section, we note that parton percolation provides a
consistent framework to study the onset of deconfinement as well as the
onset of charmonium suppression as a signature of the transition. It is
clear that the compatibility of the parton basis with other, soft hadron
production processes has to be checked; we return to this in section 6.

\bigskip

\noindent{\bf 5.4 $P_T$ Dependence}

\bigskip

In this section, we want to consider the dependence of anomalous
\J~suppression on the transverse momentum of the \J. It is well known
that the transverse momenta of secondaries from hadron-nucleus
collisions quite generally show a $p_T$-broadening. This means that the
universal parameter $\lambda \simeq 0.15$ GeV in Eq.\ (\ref{3.5}) now
depends on the size of the target nucleus. For secondary hadrons, this
is the Cronin effect \cite{Cronin}; a similar behaviour is observed
also in Drell-Yan and charmonium production \cite{NA3}. The natural
basis for all such broadening is initial state parton scattering, and
it was in fact shown some time ago \cite{G-G} - \cite{Zotov},
\cite{Gupta2} that this
describes quite well the $p_T$-dependence observed in \J~production
from $p\!-\!A$ to central $S\!-\!U$ collisions.

\par

Consider \J~production in $p-\!A$ collisions, assuming gluon fusion as
the dominant process for the creation of a $\C$ pair. Parametrizing
the intrinsic transverse momentum distribution $f(q_T)$ of a gluon in a
nucleon as
\be
f(q_T)= {1\over \pi \langle q_T^2 \rangle}~ \exp\left\{-{q_T^2\over
\langle q_T^2 \rangle}\right\}, \label{5.20}
\ee
we obtain by convolution for the transverse momentum distribution
$F_{pA}(P_T)$ of the resulting \J
\be
F_{pA}(P_T)= {1\over \pi \langle P_T^2 \rangle_{pA} }~
\exp\left\{-{P_T^2\over \langle P_T^2 \rangle_{pA}}\right\},
\label{5.21}
\ee
with
\be
\langle P_T^2 \rangle_{pA} = \langle q_T^2 \rangle_A  +
\langle q_T^2 \rangle_p. \label{5.22}
\ee
The quantity
\be
\delta_{pA} \equiv \langle P_T^2 \rangle_{pA} - \langle P_T^2
\rangle_{pp} =
\langle q_T^2 \rangle_A - \langle q_T^2 \rangle_p \label{5.23}
\ee
is thus a suitable measure for the observed nuclear broadening.
Note that the Gaussian form of Eq.\ (\ref{5.23}) is obtained from Eq.\
(\ref{3.5}) for $P_T^2 << m_{\j}^2$. For \J~production in $pp$
collisions, with the universal $\lambda^{-1} \simeq 0.15$, this leads
to $\langle P_T^2 \rangle \simeq 2m_{\j}/\lambda \simeq 1$ GeV$^2$,
which is slightly lower than the observed value of 1.23$\pm$ 0.05
\cite{NA3}.

\par

Assume now that in the passage of the projectile proton through the
nuclear target, successive interactions broaden the intrinsic momentum
distribution of the corresponding projectile gluon which will eventually
fuse with a target gluon to form a \J. If the process
of $P_T$ broadening during the passage is a random walk, then the
relevant parameter of the Gaussian distribution (\ref{5.21}) becomes
\be
\delta_{pA} = N_c^A \delta_0, \label{5.24}
\ee
where $N_c^A$ is the average number of collisions the projectile
undergoes on its passage through the target up to the fusion point,
and $\delta_0$ the average broadening of the intrinsic gluon
distribution per collision.

\par

In nucleus-nucleus collisions, a corresponding broadening occurs for
both target and projectile gluon distributions; here, however,
measurements at fixed transverse hadronic energy $E_T$ can determine the
broadening for collisions at a given centrality. Hence at fixed impact
parameter $b$ we have
\be
\delta_{AB}(b) = \langle P_T^2 \rangle_{AB}(b) - \langle
P_T^2\rangle_{pp} = N_c^{AB}(b)~ \delta_0, \label{5.25}
\ee
with $N_c^{AB}(b)$ denoting the average number of collisions for
projectile nucleons in the target and vice versa, at fixed $b$.
$N_c^{AB}(b)$ has a maximum at small $b$ and then decreases with
increasing $b$; for a hard sphere nuclear model, it would vanish when
$b=R_A+R_B$.

\par

In Glauber theory, the quantity $N_c^{AB}(b)/\sigma$ can be calculated
parameter-free from the established nuclear distributions
\cite{W-S};
here $\sigma$ denotes the cross section for the interaction of the
nucleon on its passage through the target. We shall determine $\sigma
\delta_0$ from data, so that $\sigma$ never enters explicitly. Once
$\sigma \delta_0$ is fixed, the broadening by initial state
parton scattering is given for all $p-\!A$ and $A\!-\!B$ interactions.
For Drell-Yan
production (with quarks instead of gluons in the partonic interaction),
this would be the observed effect, since the final state
virtual photon does not undergo any further (strong) interactions.
A produced nascent \J~will, however, experience pre-resonance nuclear
absorption; this suppresses \J's produced early along the path of the
projectile, since they traverse more nuclear matter and hence
are absorbed more than those produced later. As a net result, this
shifts the effective production point to a later stage. In $p-\!A$
collisions, a Drell-Yan pair will on the average be produced in the
center of the target. In contrast,
nuclear absorption shifts the average $\C$ production point further
down-stream. This effectively lengthens the path for initial state parton
scattering and hence increases the resulting broadening.

\par

The transverse momentum behaviour of normal \J~production in nuclear
collisions is thus a combination of initial state parton scattering
before the production of the basic $\C$ state, and pre-resonance
nuclear absorption afterwards; both lead to a broadening of a
$P_T$-distribution. A further broadening could come from elastic random
walk scattering of the charmonium state itself in nuclear matter;
however, the effect of this can be taken into account by fitting
$\sigma \delta_0$ to the data. The essential task is then to calculate
the number of collisions per cross section, $N_c/\sigma$, for $p-\!A$
and $A\!-\!B$ interactions, taking into account the effect of
pre-resonance nuclear absorption. The result contains as only open
parameter the quantity $\sigma \delta$; for details, see \cite{KNS1}.

\par

\begin{figure}[tbp]
\vspace*{-0mm}
\centerline{\psfig{file=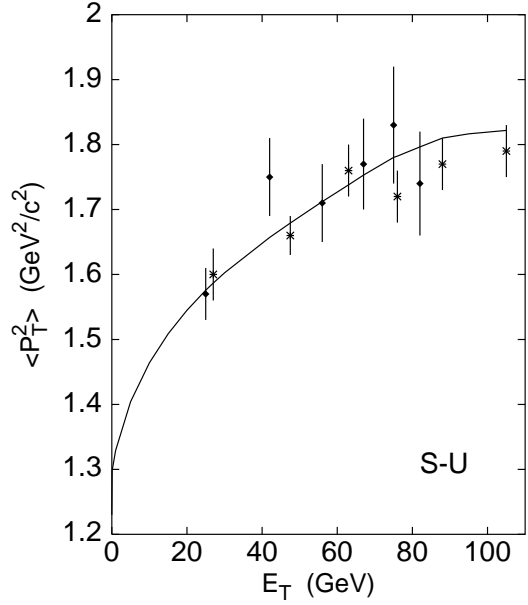,height= 80mm}}
%\vspace*{-20mm}
\caption{$P_T$ in $S-U$ collisions, compared to NA38 data \cite{NA38pt}.}
\label{P_1}
\end{figure}

Ideally, one would use $p-\!A$ data to fix $\sigma \delta_0$; the
broadening for $A\!-\!B$ interactions would then be fully predicted.
Unfortunately there are $p-\!A$ data only for three values of $A$
\cite{NA3,NA38pA}, and these have rather large errors. It is therefore
better to attempt a consistent description of all existing $p-\!A$
\cite{NA3,NA38pt} and $A\!-\!B$ data \cite{NA38pt}, up to central
$S\!-\!U$, in terms of a common $\sigma \delta_0$. Using $\langle P_T^2
\rangle_{pp} = 1.23 \pm 0.05$ GeV$^2$, a best fit to $p-\!A$,
$O\!-\!Cu$, $O\!-\!U$ and $S\!-\!U$ data \cite{KNS1} gives
$\sigma\delta_0 = 9.4 \pm 0.7$, with a $\chi^2/d.f.$ of 1.1. In Fig.\
\ref{P_1} we show the resulting description of the $S-U$ data.

\par

\begin{figure}[tbp]
\vspace*{-0mm}
\centerline{\psfig{file=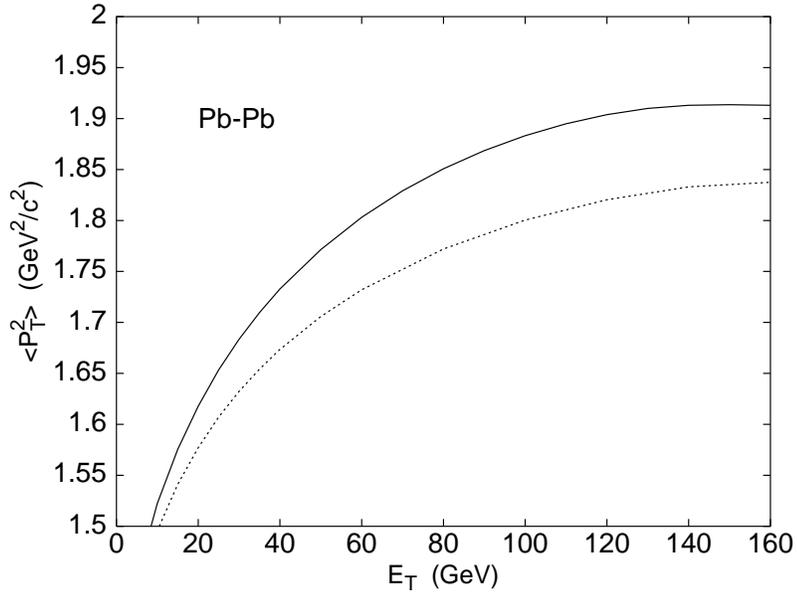,height= 80mm}}
%\vspace*{-20mm}
\caption{Normal $P_T$ broadening in $Pb-Pb$ collisions with (solid line)
and without (dashed line) pre-resonance nuclear absorption.}
\label{P_2}
\end{figure}

We then turn to $Pb\!-\!Pb$ collisions; the corresponding ``normal"
transverse momentum behaviour is shown in Fig.\ \ref{P_2}. Also shown
here is the amount of broadening obtained from initial state parton
scattering alone; we see that nuclear absorption increases
$\langle P_T^2 \rangle_{PbPb}$ by about 15 \%. Since the functional
forms in the two cases are quite similar, nuclear absorption can be
simulated by choosing a somewhat larger $\delta_0$. The basic
feature of normal absorption thus remains the monotonic increase of
$\langle P_T^2 \rangle$ with $E_T$, even though the collision geometry
makes this slightly weaker for $Pb\!-\!Pb$ than for $S\!-\!U$
interactions \cite{G-Vt}.

\par
\begin{figure}[tbp]
\vspace*{-0mm}
\centerline{\psfig{file=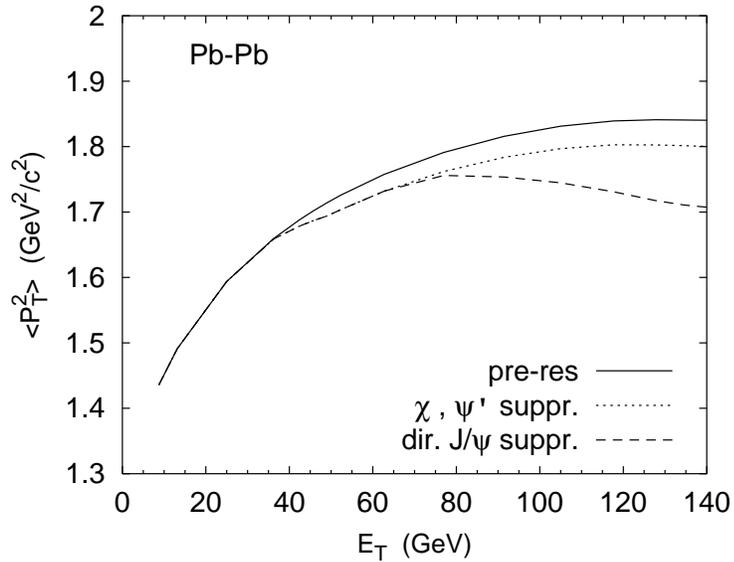,height= 80mm}}
%\vspace*{-20mm}
\caption{$P_T$ broadening in $Pb-Pb$ collisions. The solid line shows normal
broadening, the dotted line the effect of anomalous \X~and
\P~suppression, and the dashed line the effect the onset of direct
\J~suppression.}
\label{P_3}
\end{figure}

The onset of anomalous suppression results in a striking modification of
this pattern. If the \J's in the hot interior of the medium produced in
the collision are suppressed, then this will reduce their contribution
from the part of phase space leading to the most broadening. To
illustrate the effect, we assume suppression by deconfinement; in this
case the result is readily calculable \cite{Gupta2,B-O}. To be specific,
we assume deconfinement to start once the number of collisions per
wounded nucleon exceeds a certain critical value $\kappa_c$
\cite{KLNS}
\be
{N_c(b,s) \over N_w(b,s)} \geq \kappa_c. \label{5.26}
\ee
This defines for each collision configuration cool outer and hot inner
regions, resulting in charmonium survival or dissociation, respectively.
In Fig.\ \ref{P_3} we compare the resulting patterns for several
value of $\kappa_c$ to that of normal absorption; the change of
behaviour is qualitatively evident and due solely to anomalous
\J~suppression.
Moreover, abrupt onsets of anomalous \X~and direct \J~suppression
would also show up here; they have not yet been included in Fig.\
\ref{P_3}. The observation of such anomalous transverse momentum
behaviour for \J~suppression would thus constitute another tool to
study the onset of deconfinement in $Pb-Pb$ collisions.

\bigskip~\medskip

\noindent{\bf 6.\ SOFT HADRONIC PROBES}

\bigskip

In this final section, we want to address the possible supporting
evidence which can be obtained by measuring soft hadronic probes, i.e.,
mesons and baryons made up of the light $u,d$ and $s$ quarks and
produced at low transverse momenta. Since there is no large intrinsic
mass or momentum scale, their production occurs over comparatively
large space-time regions determined by $\Lambda_{\rm QCD}$ and cannot
be described perturbatively. The large intrinsic size of soft hadrons
requires relatively low energy densities to make their existence
conceptually meaningful: five mesons need a volume of at least roughly
five meson volumes in order to exist \cite{Pomeranchuk}. As discussed
in section 4.1, this implies that they are formed late in the
collision evolution, when the high initial energy density has dropped to
a sufficiently low value $\e_h~ \lsim~ 0.3 - 0.5$ GeV/fm$^3$, taking the
energy density of a typical hadron as indicator.

\par

What features of the produced medium can one then study with the help of
soft hadronic probes? It is clear, as also noted in section 4.1,
that they provide direct information about the state of the medium at the
energy density $\e_h$. In addition, one may hope to obtain
`retrospectively' also information about earlier stages at higher $\e$.
In the following,
we shall first consider two specific instances of such retrospective
probing and then discuss the study of the medium at hadronization.

\bigskip

\noindent{\bf 6.1 Retrospective Probes}

\bigskip

The onset of anomalous \J~suppression occurs at a certain associated
transverse energy of secondary hadrons. We would like to translate this
into a value of the corresponding primordial energy density. In principle,
this can be done in the same way as was used to derive Eq.\ (\ref{3.9})
in section 3.2. It is clear, however, that the result will be model
dependent. Instead of the assumed free flow leading to Eq.\
({\ref{3.9}), one could assume an isentropic flow \cite{H&K}, leading to
\be
s_0 = {1\over \pi R_A^2 \tau_0}\left( {dN_h \over dy}\right)_{y=0}^{AA}
\label{6.1}
\ee
as initial entropy density. It turns out that with Eq.\ (\ref{3.8a})
for the multiplicity and for an ideal gas of massless partons, the two
different expansion patterns result in very similar initial energy
densities up to ${\sqrt s} \simeq 200$ GeV. At much higher ${\sqrt s}$,
isentropic expansion leads to a somewhat higher energy density, but
even at LHC energy, the difference is only about 20 \% \cite{HS-Aachen}.

\par

To obtain an estimate of the energy densities in the region of anomalous
\J~suppression, we thus make use of Eq.\ (\ref{3.9}). For the 5\% most
central $Pb-Pb$ collisions, one finds on the average about 800
secondaries at central rapidity; their average energy is about 0.5 GeV
\cite{Stock}. This
leads to an initial energy density of about 2.8 GeV/fm$^3$. Note
that this value corresponds to $\e$ averaged over the collision profile;
the `hot' central value, associated to the region of highest density of
participating nucleons, will be around 30\% higher, around 3.7
GeV/fm$^3$ \cite{KLNS}. This is then the value of $\e$ at the point of
maximum \J~suppression in the SPS data. To study in detail the onset of
deconfinement, the multiplicity of soft secondaries must be known as
function of centrality. Until a detailed study of this is available,
we can estimate the result by assuming the $\e(b)$ (and hence
$\e(E_T$)) is approximately linear in $E_T$. This leads to $\e_c \simeq
2$ GeV/fm$^3$ at the onset of anomalous \J~suppression, which
presumably occurs at the onset of \X~dissociation and hence at
deconfinement. Such a critical energy density is in reasonable
agreement with the values predicted in section 2.

\begin{figure}[htb]
\vspace*{-0mm}
\centerline{\psfig{file=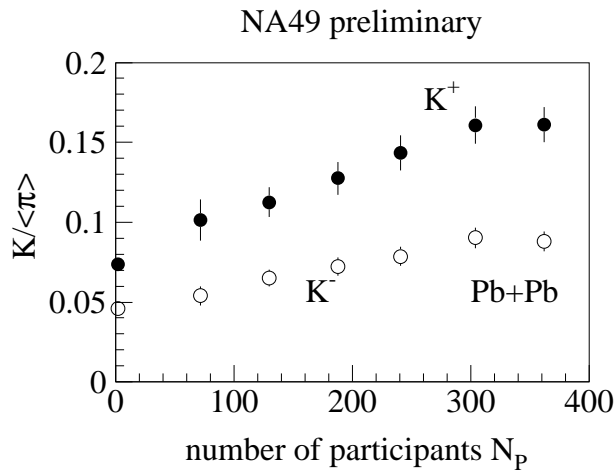,height= 65mm}}
%\vspace*{-20mm}
\caption{The $K/\pi$ ratio in $Pb-Pb$ collisions as function of the
number of participating nucleons \cite{Hoehne}.}
\label{6_1}
\end{figure}

A second and particularly interesting retrospective probe could well be
the strangeness abundance in nuclear collisions \cite{Raf-old}. We had
seen in section 3.1 that the relative abundances of the secondary
species in $p-p$ collisions were determined by the freeze-out
temperature $T_h$ of a hadronic resonance gas, with the restriction that
strangeness production was suppressed by a factor $\gamma_s$.
In nuclear collisions, the non-vanishing baryon number density
of the system requires that a corresponding analysis specifies
temperature $T_h$ and baryochemical potential $\mu_h$ at freeze-out,
and in addition a
possible strangeness suppression factor $\gamma_s$. In central
nuclear collisions, strangeness production is found to be much more
abundant than in $p-p$ collisions at the same $\sqrt s$. As
illustration, we show in Fig.\ \ref{6_1} the $K/\pi$ ratio in
$Pb-Pb$ collisions at the SPS as function of the number of
participating nucleons \cite{Hoehne,NA49-Pb}. If we assume, as discussed
in section 3.1, that the suppression factor $\gamma_s$ is determined by
the relative formation rate of the heavier strange quarks in an earlier
partonic medium of higher temperature $T > T_h$, then $\gamma_s(T)$
could provide a thermometer for this early temperature $T$. The
behaviour seen in Fig.\ \ref{6_1} would thus indicate a
significant increase with centrality for the initial temperature of the
partonic medium in $Pb-Pb$ collisions at $\sqrt s = 20$ GeV.
Comparing Fig.\ \ref{6_1} with Fig.\ \ref{3_1}, we see that
the strangeness increase with centrality in nuclear collisions is
very similar to that found as function of $\sqrt s$ or as function of
multiplicity at fixed $\sqrt s$ in the $p-p/p-{\bar p}$ interactions.
This supports an increase of the initial energy density as the
underlying mechanism for increased strangeness production; but
further comparative studies seem necessary to fully specify the origin
of the effect. If the model for $\gamma_s(T)$ discussed here is correct,
we would expect that with increasing collision energy, at RHIC and LHC,
$\gamma_s \to 1$. We also note that in accord with the interpretation of
$\gamma_s$ as thermometer, there is so far no indication of a threshold
or sudden increase in the relative strangeness production.

\par

\begin{figure}[htb]
\vspace*{6mm}
\centerline{\psfig{file=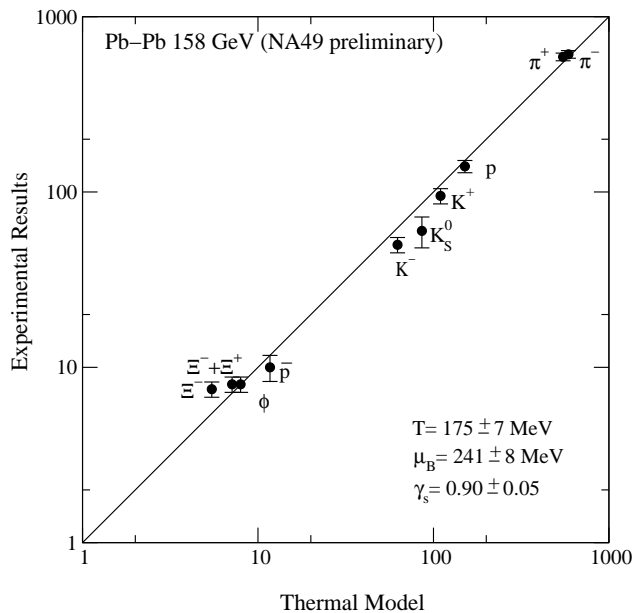,width=100mm}}
\vspace*{1mm}
\caption{Experimental hadron production multiplicities compared to those
in a resonance gas model \cite{Cley}.}
\label{6_1a}
\end{figure}

\bigskip

\noindent {\bf 6.2 Hadronisation and Freeze-Out}

\bigskip

After strongly interacting matter has cooled off to reach the
hadronization point, its freeze-out into free hadrons should be a
universal phenomenon, depending only on the freeze-out temperature and
baryon density. In view of the expansion of the produced medium, this
universality can only be `local', e.g., volume elements moving relative
to each other can lead to different effective temperatures. In general
terms, such an expansion scenario leads intuitively to three specific
predictions \cite{Heinz-Tsukuba}

\begin{itemize}
\vspace*{-0.2cm}
\item{relative hadron abundances specified by a equilibrated hadronic
resonance gas, most likely with $\gamma_s \to 1$,}
\vspace*{-0.2cm}
\item{a mass-dependent transverse momentum broadening of hadron spectra
due to radial hydrodynamic flow,}
\vspace*{-0.2cm}
\item{a source size for hadron emission which increases with
increased expansion due to increased initial energy density.}
\vspace*{-0.2cm}
\end{itemize}
All three effects are indeed observed.

For the relative abundances of the soft hadron species, there are
indications for the validity of a resonance gas distribution in $A-A$
collisions \cite{Stock} - \cite{Cley}. In Fig.\ \ref{6_1a},
this is illustrated for $Pb-Pb$ collisions at the SPS; we note that now
$\gamma_s$ is indeed quite close to unity. The latest
analysis shows some open questions concerning the universality of
the resulting values of $T_h,~\mu_h$ and $\gamma_s$ \cite{Cley}.
We note here that all studies so far have assumed that for a given
collision configuration, such a `hadrochemical' equilibrium exists
globally over the entire volume of produced matter. In view of the
collision profile of participating nuclei, howver, it seems quite
possible that such an equilibrium exists only locally, making the
triplet of values $T_h$, $\mu_h$ and $\gamma_s$ dependent on the
distance $r$ from the collision axis. Since most expansion models
start from such local equilibrium, it seems natural that species
abundances would also be affected by it.

\par

In Fig.\ \ref{6_1b}, we see that there is indeed a significant
(and mass-dependent) broadening in the transverse mass distribution
of hadronic secondaries from $Pb-Pb$ collisions at the SPS \cite{Stock}.
Clearly it would be interesting to study the kinematic counterpart of a
resonance gas freeze-out and check if the momentum distributions of
secondaries become thermal and if the expanding system leads to the
expected hydrodynamic flow \cite{Heinz-Tsukuba}. Such an endeavor is
confronted, however, by a specific problem. Besides collective final
state
effects, there are also initial state effects such as multiple state
parton scattering which can and will modify the transverse momentum
distribution Eq.\ (\ref{3.9}) of secondaries \cite{Leonidov}. Hence it
is not so easy to separate the possible origins of the observed
transverse momentum broadening in nuclear collisions \cite{Stock} and
to associate them to specific aspects of the produced state.
Nevertheless, considerable effort has been devoted to this aspect
\cite{Heinz-Tsukuba}. For the purpose of our primary question, colour
deconfinement, we note it does not seem possible to identify the initial
state of the produced medium from the hydrodynamic expansion pattern
alone \cite{Dinesh}.

\par

Because of expansion and momentum-dependent effects, the source size
situation is somewhat more complex, although an increase with incident
collision energy has been observed \cite{Stock-92}; for recent
discussions, see \cite{Heinz-Tsukuba,Stock}.

\par

All three phenomena listed above are present already in $p-p/p-{\bar
p}$ collisions for either increasing collision energy or increasing
hadron multiplicity. For hadron abundances, this was shown in
Fig.\ \ref{3_1}; transverse momentum broadening in hadronic collisions
is illustrated in Fig.\ \ref{6_2}, and the the growth of the
source size with multiplicity is shown in \cite{Giacomelli}. Evidently
this supports strongly the idea that increased initial energy density
is the common origin of the three effects.  However, to identify the
different underlying mechanisms quantitatively, it seems necessary carry
out a comprehensive systematic comparison of $p-p$, $p-A$ and $A-A$
collisions, which so far is lacking. As a result, a full understanding
of the properties of strongly interacting matter at freeze-out and
their origin seems to require considerable further experimental as well
as theoretical studies. Microscopic models based on partonic
interactions may well provide a suitable tool to formulate a consistent
picture.

\begin{figure}[htb]
\vspace*{-5mm}
\centerline{\psfig{file=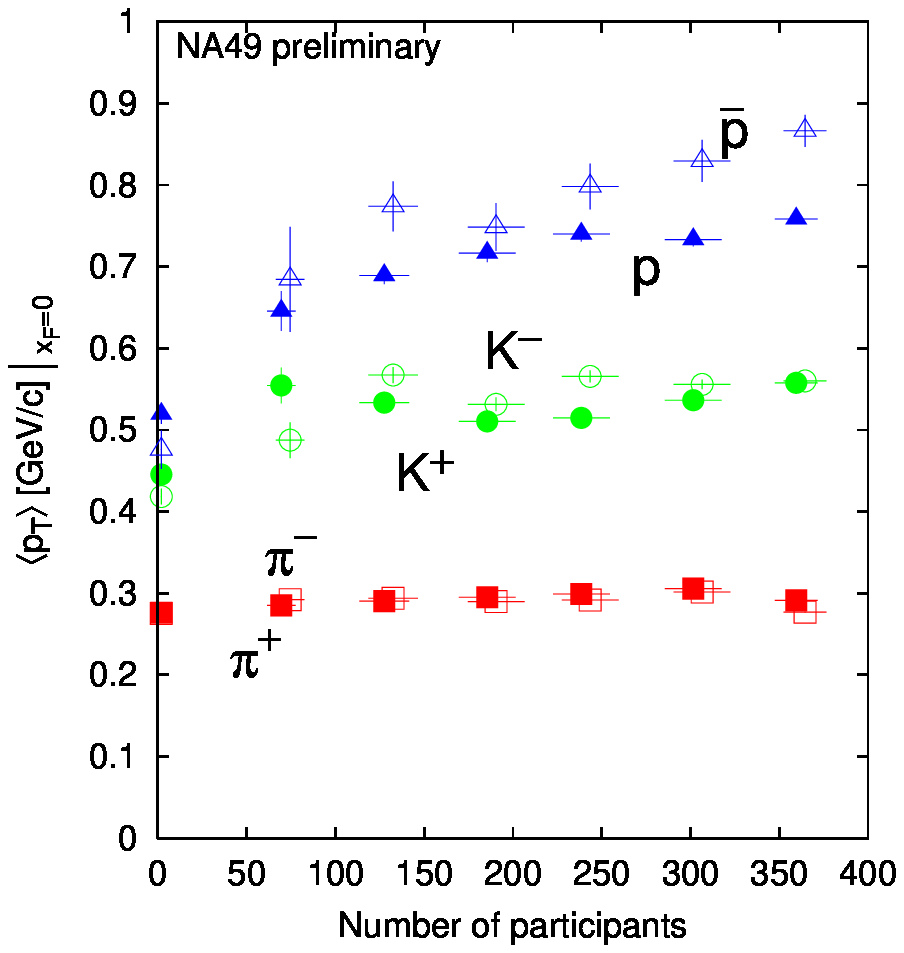,height=65mm}}
%\vspace*{-5mm}
%\vspace{32mm}
\caption{The average transverse momentum
of different hadronic secondaries in $Pb-Pb$ interactions at the SPS
\cite{Stock}.}
\label{6_1b}
%\vspace{40mm}
%\end{figure}
%\begin{figure}[htb]
%\vspace{-20mm}
\centerline{\psfig{file=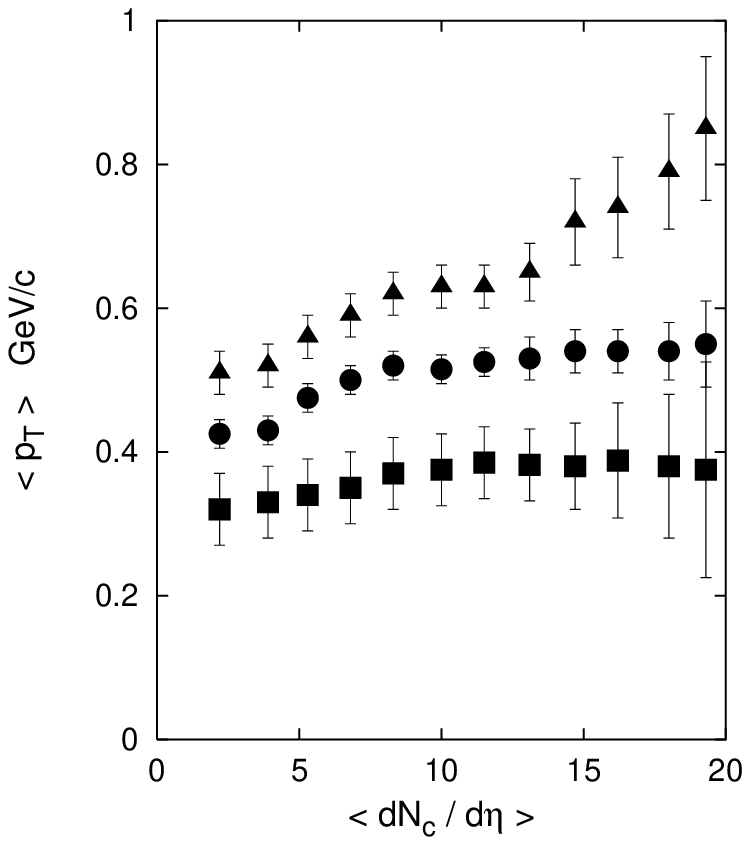,height=6cm,width=6cm}}%\hspace{12cm}}
\vspace*{-5mm}
\caption{The average transverse momenta of antiprotons, kaons and pions
produced in $p-{\bar p}$ collisions at $\sqrt s$ = 1.8 TeV and central
rapidity as function of the associated charged multiplicity
\cite{Alexo}.} 
\label{6_2}
\end{figure}

\bigskip

\noindent{\bf 6.3 Parton Cascade Models}

\bigskip

For systems evolving in thermal equilibrium, later stages carry no
information of the early history. In case of a non-equilibrium
evolution, such information can be passed on to a certain extent.
A study of the onset of deconfinement in a microscopic, non-equilibrium
space-time picture would therefore be of great help in understanding if
and to what extent the formation and the properties of the quark-gluon
plasma are related to experiments observing the different evolution
stages. The aim of partonic cascade models
\cite{cascade,Wang} is to describe through perturbative partonic
interactions embedded in a relativistic transport theory the evolution
of a high energy nucleus-nucleus collision. Hence such models seem to
be a good starting point for a dynamic study of deconfinement and the
subsequent evolution of the produced systems.

\par

We had seen above in section 5.3 that the initial state of two colliding
nucleons can be pictured as colliding beams of confined partons. The
incident partons have a distribution in intrinsic transverse momentum,
leading to some average value $\langle k_T \rangle$, which in turn
defines an average transverse parton size $r_T= 1/\langle k_T \rangle$.
When two nuclei collide at high energy, the transverse parton density
increases considerably, since now we have a superposition of
nucleon-nucleon collisions. In this medium, colour screening will
destroy the association of partons to particular hadrons,
since for sufficiently high density of colour charges, the colour
screening radius becomes much smaller than the typical hadronic scale.
Hence we expect the onset of deconfinement for some characteristic
density or for an equivalent screening scale. In perturbative QCD, the
method to calculate the colour screening mass $\mu$ (the inverse of
the screening radius $r_D$) is well-known; an extension to the
non-equilibrium medium provided by cascade models has also been
proposed \cite{BMW} and allows a calculation of the time evolution of
the screening mass $\mu(t)$ \cite{D&S}.

\par

Given $\mu(t)$, we have to specify for what value deconfinement sets
in; parton cascade models themselves does not identify such a point.
They just introduce the primary collisions between the `confined'
partons of the incident nucleons, followed by successive interactions
between primary as well as produced partons and eventually by the
hadronisation of partons according to some given scheme. There are two
possible ways  to specify deconfinement in such a scheme. One can
obtain at any given time the transverse density profile of the parton
distribution, and a natural onset of deconfinement is the percolation
point of partonic discs in the transverse plane, as discussed in section
5.3b. A more global approach is to calculate the screening mass $\mu$
and identify as critical point $\mu_c$ the value obtained from lattice
studies. These provide the temperature dependence of $\mu$ and in
particular also its value at the deconfinement point $T_c$, where one
has $\mu_c \equiv \mu(T_c) \simeq 0.4 - 0.6$ GeV \cite{HKR}. It is
not obvious that such an equilibrium value can really be used in the
non-equilibrium situation provided by the parton cascade model. A good
check would be to compare in this model the transverse parton density
at $\mu_c$ to the percolation value.

\par

We now briefly recall the most important features of the parton cascade
model; to be specific, we consider the version proposed in \cite{cascade}.
\begin{itemize}
\vspace*{-0.2cm}
\item{The initial nucleus-nucleus system is treated as two colliding
clouds of partons, whose distribution is fixed by the nucleonic parton
distribution functions determined in deep inelastic lepton-nucleon
scattering, and by the nucleon density distribution in the nuclei.}
\vspace*{-0.2cm}
\item{The parton cascade development starts when the initial parton clouds
inter-penetrate and traces their space-time development due to
interactions. The model includes multiple elastic and inelastic
interactions described as sequences of $2\, \rightarrow \,2$
scatterings, $1\, \rightarrow \, 2$ emissions, and $2 \, \rightarrow \,1$
fusions. It moreover explicitly accounts for the individual time scale
of each parton-parton collision, the formation time of the parton
radiation, the effective suppression of radiation from virtual partons
due to an enhanced absorption probability by others in regions of dense
phase space occupation, and the effect of soft gluon interference in
low energy gluon emission.}
\vspace*{-0.2cm}
\item{Finally, the hadronization in terms of a parton coalescence to
colour neutral clusters is described as a local statistical process
that depends on the spatial separation and colour of the
nearest-neighbour partons~\cite{eg}. These pre-hadronic clusters then
decay to form hadrons.}
\vspace*{-0.2cm}
\end{itemize}
This model can now be used to calculate characteristic features at
different stages of the evolution process. As an illustration we show
in Fig.\ \ref{6_3} the evolution of the colour screening mass as
function of the proper time $\tau$ of the system. It is seen that for
sufficiently early times, the hot interior of a central $Pb-Pb$
collision leads to $\mu(\tau) > \mu_c$ and thus to deconfinement
\cite{D&S}.
One can now follow the evolution of the model and study features of the
system beyond the confinement point, such as in-medium hadron mass
modifications or momentum distributions of soft secondaries
\cite{Dinesh-hadrons}.

\begin{figure}[htb]
\vspace*{2mm}
\centerline{\psfig{file=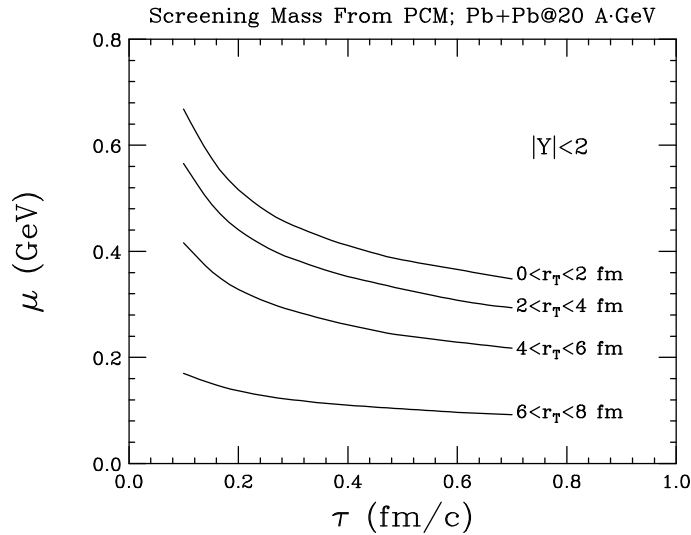,height= 70mm}}
%\vspace*{-20mm}
\caption{The time dependence of the colour screening mass in central
$Pb-Pb$ collisions at SPS energy in different radial collision zones
\cite{D&S}.}
\label{6_3}
\end{figure}

\par

It is clear that such a description is necessarily model-dependent.
Nevertheless, a coherent picture of this kind, accompanied by emprirical
cross-checks for the different evolution stages, provides an excellent
visualisation of the expansion process and could thus lead to further
predictions. It also allows an investigation of problems, such as the
comparison to $p-p$ and $p-A$ collisions or the onset of thermalisation,
for which so far no other tools exist.

\bigskip

\noindent{\bf 7.\ SUMMARY}

\bigskip

We summarize here what we take to be the main conclusions reached during
the past two decades in the study of colour deconfinement and its
manifestation in high energy nuclear collisions.

\begin{itemize}
\vspace*{-0.2cm}
\item{QCD, in particular through finite temperature lattice studies,
predicts the onset of colour deconfinement (quark-gluon plasma
formation) for strongly interacting matter of vanishing baryon number,
once energy densities in the range 1 - 3 GeV are reached.}
\vspace*{-0.2cm}
\item{High energy nuclear collisions produce in their early stages
sufficiently high energy densities for deconfinement; however, the
produced systems expand and cool rapidly. Hence the crucial problem is
the specification of probes which can unambiguously test the
confinement/deconfinement status of the produced early medium.}
\vspace*{-0.2cm}
\item{For a direct test, the high density, short distance nature of the
quark-gluon plasma requires hard probes which can distinguish between
confined and deconfined media. Two such probes are: \par
-- quarkonium dissociation, which for fully formed physical
resonances can only occur in a deconfined medium, and \par
-- jet quenching, manifested by a much higher energy loss of hard
transverse jets in a quark-gluon plasma than in hadronic matter.}
\vspace*{-0.2cm}
\item{Charmonium suppression provides a particularly well-studied
example, for which today there exist both an established theoretical
framework and quite conclusive experimental results from nuclear
collisions.\par
-- The hadroproduction of charmonium is understood as
perturbative $\C$ formation by interaction of partons whose
distribution functions are determined in deep inelastic scattering;
the colour neutralization of the $\C$ pair can be described
quantitatively in terms of colour evaporation.\par
-- Charmonium production in a confined medium is studied in $p-A$
collisions, and the resulting $A$-dependent decrease in production rate
is understood in terms of pre-resonance absorption in standard nuclear
matter.\par
-- Since about 60\% of the observed \J's are directly produced, while
the rest comes from decay of higher $\C$ excitations (mainly \X),
the onset of deconfinement in nucleus-nucleus collisions should lead
to a suppression hierarchy. In a first step, essentially at the
deconfinement point, the higher excitations are suppressed. In a
second step at clearly higher energy density, the smaller direct \J's
are dissociated.\par
-- \J~production in nucleus-nucleus collisions up to central $S-U$
collisions shows only the normal pre-resonance absorption known from
$p-A$ collisions. While peripheral $Pb-Pb$ collisions also follow this
pattern, there is with increasing centrality an abrupt first onset of
some 30\% further `anomalous' suppression, followed by a later second
further drop of some 20\% in the observed \J~suppression. This step-wise
sudden onset of anomalous \J~suppression disagrees with all conventional
hadronic suppression models and is understandable only as the onset of
colour deconfinement.}
\vspace*{-0.2cm}
\item{The onset of deconfinement is supported by multiplicity data of
soft hadronic secondaries, which can be used to estimate the initial
energy densities of the collision. With values of up to 3.7 GeV/fm$^3$
in the inner region of central $Pb-Pb$ collisions, the deconfinement
range of energy densities is indeed attained.\par
The relative strangeness abundance could provide another check of the
initial energy density. If the strangeness suppression observed in
$p-p$ collisions of average associated multiplicity is due to the
relatively low energy density of the early medium, it can
be expected to end when in $A-A$ interactions the initial energy density
becomes high enough. There are indications that this is happening.}
\vspace*{-0.2cm}
\end{itemize}
\J~production experiments at the CERN-SPS \cite{NA50,NA50last} thus
provide first evidence for the predicted new state of deconfined quarks
and gluons. Future experiments at the RHIC in Brookhaven and the LHC at
CERN, making use of charmonium, bottonium and jet production, are
expected to extend these results and study the quark-gluon plasma in
more detail and up to much higher energy densities. Soft probes, in
particular the strangeness abundance, can serve to determine these
initial energy densities quantitatively and study their effects on the
final hadronic state of matter.

\bigskip~

\bigskip\noindent

{\bf ACKNOWLEDGEMENTS}

\bigskip

This report owes very much to stimulating discussions with numerous
colleagues, and I want to  express my sincere gratitude to all of them.
Some have contributed decisively to different aspects of the work
presented here, and so it is a particular pleasure to thank R.\ Baier,
P.\ Braun-Munzinger, J.\ Cleymans, M.\ Gonin, F.\ Karsch, D.\ Kharzeev,
L.\ Kluberg, C.\ Louren{\c c}o, M.\ Nardi, J.\ Schukraft, J.\ Stachel,
R.\ Stock, D.\ Srivastava and R.\ Vogt for help, challenge and
inspiration -- without implying that they agree with all points of my
interpretation of the state of the field. --  For different parts of
this work, financial support by DFG contract Ka 1198/4-1, BMBF contract
06BI804/5 and GSI contract BISATT is gratefully acknowledged.

\end{document}